\documentclass[11pt]{article}
\pdfoutput=1
\usepackage[top=2cm, bottom=2cm, left=2cm, right=2cm]{geometry}
\usepackage{jcappub}
\usepackage{float}
\usepackage{color}
\usepackage{graphicx}

\newcommand{\be}{\begin{equation}}
\newcommand{\ee}{\end{equation}}
\renewcommand{\d}{\mathrm{d}}
\newcommand{\gf}{\phi}
\newcommand{\gs}{\sigma}
\renewcommand{\ge}{\epsilon}
\newcommand{\getpa}{\eta^\parallel}		
\newcommand{\getpe}{\eta^\perp}
\newcommand{\gxpa}{\xi^\parallel}
\newcommand{\gxpe}{\xi^\perp}
\newcommand{\gc}{\chi}
\newcommand{\gw}{\tilde{W}}
\newcommand{\gU}{\tilde{U}}
\newcommand{\gV}{\tilde{V}}
\newcommand{\gv}{\bar{v}}
\newcommand{\gk}{\kappa}
\newcommand{\ga}{\alpha}
\newcommand{\gb}{\beta}

\newcommand{\lh}{\left(}
\newcommand{\rh}{\right)}
\newcommand{\half}{{\textstyle\frac{1}{2}}}

\newcommand{\fnl}{f_{\mathrm{NL}}}
\newcommand{\gint}{g_{\mathrm{int}}}
\newcommand{\ns}{n_{s}}
\newcommand{\siml}{\raisebox{-.1ex}{\renewcommand{\arraystretch}{0.3}
$\begin{array}{@{}c} \scriptstyle < \\ \scriptstyle \sim \end{array}$}}

\title{Non-Gaussianity in two-field inflation beyond the slow-roll approximation}
\author{Gabriel Jung}
\author{and Bartjan van Tent}
\affiliation{Laboratoire de Physique Th\'eorique (UMR 8627), CNRS, Univ. Paris-Sud,
Universit\'e Paris-Saclay, 91405 Orsay Cedex, France}
\emailAdd{gabriel.jung@th.u-psud.fr}
\emailAdd{bartjan.van-tent@th.u-psud.fr}

\abstract{We use the long-wavelength formalism to investigate the level of 
bispectral non-Gaussianity produced in two-field inflation models with standard
kinetic terms. Even though the Planck satellite has so far not detected any
primordial non-Gaussianity, it has tightened the constraints significantly,
and it is important to better understand what regions of inflation model space
have been ruled out, as well as prepare for the next generation of experiments
that might reach the important milestone of $\Delta\fnl^\mathrm{local}=1$.
We derive an alternative formulation of the previously derived integral 
expression for $\fnl$, which makes it easier to physically interpret the result
and see which types of potentials can produce large non-Gaussianity. We apply
this to the case of a sum potential and show that it is very difficult to
satisfy simultaneously the conditions for a large $\fnl$ and the observational
constraints on the spectral index $\ns$. In the case of the sum of two monomial
potentials and a constant we explicitly show in which small region of parameter 
space this is possible, and we show how to construct such a model. Finally, the
new general expression for $\fnl$ also allows us to prove that for the sum
potential the explicit expressions derived within the slow-roll approximation
remain valid even when the slow-roll approximation is broken during the turn
of the field trajectory (as long as only the $\ge$ slow-roll parameter remains 
small).}


\arxivnumber{}

\begin{document}

\maketitle
\flushbottom

\section{Introduction}

The theory of inflation \cite{Starobinsky:1980te, Guth:1980zm,
  Linde:1983gd} describes a period of rapid and accelerated expansion
which takes place in the very early universe. It solves several issues
of the pre-inflationary standard cosmology like the horizon and the
flatness problems. More remarkably, inflation also gives an
explanation for the origin of the primordial cosmological
perturbations which are the seeds of the large-scale structure in the
universe observed today.

The Cosmic Microwave Background radiation (CMB) is an almost direct window on
these primordial fluctuations and its temperature and polarization
anisotropies have been observed by several missions. The most recent
results come from the Planck satellite \cite{Ade:2015xua, Ade:2015lrj,
Ade:2015ava}, which, like its predecessors, found no disagreement
with the basic inflationary predictions: the
distribution of primordial density perturbations is almost but not
exactly scale-invariant and it is consistent with Gaussianity. The
main information is encoded in the power spectrum which is the Fourier
transform of the two-point correlation function of CMB
temperature/polarization fluctuations. The most interesting observable
from the point of view of inflation is the spectral index $\ns$ that
describes its slope, or in other words the deviation from exact
scale invariance.  

The Planck satellite also significantly improved the constraints on
any potential deviations from a Gaussian distribution (i.e.\ on
non-Gaussianity) \cite{Ade:2015ava}. Primordial non-Gaussianity is
generally parametrized by the amplitude parameters $\fnl$ of a number
of specific bispectrum shapes that are produced in generic classes of
inflation models. The bispectrum is the Fourier transform of the
three-point correlator and in the case of standard single-field
slow-roll inflation it is known to be unobservably small
\cite{Maldacena:2002vr}. However, this result does not hold in more
general situations and many extensions of that simple case have been
proposed with different predictions for non-Gaussianity, meaning that
observations can in principle be used to constrain them.\footnote{It has 
been pointed out \cite{Tanaka:2011aj, Pajer:2013ana} that the finite size of 
the observable universe leads to gauge corrections, which have to be taken 
into account to convert the inflationary bispectrum to actual observations. 
Indeed in single-field inflation the squeezed limit of the bispectrum vanishes 
identically for a local observer today. In multiple-field inflation, on the 
other hand, these corrections are also of order $1-\ns$ \cite{Tada:2016pmk} 
and hence are expected to be negligible in the case of large $\fnl$.}
For example, models with
higher derivative operators based on the Dirac-Born-Infeld action
\cite{Alishahiha:2004eh, Silverstein:2003hf, Mizuno:2009cv,
Mizuno:2010ag,Tzavara:2013wxa} can produce large non-Gausianity of the so-called
equilateral type. Another possibility is to consider multiple fields
during inflation, which adds isocurvature perturbations to the usual
adiabatic perturbation. The isocurvature perturbations can interact with
the adiabatic one on super-Hubble scales (while in single-field inflation the
adiabatic perturbation is constant on super-Hubble scales) which can lead
to so-called local non-Gaussianity. In this case non-Gaussianity can be 
generated long after inflation as in the curvaton scenario
\cite{Lyth:2001nq,Bartolo:2003jx,Kobayashi:2013bna,Enqvist:2013paa,
Byrnes:2014xua,Vennin:2015egh,Hardwick:2016whe}, or directly after inflation during (p)reheating
\cite{Zaldarriaga:2003my,Lyth:2005qk,Bernardeau:2004zz,Barnaby:2006km,
Enqvist:2004ey, Jokinen:2005by, Elliston:2014zea}. 
However, in this paper we will be interested in the case where this
local non-Gaussianity is produced on super-Hubble scales during
inflation. Since we will only talk about local non-Gaussianity in the
rest of this paper, $\fnl$ should always be understood as
$\fnl^\mathrm{local}$.  

A large amount of work has been done to study if observably large
non-Gaussianity can be produced during multiple-field inflation. This
involves studying the
large-scale evolution of the perturbations which can be done using different 
formalisms, the
$\delta N$ formalism \cite{Starobinsky:1986fxa, Sasaki:1995aw,
Lyth:2005fi} being the most popular but the long-wavelength
formalism \cite{Rigopoulos:2005xx, Rigopoulos:2005ae,
Rigopoulos:2005us, Tzavara:2010ge, Tzavara:2011hn, Tzavara:2013yca}
offering an interesting alternative. Many results have been obtained
for two fields, a number sufficient to highlight multiple-field
effects (some of them have then been generalized to more fields). In
the slow-roll approximation, the sum-separable \cite{Vernizzi:2006ve}
as well as the product-separable potential \cite{Choi:2007su} have
been solved analytically, while more general separable potentials have
been studied in \cite{Meyers:2010rg,Tzavara:2010ge}.  The solution
beyond slow-roll for Hubble-separable models was given in
\cite{Byrnes:2009qy, Battefeld:2009ym}. Different conditions for large
non-Gaussianity have been found \cite{Elliston:2011dr,Elliston:2011et}
depending on whether the isocurvature modes have vanished before the
end of inflation or not, the latter case requiring a proper treatment
of the reheating phase to be sure that the results actually persist
until the time of recombination and the CMB, which is generally not
done. The scale dependence of the bispectrum is also an important
topic of study of the last few years. Different aspects have been
studied, like the computation of the bispectrum in the squeezed limit,
the scale-dependence of $\fnl$ or the possible observational effects
\cite{Byrnes:2010ft,Byrnes:2012sc,Tzavara:2012qq,Kenton:2015lxa,
  Byrnes:2015dub,Kenton:2016abp}.
Another related subject that has received much attention in recent years
is the study of features in the effective inflaton potential or kinetic terms
(like changes in the sound speed for the inflaton interactions), possibly 
due to the presence of massive fields, which lead to correlated oscillations
in the power spectrum and the bispectrum \cite{Chluba:2015bqa,Achucarro:2010da,
Flauger:2016idt,Achucarro:2012fd,Hotchkiss:2009pj,Achucarro:2014msa}.
Two codes \cite{Dias:2016rjq, Mulryne:2016mzv} for numerical evaluation of 
the bispectrum have been recently released.

The aims of this paper are threefold. The first is a continuation of the
work on the long-wavelength formalism, in particular of \cite{Tzavara:2010ge}.
In that paper a completely general expression for the $\fnl$ produced 
in two-field inflation on super-Hubble scales was derived. However, this
expression involves an integral and two different time variables, which makes
it hard to fully understand its implications, and to see which types of 
potentials could give large non-Gaussianity. In this paper we derive
an alternative formulation of that expression and discuss its consequences 
for certain classes of potentials. Since Planck has excluded the possibility 
of large local non-Gaussianity (of order 10), the reader might wonder what
the interest is of looking for models with large non-Gaussianity. However,
it is very important in order to understand if Planck actually ruled out any 
significant parts of the multiple-field model space, or if these models generically
predict small non-Gaussianity. Moreover, with large non-Gaussianity in this 
paper we often mean an $\fnl$ of order 1, which has not yet been ruled out by 
Planck but which might be observable by the next generation of experiments.

The second aim is to understand if it is possible to have large
non-Gaussianity while staying within the slow-roll approximation. For
explicitness we assume a two-field sum potential (with standard
kinetic terms), where explicit analytical results within the slow-roll
approximation are possible (and have been derived before). In
particular this question was studied within the $\delta N$ formalism
by the authors of \cite{Elliston:2011dr, Elliston:2011et}, who
concluded that with enough fine-tuning an arbitrarily large $\fnl$ is
possible. However, apart from rederiving those results in another
formalism, the new ingredient here is that we take into account the
constraints from Planck on the other inflationary observables, in
particular $\ns$. And it turns out that satisfying the observational
constraints on $\ns$ while having a large $\fnl$ and staying within
the slow-roll approximation is very hard. In the case of a sum of two monomial
potentials and a constant we explicitly work out the region of the
parameter space (in terms of the powers of the two potentials) where
this is possible. Note that we assume everywhere that the isocurvature
mode has disappeared by the end of inflation. Otherwise it would be
easy to get large non-Gaussianity by ending inflation in the middle of
a turn of the field trajectory, but we feel that in that case the
results at the end of inflation would be meaningless, since they could
not be extrapolated to the time of recombination and the CMB without
properly treating the end of inflation and the consecutive period of
(p)reheating.

Finally, the third aim of the paper is to understand the, at first sight
very surprising, numerical observation that even in the case where the
slow-roll approximation is broken during the turn of the field
trajectory, the analytical slow-roll expression for $\fnl$ is often still
a very good approximation of the final exact result. It turns out that we
can understand this using the new formulation mentioned above. In that
formulation $\fnl$ is given by a differential equation and the solution
can be written as the sum of a homogeneous and a particular solution. As
we will show, the homogeneous solution can be given analytically in an exact 
form (without any need of the slow-roll approximation), while the particular
solution is negligible exactly in the regions where slow roll is broken and
we cannot compute it analytically.

This paper is organized as follows. Section~\ref{Definitions} defines
the slow-roll parameters and other quantities used in the rest of the paper. 
It also recalls some elements of the long-wavelength formalism, in particular 
the Green's functions used to solve the perturbation equations and some of their
properties, and the expressions for the different observables. This section is
also where we derive the new formulation mentioned above.
Section~\ref{Slow-roll} treats the slow-roll results mentioned in aim two 
above. It uses increasing levels of approximation. First, the slow-roll 
approximation is discussed. Then we add the hypothesis that the potential is
sum-separable to solve the Green's function equations and to obtain
simple expressions for the observables. Then they are applied to the specific 
class of monomial potentials, where the effects of the spectral index 
constraint on the region of the parameter space where $\fnl$ is large are computed.
In section~\ref{Beyond slow-roll}, we keep the sum-separable potential 
hypothesis to compute $\fnl$ beyond the slow-roll approximation. Two different 
types of generic field trajectories with a turn are discussed. We show that
in the end the slow-roll expression from the previous section also gives a
very good approximation of the exact result for $\fnl$ in this case.
Section~\ref{numerical section} contains several specific
examples to illustrate the different results of the paper. The method
to build a monomial potential that produces a large $\fnl$ while satisfying 
all constraints is detailed, 
while some examples from existing literature are also discussed. Each time 
we compare the exact numerical results in the 
long-wavelength formalism to the approximated analytic expressions derived
in this paper. Finally we conclude in section \ref{conclusion}, while some
additional details are treated in the appendices, including some results
about product potentials.

\section{Definitions and set-up}
\label{Definitions}

This section sets up the basic equations and definitions used in the rest
of the paper. Most of this section summarizes results derived in previous 
papers, but the final section~\ref{secgint} contains an important new result.

\subsection{Background dynamics}

The models we will consider are two-field inflation models with standard kinetic
terms and a potential $W(\gf,\gs)$ in the framework of general relativity. 
Here $\gf(t,\vec{x})$ and $\gs(t,\vec{x})$ denote the 
two fields, which we will often combine into the vector $\gf^A$ with $A=1$ for
$\gf$ and $A=2$ for $\gs$. Since we have standard kinetic terms (trivial field
metric), there is no difference between upper and lower field indices. 
For the moment we keep $W$ completely general,
although in the later sections we will often have to assume some specific
form of the potential in order to solve the equations.

As time coordinate $t$ we will use the number of e-folds $t \equiv \ln a$, 
where $a(t)$ is the scale factor of the universe, and we denote derivatives 
with respect to this time coordinate by overdots. The Hubble parameter of the 
universe is denoted by $H(t)$. (Unlike in the case of cosmic time, where
the expansion information of the universe is encoded in $a$ and $H$ is directly
derived from it, when using the number of e-folds as time coordinate, $a$ is
a trivial function, and the expansion information is encoded in $H$, which
can in this case not be derived from $a$.)

In terms of the number of e-folds the background field equation for $\gf^A$
and the Friedmann equation for $H$ take the following form:
\be
\ddot{\gf}^A + (3-\ge)\dot{\gf}^A + \frac{W_A}{H^2} = 0, \qquad\qquad
H^2 = \frac{\gk^2 W}{3-\ge}.
\label{fieldeq}
\ee
Here $\gk^2 \equiv 8\pi G$ and the index on $W$ denotes a derivative with 
respect to the fields: we define 
$W_{A_1\ldots A_n} \equiv \partial^n W/(\partial\gf^{A_1}\cdots\partial\gf^{A_n})$.
The quantity $\ge$ is a short-hand notation of which the physical interpretation 
will be discussed in the next section. It is defined as
\be
\ge \equiv - \frac{\dot{H}}{H} = \frac{\gk^2}{2} \lh \dot{\gf}^2 + \dot{\gs}^2 \rh
\label{defeps}
\ee
(where the second equality follows from the Friedmann equation for $\dot{H}$,
which we have not given explicitly here but which is easily deduced).

As we have a two-dimensional field space, we need a basis, and as usual we will
define the basis vectors with respect to the field trajectory 
\cite{GrootNibbelink:2000vx, GrootNibbelink:2001qt, Gordon:2000hv}:
\be
e_1^A = (e_{1\gf}, e_{1\gs}), \qquad
e_2^A = (e_{1\gs}, -e_{1\gf}), \qquad
e_{1\gf} = \frac{\dot{\gf}}{\sqrt{\dot{\gf}^2+\dot{\gs}^2}}, \qquad
e_{1\gs} = \frac{\dot{\gs}}{\sqrt{\dot{\gf}^2+\dot{\gs}^2}}.
\label{basis}
\ee
So the first basis vector is always along the field trajectory as it is
defined as the direction of the field velocity. The second basis vector
is perpendicular to the first, and since we have only two dimensions it can
be completely expressed in terms of the components of the first basis vector
(see appendix A of \cite{Tzavara:2010ge} for some refinements of this basis originally
introduced in \cite{GrootNibbelink:2000vx}).

For later use we will define the following quantities:
\be
\gw_{A_1\ldots A_n}=
\lh\frac{\sqrt{2\ge}}{\gk}\rh^{n-2}\frac{W_{A_1\ldots A_n}}{3H^2}, \qquad\qquad
\gw_{m_1\ldots m_n}=\gw_{A_1\ldots A_n}e_{m_1}^{A_1}\cdots e_{m_n}^{A_n},
\label{defW}
\ee
where the $m$ indices denote the components of the basis and the Einstein
summation convention is implied. In order to 
distinguish explicit components of these two different quantities, indices
1 and 2 will indicate components in the basis (\ref{basis}) (e.g.\ $\gw_{21}$),
while indices $\gf$ and $\gs$ will be used to indicate components in terms of
the original fields (e.g.\ $\gw_{\gs\gs}$).

\subsection{Slow-roll parameters}
\label{srparsec}

If the potential is almost flat and the field slowly rolls down,
certain terms in the equations can be neglected. To quantify this we
can introduce a set of slow-roll parameters. It is important to keep
in mind that the introduction of these parameters is not yet an
approximation: the equations are still completely exact and the
slow-roll parameters can be considered as just a short-hand
notation. It only becomes an approximation (the slow-roll
approximation) 
if we then say that some of these
parameters are small and start neglecting certain terms. We will do
that in certain later sections, but not here.

The first slow-roll parameter is $\ge$ defined in (\ref{defeps}). It will be
small if the kinetic energy of the fields is small compared to their potential 
energy. The other slow-roll parameters are vectors in field space and can be 
defined as follows with $n\geq 2$~\cite{GrootNibbelink:2001qt,Tzavara:2010ge}:
\be
\eta^{(n)A}\equiv\frac{1}{H^{n}\sqrt{\dot{\gf}^2+\dot{\gs}^2}}
\lh H\frac{\d}{\d t}\rh^{n-1} \lh H \dot{\gf}^A \rh.
\label{defeta}
\ee
As usual the most important ones are for $n=2$ (simply called $\eta^A$) and
$n=3$ (called $\xi^A$). For example, for $\eta^A$ the expression above becomes
\begin{equation}
\eta^{A} = \frac{1}{\sqrt{\dot{\gf}^2+\dot{\gs}^2}}
\lh \ddot{\gf}^A - \ge \dot{\gf}^A \rh.
\label{defeta_detail}
\end{equation}
We will usually consider the parallel and perpendicular components
of these as defined in the basis (\ref{basis}):
\begin{equation}
\getpa\equiv
\eta^{A}e_{1A},\qquad \getpe\equiv \eta^{A}e_{2A},\qquad \gxpa\equiv
\xi^{A}e_{1A},\qquad\gxpe\equiv \xi^{A}e_{2A}.
\end{equation}
The parameters $\getpa$ and $\getpe$ will be small if the components
of the field acceleration parallel and perpendicular to the field
velocity, respectively, are small compared to the field
velocity.\footnote{This remark is exact when acceleration in terms of
  cosmic time is considered. When using the number of e-folds as time
  coordinate, as we do here, there is a correction term as seen in
  (\ref{defeta_detail}). However, that correction disappears for
  $\getpe$.}  The parameter $\getpe$ is quite fundamental to anything
concerning multiple-field inflation: as long as it is negligible we
are in an effectively single-field situation, but as soon as it
becomes significant we have true multiple-field effects. This will be
illustrated quite clearly by the results of this paper. 

In the context of the slow-roll approximation, $\ge, \getpa, \getpe$ are
called first-order slow-roll parameters, while $\gxpa,\gxpe$ are
second-order slow-roll parameters. Now one might wonder about the fact
that we call $\getpe$ a slow-roll parameter, given that the actual 
slow-roll approximation (in the spirit of a field slowly
rolling along its trajectory) would only require $\ge$, $\getpa$
and higher-order parallel slow-roll parameters to be small, and say nothing
about the perpendicular parameters. However, in order to be able to derive 
the analytical expressions in section~\ref{Slow-roll}, where we treat the 
slow-roll regime, we need to assume a stronger version of the standard
slow-roll approximation where all parameters, including $\getpe$ 
and even $\gc$ (defined in \eqref{defchi}) are small. And as we will later
see, in the models considered in this paper it is anyway not possible to have
a large $\getpe$ while $\getpa$ stays small. Hence we will call all these
parameters slow-roll parameters, and assume all of them to be small
in the slow-roll approximation (sometimes adding the word ``strong'' to be 
explicit). On the other hand, when talking about breaking the slow-roll regime 
in section~\ref{Beyond slow-roll}, we consider situations where $\ge$ or 
$\getpa$ becomes large during inflation (in addition to $\getpe$), which breaks 
slow roll according to anyone's definition.
In the current section, however, we are not assuming anything to be small and 
not making any approximations.

From their definition and using the field equation (\ref{fieldeq}) and its
derivative, one can show that
\begin{align}
  \getpa & =-3-3\gw_{1}, & \getpe & =-3\gw_{2},\nonumber\\
  \gxpa & =-3\gw_{11}+3\ge-3\getpa, & \gxpe & =-3\gw_{21}-3\getpe.
\label{srpareq}
\end{align}
We also introduce the parameter
\be
\gc \equiv \gw_{22}+\ge+\getpa.
\label{defchi}
\ee
Despite its similarity to the expressions for $\gxpa$ and $\gxpe$, the parameter
$\gc$ is a first-order slow-roll parameter and not a second-order one. The 
reason is that within the slow-roll approximation cancellations occur in
the right-hand sides of (\ref{srpareq}), making the slow-roll parameters on
the left-hand side one order smaller than the individual terms on the 
right-hand side. However, no such cancellation occurs in (\ref{defchi}).

We can compute the time derivatives of the basis vectors and the slow-roll
parameters and find:
\be
\begin{split}
\dot{e}_{1\gf}&=\getpe e_{1\gs},\qquad \dot{e}_{1\gs}=-\getpe e_{1\gf},\qquad
\dot{\ge}=2\ge(\ge+\getpa),\\
\dot{\eta}^{\parallel}&=\gxpa+(\getpe)^{2}+(\ge-\getpa)\getpa,\qquad
\dot{\eta}^{\perp}=\gxpe+(\ge-2\getpa)\getpe,\\
\dot{\gc}&=\ge\getpa+2\ge\gc-(\getpa)^{2}+3(\getpe)^{2}+\gxpa+\frac{2}{3}\getpe\gxpe+\gw_{221},\\
\dot{\xi}^{\parallel}&=-3 \gw_{111}+2\getpe\gxpe+(2\ge-3)\gxpa+9\ge\getpa+3(\getpa)^{2}+3(\getpe)^{2},\\
\dot{\xi}^{\perp}&=-3 \gw_{211}-\getpe\gxpa+(2\ge-3)\gxpe+9\ge\getpe+6\getpa\getpe-3\getpe\gc.
\end{split}
\label{srderivatives}
\ee

\subsection{Perturbations}
\label{perturbsec}

We are in this paper interested in predictions of non-Gaussianity from 
inflation, so we need to consider not only first-order but also second-order
perturbations on top of the homogeneous background. For their computation
we will use the long-wavelength formalism developed in \cite{Rigopoulos:2005xx, Rigopoulos:2005ae, Rigopoulos:2005us, Tzavara:2010ge, Tzavara:2011hn, Tzavara:2013yca}. In fact we will directly 
use the final results of that formalism for the non-Gaussianity parameter
$\fnl$ as our starting point, referring the reader to in particular
\cite{Tzavara:2010ge, Tzavara:2013yca} for the derivation.

The most important (potential) observables predicted by inflation are
the amplitude of the scalar power spectrum $\mathcal{P}_s$ of the adiabatic 
curvature perturbation\footnote{At first order and in the flat
gauge we are using here (where the scale factor $a$ is homogeneous),
$\zeta_m = - \frac{\gk}{\sqrt{2\ge}} e_{mA} \delta\gf^A$. For the generalized
definition at higher order and various gauge issues, see the cited 
literature.} $\zeta_1$, its spectral index 
$\ns - 1 \equiv\frac{\d \ln{\mathcal{P}_{s}}}{\d \ln k}$, the tensor-to-scalar
ratio $r \equiv \mathcal{P}_t/\mathcal{P}_s$ and the non-Gaussianity
parameters $\fnl$ of a few specific bispectrum shapes (local, equilateral, 
orthogonal). The first two have been measured quite accurately
by the Planck satellite, while for the latter two we have so far only upper
limits. Of course there are more predicted parameters, especially in the case
of multiple-field inflation, for example the running of the power spectrum,
the spectral index of the tensor power spectrum, the power spectrum of 
isocurvature modes, and non-Gaussianity parameters of many more bispectrum 
shapes, but none of these
have been detected so far. In this paper we will focus on the local
non-Gaussianity parameter $\fnl$ of certain quite general classes
of two-field inflation models. We will in particular investigate if these
models can give an $\fnl$ of order unity (which is large compared
to the prediction of standard single-field slow-roll inflation of 
$\mathcal{O}(10^{-2})$) or even larger. In other words, does the Planck
constraint of $\fnl=0.8 \pm 5.0$ \cite{Ade:2015ava} rule out some of the
parameter regions of these models, or is everything still allowed? 
The observational constraints on $\ns$ will turn out to be
an important ingredient of our considerations. The current Planck result is
$\ns=0.968 \pm 0.006$ \cite{Ade:2015lrj}, while the planned next-generation
satellite experiment CORE expects to reach error bars that are about four
times smaller, of about $0.0015$. On the other hand, it turns
out that the current observational constraint on $r$ does not give any 
additional information compared to $\ns$ for our purposes, so we will ignore
it in the rest of the paper.\footnote{The current upper bound $r<0.12$ 
\cite{Ade:2015lrj} starts to be constraining for some models of single-field 
inflation. However, as explained later, we are interested in two-field models 
where the value of $\gv_{12}$ in \eqref{r} is at least around 4, which makes 
$r$ easily one order of magnitude smaller than in those single-field models.}

The explicit expressions for the first three quantities in the case of
two-field inflation are (see e.g.~\cite{Tzavara:2010ge}):
\be
\mathcal{P}_{s}=\frac{\gk^2 H_{*}^{2}}{8\pi^2\ge_*}\left[1+(\gv_{12})^2\right],
\ee
\be
\begin{split}
\ns-1=\frac{1}{1-\ge_*}&\left[ -4\ge_* - 2\getpa_* + 2 \frac{\gv_{12}}{1+(\gv_{12})^2} \lh-2\getpe_* + \gc_*\gv_{12}\right.\right. \\
&\left.\left. + G_{13}(t,t_*)\lh -\gw_{221*} + 2\ge^2_* + (\getpa_*)^2 + (\getpe_*)^2 + 3\ge_{*}(\getpa_* - \gc_{*})-2\getpa_* \gc_* + \gc_*^2\rh\rh\right],
\end{split}
\label{ns}
\ee
and
\be
r = \frac{16 \ge_*}{1 + (\gv_{12})^2}.
\label{r}
\ee
The asterisk subscript indicates that quantities are evaluated at the time
of horizon crossing ($t_*$). One of the most important differences between
multiple-field and single-field inflation is that the curvature perturbation
$\zeta_1$ is not necessarily frozen on super-horizon scales, but 
can evolve under the influence of the isocurvature mode $\zeta_2$. In fact 
this is described by the very simple but exact equation (see \cite{Tzavara:2012qq} for 
the proof that it is valid fully nonlinearly on super-horizon scales)
\be
\dot{\zeta}_1 = 2 \getpe \zeta_2.
\ee
Hence if we have both a non-zero $\getpe$ and a non-zero isocurvature mode,
then the adiabatic perturbation will still evolve on super-horizon scales,
and not be fully determined at horizon-crossing. In the above expressions this
influence of the isocurvature mode on the adiabatic mode on super-horizon scales
is encoded in $\gv_{12}$ and $G_{13}(t,t_*)$, which will be defined in 
section~\ref{secgreen}. Both 
these quantities still depend on time and in principle have to be evolved all 
the way to recombination in order to compute the CMB observables. However, we 
will impose on all our models that the isocurvature modes have disappeared by 
the end of inflation, so that we have returned to an effectively single-field
situation by then and $\gv_{12}$ and $G_{13}(t,t_*)$ have become constant and no
longer evolve. In that case we can pick the end of inflation as the time to 
evaluate those two quantities and compute the observables without needing to 
know any details about the evolution of the universe after inflation.

\begin{figure}
\centering
\includegraphics[width=0.5\textwidth]{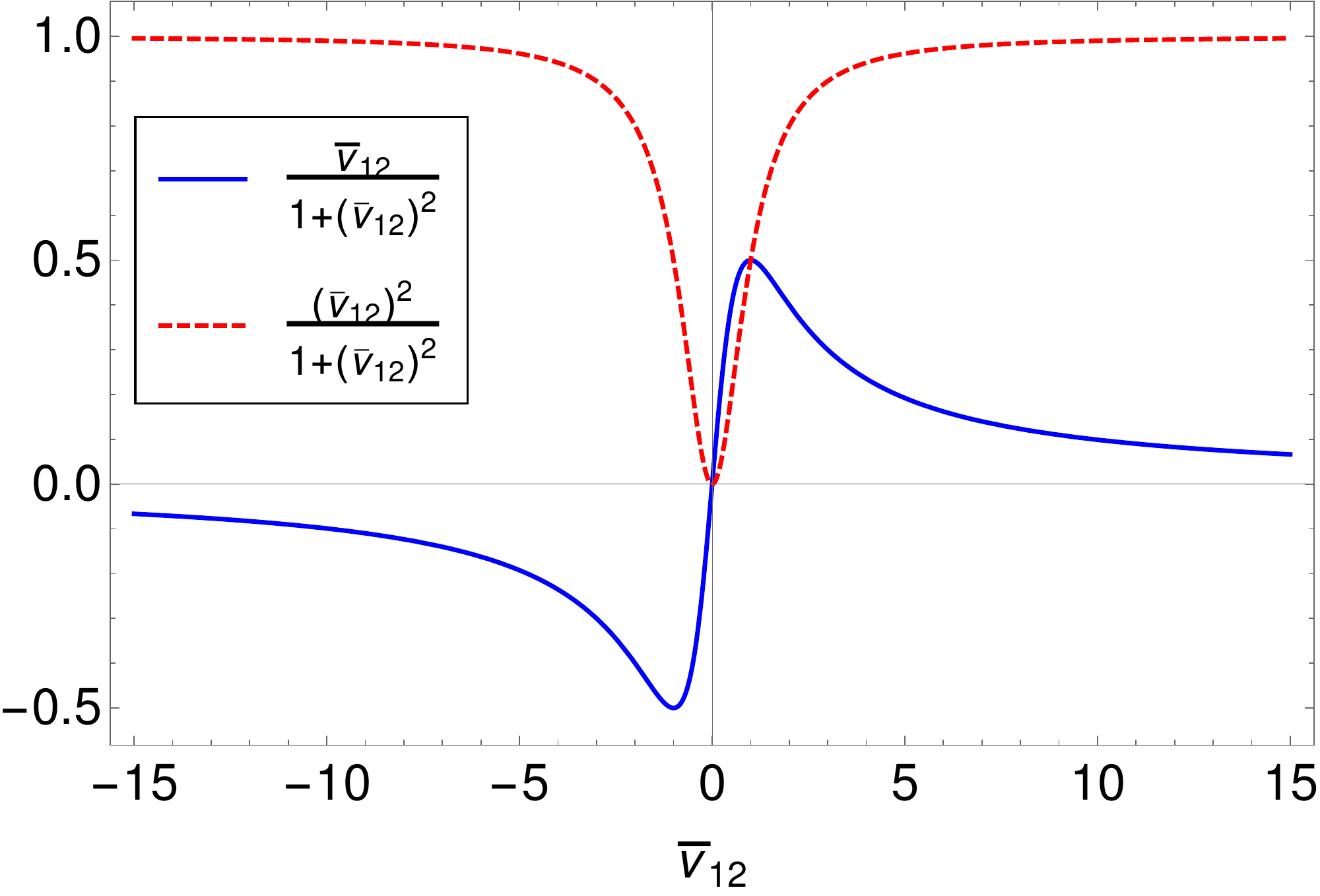}
\caption{$\gv_{12}/(1+\gv_{12}^2)$ and 
$\gv_{12}^2/(1+\gv_{12}^2)$ as a function of $\gv_{12}$.}
\label{fig:factors2}
\end{figure}

An important conclusion can be drawn from the expression of the spectral index.
Given that $G_{13}(t,t_*) \approx \gv_{12}/3$ as we will later show in 
\eqref{proportionality}, 
the relevant factors to study are $\gv_{12}/(1+\gv_{12}^2)$ and 
$\gv_{12}^2/(1+\gv_{12}^2)$, which are shown in figure~\ref{fig:factors2}. 
We see that they are never larger than unity in absolute value and are in fact
of order unity unless $\gv_{12}\approx 0$, which is when multiple-field effects
are negligible and which is not interesting from the point of view of this 
paper.\footnote{The factor $\gv_{12}/(1+\gv_{12}^2)$ also goes to zero for
$|\gv_{12}|\rightarrow\infty$. However, while this term in (\ref{ns}) would then
be compatible with a large $\getpe_*$, that is forbidden by the other terms.}
So barring any fine-tuned cancellations between terms, the observed value of
$\ns$ allows us to conclude that slow roll is a good approximation at horizon
crossing with all first-order slow-roll parameters at $t_*$ at most of order 
$10^{-2}$. However, it is certainly possible for slow roll to be broken
afterwards.

The final result from the long-wavelength formalism for the local
adiabatic bispectral non-Gaussianity parameter $\fnl$ is
\cite{Tzavara:2010ge}:\footnote{It should be noted that this is only
  the part of $\fnl$ that comes from the three-point correlator of two
  first-order perturbations and one second-order perturbation
  (expressed as products of two first-order ones), sometimes called
  $\fnl^{(4)}$ in the literature (see e.g.~\cite{Vernizzi:2006ve}),
  which is the only contribution on super-horizon scales. It does not
  include the so-called intrinsic non-Gaussianity $\fnl^{(3)}$ due to
  interaction terms in the cubic action, which only play a role before
  and at horizon crossing and are necessarily slow-roll suppressed in
  models with standard kinetic terms.}  \be
-\frac{6}{5}\fnl=\frac{-2(\gv_{12})^2}{(1+(\gv_{12})^2)^2} \lh
g_\mathrm{iso} + g_\mathrm{sr} + g_\mathrm{int} \rh.
\label{fNLexpression}
\ee
Here the only approximation made is that slow roll is a good approximation
at horizon crossing (but can be broken afterwards), as we will assume
throughout the paper and which is motivated by the observed value of the 
spectral index as discussed above.\footnote{Some small
additional momentum dependence in $\fnl$ is also neglected, see 
\cite{Tzavara:2010ge,Tzavara:2012qq} for an investigation of that effect.}
The factor $-6/5$ in the definition is a historical artifact due to the
way $\fnl$ was originally defined in terms of the gravitational potential
$\Phi$ and not the adiabatic curvature perturbation $\zeta_1$.
The isocurvature, slow-roll, and integral contributions are given by
\be 
\begin{split}
  &g_\mathrm{iso} = (\ge +\getpa) (\gv_{22})^2 + \gv_{22} \gv_{32}, \qquad g_\mathrm{sr}=-\frac{\ge_*+\getpa_*}{2\gv_{12}^2}+\frac{\getpe_*\gv_{12}}{2}-\frac{3}{2} \lh \ge_* + \getpa_* - \gc_* + \frac{\getpe_*}{\gv_{12}}\rh,\\
  &g_\mathrm{int} = -\int_{t_*}^{t} \d t'\left[2(\getpe)^2
    (\gv_{22})^2 + (\ge +\getpa)\gv_{22}\gv_{32} + (\gv_{32})^2 -
    G_{13}(t,t') \gv_{22} (\Xi\gv_{22}+9\getpe \gv_{32})\right],
\end{split}
\label{defgint}
\ee
where we have defined
\be \Xi \equiv 12\getpe\gc - 6\getpa\getpe +
6(\getpa)^{2}\getpe + 6(\getpe)^{3}-2\getpe\gxpa - 2\getpa\gxpe -
\frac{3}{2}(\gw_{211}+\gw_{222}).
\label{defC}
\ee
The explicit time dependence of all functions has been omitted, except for 
$G_{13}$ since it depends on two times. The various $\gv$ and $G$ terms will
be properly defined just below in section~\ref{secgreen}, but let us say here 
that $\gv_{22}$ and
$\gv_{32}$ are proportional to the isocurvature mode and hence will go to zero
at the end of inflation by our assumption, so that $g_\mathrm{iso}$ vanishes
there. If we relaxed our assumption of
the isocurvature mode going to zero by the end of inflation, it would be easy
to get huge non-Gaussianity at the end of inflation from the $g_\mathrm{iso}$
term, but it would be meaningless since one would have to follow its evolution
explicitly through the rest of the evolution of the universe to get a 
prediction for the observable.
In the single-field limit, a small, slow-roll suppressed part of $g_\mathrm{sr}$
is all that survives and it gives back the $\fnl^{(4)}$ part of the usual 
single-field result of Maldacena \cite{Maldacena:2002vr}. In the
two-field case all terms of $g_\mathrm{sr}$ are also slow-roll suppressed since
they are proportional to slow-roll parameters at horizon crossing. (It is
easy to check that the various functions of $\gv_{12}$ can never become
large, independent of the value of $\gv_{12}$.)
Hence the only persistent large non-Gaussianity can come from the integrated
contribution $g_\mathrm{int}$. We will come back to it in section~\ref{secgint}.

\subsection{Green's functions}
\label{secgreen}

The functions $G_{xy}(t,t')$ (with $x,y=1,2,3$ and $t \geq t'$) are Green's 
functions introduced to solve the first-order perturbation equations (and
then the same functions also serve to solve the second-order equations). 
Here we only give their final equations; see \cite{Rigopoulos:2005us, 
Tzavara:2010ge} for the derivation. 
They satisfy the following differential equations:
\be
\begin{split}
	&\frac{\d}{\d t} G_{1y}(t,t') = 2\getpe(t) G_{2y}(t,t'),\\
	&\frac{\d}{\d t} G_{2y}(t,t') = G_{3y}(t,t'),\\
	&\frac{\d}{\d t} G_{3y}(t,t') = - A_{32}(t) G_{2y}(t,t') - A_{33}(t) G_{3y}(t,t'),
\end{split}
\label{greent}
\ee 
with 
\be
A_{32}= 3\gc + 2\ge^2 + 4\ge \getpa + 4(\getpe)^2 + \gxpa, \qquad\qquad
A_{33} = 3 + \ge + 2\getpa,
\label{defA32A33}
\ee
as well as the following differential equations in terms of the time $t'$:
\be
\begin{split}
	&\frac{\d}{\d t'} G_{x2}(t,t') = -2\getpe(t')\delta_{x1} + A_{32}(t') G_{x3}(t,t'),\\
	&\frac{\d}{\d t'} G_{x3}(t,t') = -G_{x2}(t,t') + A_{33}(t') G_{x3}(t,t').
\end{split}
\label{greentprime}
\ee
The initial conditions are $G_{xy}(t,t)=\delta_{xy}$. We can also combine
the equations~(\ref{greent}) into a second-order differential equation for
$G_{2y}$ in closed form:
\be
\frac{\d^2}{\d t^2} G_{2y}(t,t') + A_{33}(t) \frac{\d}{\d t}
G_{2y}(t,t') + A_{32}(t) G_{2y}(t,t') = 0.
\label{G22eq}
\ee

For $y=1$, the solutions are: $G_{11} = 1$, $G_{21} = G_{31} = 0$. For $y=2,3$ 
we need to make some approximations to solve the equations analytically.
We further introduce the short-hand notation
\be
\gv_{x2}(t) \equiv G_{x2}(t,t_*)-\gc_* G_{x3}(t,t_*).
\label{defvbar}
\ee
This means that $\gv_{12*}=0$, $\gv_{22*}=1$ and $\gv_{32*}=-\gc_*$. 
The functions $\gv_{x2}$ satisfy the same differential equation (\ref{greent}) in terms 
of $t$ as the $G_{x2}$.

In the general case, these equations cannot be solved analytically. Hence, to go further, we will focus on the case $t'=t_*$ and we assume that at horizon-crossing the slow-roll approximation is valid for at least a few e-folds. This means that during these few e-folds, the different slow-roll parameters, which evolve slowly, can be considered as constants at the lowest order. Under these conditions, the differential equation \eqref{G22eq} takes the form:
\be
\ddot{g}(t) + A_{33} \dot{g}(t) + A_{32}g(t)=0,
\ee
where $g$ can be either $G_{22}$, $G_{23}$ or $\gv_{22}$, differing only in initial condition. Here, $A_{32}$ and $A_{33}$ are now constants. The solution of this equation is:
\be
g(t)=\frac{1}{\lambda_- - \lambda_+} \left[ (\lambda_- g_0 - \dot{g}_0) e^{\lambda_+ t} + (-\lambda_+ g_0 + \dot{g}_0)e^{\lambda_- t}\right],
\label{solconstant}
\ee
where $\lambda_+= \half\lh-A_{33}+\sqrt{(A_{33})^2-4A_{32}}\rh$, $\lambda_-= \half\lh-A_{33}-\sqrt{(A_{33})^2-4A_{32}}\rh$ and $g_0$, $\dot{g}_0$ are the initial values of $g$ and $\dot{g}$. In the slow-roll regime, $|A_{32}|\ll 1$ while $A_{33}\approx 3$. The direct consequence is that $|\lambda_+|\ll 1$, which implies that the $e^{\lambda_+ t}$ mode does not change much in a few e-folds, while $\lambda_-\approx -3$, which means that the other mode decays exponentially and can be neglected after a few e-folds (three is sufficient).

For two different sets of initial conditions, the ratio between the solutions becomes:
\be
\frac{g_1}{g_2}=\frac{\dot{g}_1}{\dot{g}_2}=\frac{\lambda_- g_{1_0} - \dot{g}_{1_0}}{\lambda_- g_{2_0} - \dot{g}_{2_0}},
\label{propor}
\ee
which is a constant. Hence, $G_{22*}$ (defined as $G_{22}(t,t_*)$), $G_{23*}$ and $\gv_{22}$ become proportional after a few e-folds of slow-roll. Then, after a few more e-folds of inflation, the approximation of constant slow-roll parameters stops to be valid and we can no longer consider $A_{32}$ and $A_{33}$ to be constants. However, by this time the proportionality between $G_{22*}$, $G_{23*}$, $\gv_{22}$ and their derivatives $G_{32*}$, $G_{33*}$, $\gv_{32}$ has been established, and because of the linearity of the differential equation \eqref{G22eq}, they will stay proportional until the end of inflation.

The case of $G_{12}$, $G_{13}$ and $\gv_{12}$ is a little trickier. With $\getpe$ being a constant, these functions are the primitives of $G_{22}$, $G_{23}$, $\gv_{22}$ according to \eqref{greent}. However, one does not obtain the same factor of proportionality \eqref{propor} with a simple integration of \eqref{solconstant} because of the constant of integration. On the other hand, from \eqref{greent} we know these functions stay small compared to one before the turn of the field trajectory, because $\getpe$ is negligible compared to other slow-roll parameters. During the turn, while $\getpe$ is of the same order as other slow-roll parameters or even larger, they can become large. We will see later that typical and interesting values of $\gv_{12}$ are larger than order unity. Hence, the only relevant part of the integral is after the beginning of the turn. To compute it, one can just integrate the first equation of \eqref{greent} starting at the beginning of the turn instead of at horizon-crossing. Moreover, once the turn has started, we know that the relations of proportionality between $G_{22*}$, $G_{23*}$ and $\gv_{22}$ are already established, which means that from \eqref{greent} the same relations exist between $\dot{G}_{12*}$, $\dot{G}_{13*}$ and $\dot{\gv}_{12}$ on the only relevant part of the integration interval. Then the common factor is conserved by the integration. During the turn, \eqref{propor} becomes valid for the Green's functions $G_{12*}$, $G_{13*}$ and $\gv_{12}$. In particular this is true for the final values of these functions, which will play an important role in the next sections.
If these functions stay negligible during the turn, or vanish at the end, the result does not hold. However, as already mentioned, this case is not interesting as multiple-field effects will play no role. To summarize, the explicit proportionality relations are:
\be 
\frac{G_{12*}}{G_{13*}}=\frac{G_{22*}}{G_{23*}}=\frac{G_{32*}}{G_{33*}}=-\lambda_-\approx 3 \qquad \text{and} \qquad \frac{\gv_{12}}{G_{12*}}=\frac{\gv_{22}}{G_{22*}}=\frac{\gv_{32}}{G_{32*}}=\frac{\lambda_-+\gc_*}{\lambda_-}\approx 1.
\label{proportionality}
\ee

\subsection{The $g_\mathrm{int}$ equation}
\label{secgint}

As discussed at the end of section~\ref{perturbsec}, the only persistent
large non-Gaussianity can come from the integral term $g_\mathrm{int}$
(\ref{defgint}), first derived in \cite{Tzavara:2010ge}. 
So to answer our question if
large non-Gaussianity is at all possible and if so in which models, we need
to investigate this term. Unfortunately, the fact that it is an integral, and that
the time dependence is not only in the upper limit of the integral but also
in the $t$ dependence of $G_{13}$, makes it rather hard to get a handle 
analytically on its behaviour in general.

However, as is shown in appendix~\ref{appendix}, by taking several derivatives
of (\ref{defgint}) it is possible to derive a second-order differential 
equation for the derivative of $g_\mathrm{int}(t)$ in closed form in terms 
of $t$ only:
\be
\begin{split}
&(\getpe)^2\, \dddot{g}_{int} + \getpe\left[3\getpe -\ge \getpe + 6 \getpa\getpe - 2 \gxpe\right]\,\ddot{g}_{int} \\
& \quad + \left[(\getpe)^2 \lh - 12\ge + 6\gc + 6(\getpa)^2 + 6 (\getpe)^2 + 4 \gxpa \rh + \getpe \lh 3 \gw_{211} - 8 \getpa\gxpe \rh  + 2(\gxpe)^2 \right]\,\dot{g}_{int}\\
&= K_{22} (\gv_{22})^2 + K_{23} \gv_{22}\gv_{32} + K_{33} (\gv_{32})^2,
\end{split}
\label{equadiff}
\end{equation}
where the $K_{xy}$ are explicit (long) expressions in terms of products of 
slow-roll parameters and are defined in (\ref{defKxy}). This differential
equation and its general solution discussed below is one of the central
new results of this paper.

Despite its complicated looks, (\ref{equadiff}) actually admits a completely
exact analytical homogeneous solution:
\be
\dot{g}_\mathrm{int}(t) = 2 A \, \getpe(t) G_{22}(t,t_*) 
+ 2 B \, \getpe(t) G_{23}(t,t_*) + P(t),
\label{solgintdot}
\ee
where $A$ and $B$ are integration constants to be determined from the 
initial conditions and $P(t)$ is a particular solution of the equation.
This expression can then be integrated to give
\be
g_\mathrm{int}(t) = A \, G_{12}(t,t_*) + B \, G_{13}(t,t_*)
+ \int_{t_*}^t \d t' P(t').
\label{solgint}
\ee
Here we used the fact that $g_\mathrm{int}(t_*)=0$ to eliminate the additional
integration constant. Note that instead of $2\getpe G_{22}$ we can also use
$2\getpe\gv_{22}$ as independent homogeneous solution, which integrates to
$\gv_{12}$.

Now one might wonder if we have made any progress here, since in 
(\ref{solgint}) $g_\mathrm{int}$ is still expressed in terms of an integral, 
and while there is only a single time now, it does involve an a priori
unknown function $P(t)$. However, as we will show in the next section,
for certain classes of potentials and within the slow-roll approximation,
we can find an explicit analytical expression both for $P(t)$ and for its
integral. Since slow roll is a good approximation at horizon crossing, as
discussed before, where the initial conditions are given, this then allows us 
to determine the constants $A$ and $B$ for those models. Finally we will show 
in a later section that in the regions where the slow-roll approximation
for $\getpa$ and $\getpe$
breaks down (with the only condition that $\ge$ remains small) and we do not 
have an explicit analytical solution for $P(t)$, we do not actually need it 
since its contribution is negligible compared to the homogeneous solution.
This will finally allow us to write down the exact analytical result for the
observable $\fnl$ in those models, even if slow roll is broken
during some part of the inflationary evolution.

\section{Slow roll}
\label{Slow-roll}

In this section, we use several consecutive levels of approximations to 
simplify the expressions of the previous section. We start by applying 
only the (strong) slow-roll approximation to general two-field potentials. 
As discussed in section~\ref{srparsec}, this means that all slow-roll 
parameters, including $\getpe$ and $\gc$, are assumed to be small, which is a 
stronger approximation than the standard slow-roll approximation where only
parallel slow-roll parameters are assumed to be small. Then, in the next 
subsection, we focus on 
sum-separable potentials where the Green's functions can be computed as well 
as the different observables. Finally, in the last two subsections, we 
specialize to the case of monomial sum potentials.

\subsection{General case}

We apply the slow-roll approximation to the equations of the previous section, starting by the slow-roll parameters. Using the field equation, we obtain explicit expressions for the basis components. We then perform a first-order slow-roll expansion on the second line of \eqref{srpareq} to obtain $\getpa$ and $\getpe$. For $\gxpa$ and $\gxpe$ we proceed in a similar way on \eqref{srderivatives}. The results are:
\be
\begin{split}
\label{sreta}
&e_{1A} = -\gw_{A},\qquad \getpa = \ge - \gw_{11},\qquad \getpe = -\gw_{21},\\
&\gxpa = 3\ge\getpa + (\getpa)^2 + (\getpe)^2 - \gw_{111}, \qquad \gxpe = 3\ge \getpe + 2\getpa\getpe - \getpe\gc - \gw_{211}.
\end{split}
\ee
The same slow-roll expansion applied to the differential equations for the Green's functions \eqref{greent} and \eqref{greentprime} gives:
\be
\label{greensr}
\frac{\d}{\d t}G_{22}(t,t') + \gc(t) G_{22}(t,t') = 0,
\ee
\be
G_{32}(t,t')=-\gc(t)G_{22}(t,t'),\qquad G_{x3}(t,t')=\frac{1}{3}G_{x2}(t,t').
\label{greensr2}
\ee
For the observables, from \eqref{ns} we get:
\be
\ns - 1  = -4 \ge_* - 2\getpa_* + 2 \frac{\gv_{12}}{1+\gv_{12}^2} \lh -2 \getpe_* + \gv_{12} \chi_* \rh
\ee 
and for the different terms of $\fnl$ in \eqref{fNLexpression}:
\be
g_{\mathrm{iso}} = (\ge +\getpa -\gc)\gv_{22}^2, \qquad \qquad g_{\mathrm{sr}}=-\frac{\ge_*+\getpa_*}{2\gv_{12}^2}+\frac{\getpe_*\gv_{12}}{2}-\frac{3}{2} \lh \ge_* + \getpa_* - \gc_* + \frac{\getpe_*}{\gv_{12}}\rh.
\ee
For $\gint$, the slow-roll approximation is not sufficient to compute the integral. However, we can simplify the differential equation \eqref{equadiff} to
(see appendix \ref{appendix} for the details of the computation):
\be
\label{equadiffsr}
\getpe\,\ddot{g}_{\mathrm{int}} - \left[\getpe(\ge-2\getpa-\gc)+\gxpe\right]\,\dot{g}_{\mathrm{int}}= K_{\mathrm{sr}}(\gv_{22})^2,
\ee
with
\be
\begin{split}
K_{\mathrm{sr}} =&~\getpa \getpe \gxpa+3 (\getpa)^2 \getpe \gc -3 \getpa \getpe \gc ^2- (\getpa)^3 \getpe + \getpa (\getpe)^3-\getpa \gxpe \gc -\getpe \gxpa \gc -(\getpe)^2 \gxpe \\
&+\gxpe \gc ^2+\getpa \getpe \gw_{221}-2 \getpe \gc  \gw_{221} + \ge\getpe\gw_{221} - (\getpe)^2 \gw_{222} + 4 \ge\getpa \getpe \gc   + \ge^2\getpa \getpe   \\
&-4 \ge\getpe \gc ^2  +3\ge^2 \getpe \gc  -2 \ge(\getpe)^3  -\ge\gxpe \gc +\getpe \gc^3 +\ge\getpe \gxpa.
\end{split}
\ee
This equation can be solved for certain classes of potentials. We will look at the simple case of a sum potential, which was solved initially in \cite{Vernizzi:2006ve, Battefeld:2006sz} and discussed in detail in \cite{Byrnes:2008wi,Elliston:2011et,Elliston:2011dr}. The case of a product potential is treated in appendix \ref{appendix:product}.

\subsection{Sum potential}
\label{Sum Potential Section}

A sum potential has the form
\be
W(\gf,\gs) = U(\gf)+V(\gs).
\ee
An immediate consequence of this form is that all mixed derivatives of the potential are zero. Using this and by writing out 
$\gw_{11}, \gw_{22}, \gw_{21}$ (defined in (\ref{defW})) explicitly in terms
of $\gw_{\gf\gf}, \gw_{\gs\gs}, \gw_{\gf\gs}$ and using the normalization of the 
basis $e_{1\gf}^2 + e_{1\gs}^2=1$, one can show that
\be
e_{1\gf}e_{1\gs}(\gw_{11}-\gw_{22})=(e_{1\gf}^{2}-e_{1\gs}^{2})\gw_{21},
\label{W11eq}
\ee
which using (\ref{srpareq}) and (\ref{defchi}) is equivalent to 
\be
e_{1\gf}e_{1\gs}(\gxpa+3\gc-6\ge)=(e_{1\gf}^{2}-e_{1\gs}^{2})(\gxpe+3\getpe) .
\label{W11eq2}
\ee 
Similarly for third-order derivatives, we can write:
\be
\begin{split}
\label{sumequation}
&e_{1\gf}e_{1\gs}\gw_{221} =  e_{1\gf}e_{1\gs}\gw_{111} + (e_{1\gs}^2-e_{1\gf}^2)\gw_{211},\\
&e_{1\gf}e_{1\gs}\gw_{222} =  e_{1\gf}e_{1\gs}\gw_{211} + (e_{1\gs}^2-e_{1\gf}^2)\gw_{221}.
\end{split}
\ee
Using \eqref{W11eq2}, they are equivalent to
\be
\begin{split}
&(\gxpe+3\getpe)\gw_{221} = (\gxpe+3\getpe)\gw_{111} -(\gxpa +3\gc - 6\ge)\gw_{211},\\
&(\gxpe+3\getpe)\gw_{222} = (\gxpe+3\getpe)\gw_{211} -(\gxpa +3\gc - 6\ge)\gw_{221}.
\end{split}
\ee
Note that these equations are general and not only slow-roll. After a first-order slow-roll expansion, they become:
\be
\label{sumequationsr}
\begin{split}
&\getpe\gw_{221} = \getpe\gw_{111} - (\gc - 2\ge)\gw_{211},\\
&\getpe\gw_{222} = \getpe\gw_{211} - (\gc - 2\ge)\gw_{221}.
\end{split}
\ee
We use this to rewrite the right-hand term of \eqref{equadiffsr} as
\be
K_{\mathrm{sr}}= 2\ge \lh -3 \ge^2 \getpe + 3(\getpa)^2 \getpe - 3 (\getpe)^3 + \ge\getpe\gc - 3\getpa\getpe\gc + \getpe \gxpa +\ge \gxpe - \getpa\gxpe \rh.
\ee
Then, one can show that a particular solution of this equation is $\dot{g}_{\mathrm{int}}=2\ge(\ge+\getpa-\gc)\gv_{22}^2$, which can be integrated into $\gint = \ge\gv_{22}^2 -\ge_*$.

We also know that $\dot{g}_{\mathrm{int}*}=-2(\getpe_*)^2 + (\ge_*+\getpa_*-\gc_*)\gc_*$ from \eqref{derivgint} and the initial conditions of the Green's functions. Combining this particular solution with the homogeneous solution, we get the full solution for $\dot{g}_{\mathrm{int}}$ and then $\gint$ after integration, in agreement with the known result from~\cite{Tzavara:2010ge}:
\be
\label{gint}
\begin{split}
\dot{g}_{\mathrm{int}}&= 2\ge(\ge+\getpa-\gc)(\gv_{22})^2 - \frac{e_{1\gf*}^2\gV_{\gs\gs*} - e_{1\gs*}^2\gU_{\gf\gf*}}{e_{1\gf*}e_{1\gs*}} \getpe\gv_{22}, \\
\gint &= \ge \gv_{22}^2 - \ge_* - \left[ \getpe_* - \frac{1}{2\getpe_*} (\ge_* + \getpa_* - \gc_*)(\gc_* - 2\ge_*)\right]\gv_{12},\\
        &= \ge \gv_{22}^2 - \ge_* - \frac{e_{1\gf*}^2\gV_{\gs\gs*} - e_{1\gs*}^2\gU_{\gf\gf*}}{2e_{1\gf*}e_{1\gs*}} \gv_{12}.
\end{split}
\ee
Here the first two terms on the last line are the particular solution, and the last term the homogeneous solution. It is possible to show that the particular solution and the homogeneous solution are generally of the same order during inflation (this is discussed later in section \ref{particular solution section}). However, we are only interested in the final values of the observables $\ns$ and $\fnl$. As discussed before, the only large contribution in $\fnl$ can come from $\gint$, if we suppose isocurvature modes vanish before the end of inflation, which means in terms of Green's functions that $\gv_{22}$ and $\gv_{32}$ vanish while $\gv_{12}$ becomes constant. Hence in that case, the integrated particular solution is also slow-roll suppressed and only the homogeneous solution matters at the end of inflation.
From now on, the different expressions for the observables are only given at the end of inflation. For every other parameter (like the Green's functions and the slow-roll parameters), if they are evaluated at the end of inflation, it is indicated by the subscript $e$.

Using the result \eqref{gint} with $\gv_{22e}=0$, we can write:
\be
\begin{split}
-\frac{6}{5}\fnl &= \left[ \getpe_* - \frac{1}{2\getpe_*} (\ge_* + \getpa_* - \gc_*)(\gc_* - 2\ge_*)\right]\frac{2(\gv_{12e})^3}{\lh 1+(\gv_{12e})^2\rh ^2} + \mathcal{O}(10^{-2})\\
 &=  \frac{e_{1\gf*}^2\gV_{\gs\gs*} - e_{1\gs*}^2\gU_{\gf\gf*}}{e_{1\gf*}e_{1\gs*}} \frac{(\gv_{12e})^3}{\lh 1+(\gv_{12e})^2\rh ^2} + \mathcal{O}(10^{-2}).
\label{fnlapp}
\end{split}
\ee
This depends on the final value of the Green's function $\gv_{12}$, which describes the contribution of the isocurvature mode to the adiabatic mode. Without computing it, it is possible to determine a necessary condition for $\fnl$ to be of order unity or larger. Indeed it is easy to show that, for any value of $\gv_{12e}$: 
\be
\left|\frac{(\gv_{12e})^3}{\lh 1+(\gv_{12e})^2\rh ^2}\right|\leq\frac{3^{3/2}}{16} \approx 0.325.
\label{numfactor}
\ee
If the slow-roll approximation is valid at horizon-crossing, which is the main assumption in the computation of $\fnl$, we expect that $\gV_{\gs\gs*}$ and $\gU_{\gf\gf*}$ are of order slow-roll (small compared to one). Then, the only possibility to get $\fnl$ of order unity is that one of the basis components is negligible at horizon-crossing. This means one of the fields is dominating at that time, by definition we choose it to be $\gf$. Hence, at horizon-crossing $e_{1\gf*}^2 \approx 1$ and $e_{1\gs*}^2\ll 1$. Using \eqref{sreta}, this also implies that $|U_{\gf*}|\gg |V_{\gs*}|$ and we can simplify:
\be
\frac{e_{1\gf*}^2\gV_{\gs\gs*} - e_{1\gs*}^2\gU_{\gf\gf*}}{e_{1\gf*}e_{1\gs*}} = \frac{e_{1\gf*}\gV_{\gs\gs*}}{e_{1\gs*}} = \frac{\sqrt{2\ge_*}}{\gk} \frac{V_{\gs\gs*}}{V_{\gs*}}.
\label{fnlsrfactor}
\ee
This has to be large to have $\fnl$ non-negligible, which means that the second-order derivative $V_{\gs\gs*}$ is large compared to the first-order derivative $V_{\gs*}$. Hence around $\gs_*$, the potential is very flat in the $\gs$ direction. In terms of slow-roll parameters, this means that $|\getpe_*|\siml |(\ge_* + \getpa_* - \gc_*)(\gc_* - 2\ge_*)|$. For the usual slow-roll order values of $10^{-2}$, $\getpe_*$ is at most of order $10^{-4}$.

\begin{figure}
\centering
\includegraphics[width=0.5\textwidth]{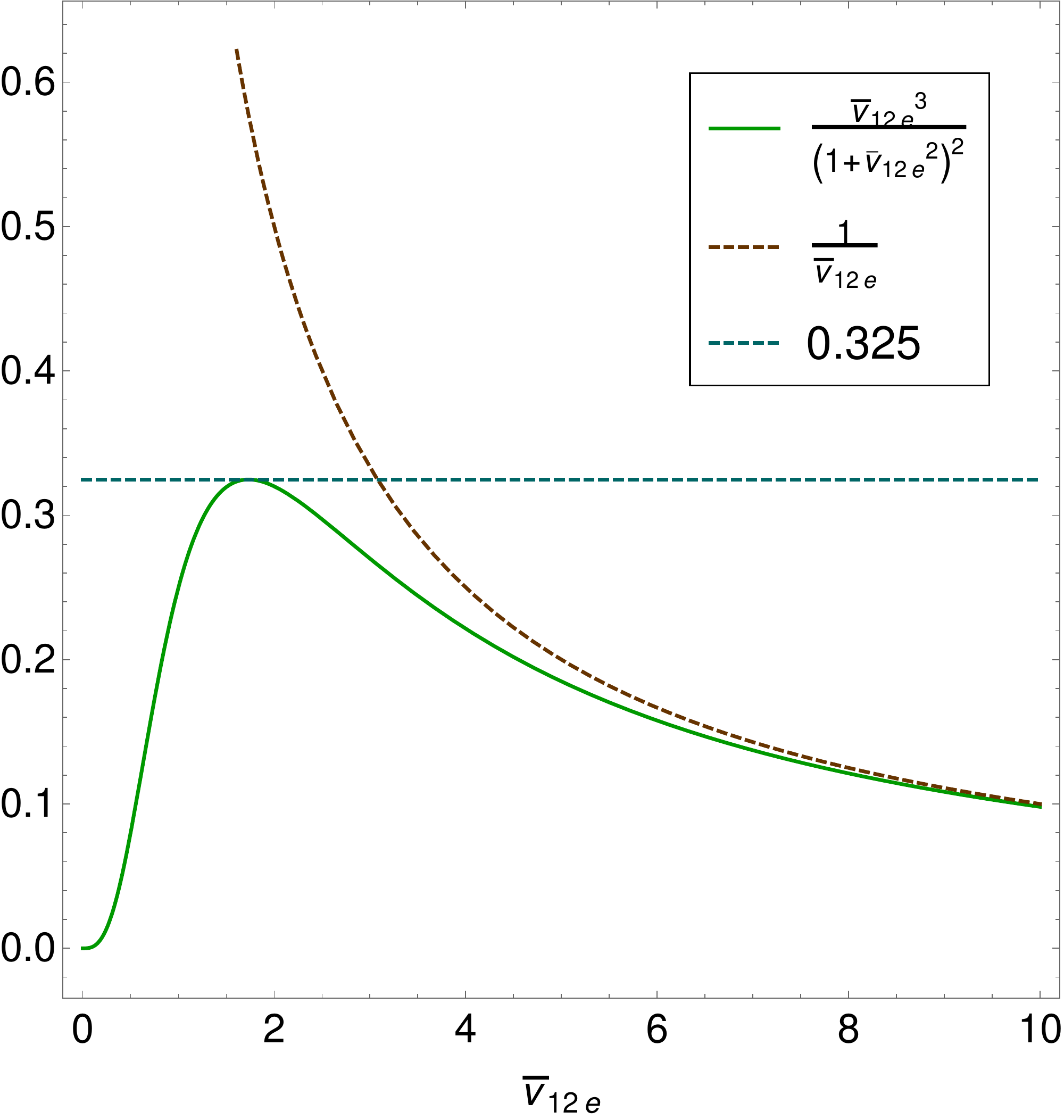}
\caption{$\frac{(\gv_{12e})^3}{\lh(1+(\gv_{12e})^2\rh^2}$ as a function of $\gv_{12e}$ and its two upper limits 0.325 and $1/\gv_{12e}$.}
\label{fig:factors}
\end{figure}

Another useful limit is:
\be 
\label{v12limit}
\left|\frac{\gv_{12e}^3}{(1+\lh\gv_{12e})^2\rh^2}\right|<\left|\frac{1}{\gv_{12e}}\right|,
\ee
which becomes a very good approximation if $|\gv_{12e}|>4$ . These two limits are shown explicitly in figure~\ref{fig:factors}. From \eqref{greent}, if $\gv_{12e}$ is of order unity, this implies that at some time there was a turn of the field trajectory where both the isocurvature mode and $\getpe$ are non-negligible. This turn is then a necessary condition of large non-Gaussianity.

Still using the slow-roll approximation, we can go further by computing the Green's functions. From \eqref{W11eq2}, we get:
\be
\gc = 2\ge +\getpe \frac{e_{1\gf}^2-e_{1\gs}^2}{e_{1\gf}e_{1\gs}}=-\frac{\d}{\d t}\ln{\lh H^2 e_{1\gf}e_{1\gs}\rh}.
\ee
We can then solve \eqref{greensr}:
\be
G_{22}(t,t')=\frac{H(t)^2 e_{1\gf}(t)e_{1\gs}(t)}{H(t')^2 e_{1\gf}(t')e_{1\gs}(t')}.
\ee
Moreover, we have:
\be
\getpe H^2 e_{1\gf}e_{1\gs}= \frac{\gk^2}{6}\frac{\d Z}{\d t},
\label{Z}
\ee
with $Z\equiv V e_{1\gf}^2- U e_{1\gs}^2$ \cite{Vernizzi:2006ve,Tzavara:2010ge}, which gives us:
\be
\gv_{12} = \frac{Z-Z_*}{W_* e_{1\gf*}e_{1\gs*}},\qquad \gv_{22}=\frac{W e_{1\gf} e_{1\gs}}{W_* e_{1\gf*} e_{1\gs*}}.
\label{Green srsolution}
\ee
At the end of inflation, when the fields reach the minimum of the potential, $Z$ tends to zero. Obviously, this can only happen if there is a turn of the field trajectory at some time after horizon-crossing to make both fields evolve. Moreover, if $e_{1\gs*}^2\ll 1$ (necessary condition for $\fnl$ of order unity), $Z_*=V_* e_{1\gf*}^2$. We then obtain, using \eqref{sreta}:
\be
\gv_{12e}=-\frac{V_* e_{1\gf*}}{W_* e_{1\gs*}}= \text{sign}(e_{1\gf*})\sqrt{2\ge_*}\frac{\gk V_*}{V_{\gs*}}.
\label{v12e}
\ee
With a small enough $e_{1\gs*}$, it is easy to obtain $\gv_{12e}$ larger than four or five. In figure~\ref{fig:factors}, this places us on the right where $\frac{(\gv_{12e})^3}{\lh 1+(\gv_{12e})^2\rh^2} \approx \frac{1}{\gv_{12e}}$. The consequence for the potential $V$ is that $\gk V_*\gg V_{\gs*}$.

Substituted into \eqref{fnlapp}, in the case where the slow-roll parameters factor is large, we obtain:
\be
-\frac{6}{5}\fnl \approx \frac{\gV_{\gs\gs*}e_{1\gf*}}{e_{1\gs*}} \frac{1}{\gv_{12e}} =\frac{\gV_{\gs\gs*}e_{1\gf*}}{e_{1\gs*}}\frac{W_*  e_{_1\gs*}}{- V_* e_{1\gf*}} = -\frac{V_{\gs\gs*}}{\gk^2 V_*}.
\label{fnllimit}
\ee
This directly shows that $\fnl$ is of order unity when the second derivative of $V_*$ and $V_*$ itself are of the same order, while its first-order derivative $V_{\gs*}$ is small compared to the two previous quantities because of \eqref{fnlsrfactor} and \eqref{v12e}, a result already highlighted in \cite{Elliston:2011et,Elliston:2011dr}. Larger $\fnl$ is a priori possible, but requires a fine-tuning of the model. Moreover, the sign of $\fnl$ is the sign of $V_{\gs\gs*}$. A negative $\fnl$ corresponds to a potential in the form of a ridge at $t_*$, where $\gs_*$ is very close to the maximum for the potential to be flat enough in the $\gs_*$ direction, while a positive $\fnl$ corresponds to a valley potential.

In the same limit of large $\gv_{12e}$, the spectral index takes the form 
\be
\ns - 1  = -4 \ge_* - 2\getpa_* + 2\chi_*  = -2\ge_* + 2\gV_{\gs\gs*}.
\label{nstwofield}
\ee
The spectral index is close to 1, hence $\gV_{\gs\gs*}=\frac{V_{\gs\gs*}}{\gk^2 W_*}$ is at most of order $10^{-2}$. If it is smaller, this requires a fine-tuning of $\ge_*$. If $\fnl$ is of order unity, then $\frac{V_*}{W_*}$ is also of order $10^{-2}$. 

To summarize, at horizon-crossing, the conditions are $U_* \gg V_*$ and $|U_{\gf*}|\gg |V_{\gs*}|$. The second-order derivative $V_{\gs\gs*}$ is not negligible and can be either smaller, equal or larger than $U_{\gf\gf*}$ but it is not hugely larger or smaller. To be precise, we make a quite general assumption that $|V_{\gs\gs*} U_{\gf*}^2|\gg |U_{\gf\gf*}V_{\gs*}^2|$ and $|V_{\gs\gs*} V_{\gs*}^2|\ll |U_{\gf\gf*}U_{\gf*}^2|$. With these different assumptions for the potential, the expressions for the slow-roll parameters and basis vectors become:
\be
\label{srparametersflat}
\begin{split}
&\ge = \frac{1}{2\gk^2} \lh \frac{U_{\gf}}{U} \rh^2, \qquad
\getpa = - \frac{U_{\gf\gf}}{\gk^2 U} + \ge, \qquad
\getpe = - \frac{V_{\gs}}{U_{\gf}} \frac{U_{\gf\gf} - V_{\gs\gs}}{\gk^2 U},\\
&e_{1\gf}= -\text{sign}(U_{\gf}), \qquad e_{1\gs}=-\frac{V_\gs}{\gk\sqrt{2\ge}\, U}, \qquad \chi = \frac{V_{\gs\gs}}{\gk^2 U} + \ge + \getpa = \frac{U_{\gf}}{V_{\gs}} \getpe + 2\ge.
\end{split}
\ee

At horizon-crossing, the situation is very close to single-field inflation. In the slow-roll regime, by definition everything evolves slowly, hence a legitimate question is to ask when these conditions will stop to be valid. In fact, they will break at the turn of the field trajectory. At that time $V_{\gs}$ stops to be negligible compared to $U_{\gf}$ (or equivalently, $e_{1\gs}$ is not small compared to one). As already discussed, the turn is mandatory to have $\gv_{12e}$ large enough. However, they will also break if $V$ stops to be negligible compared to $U$, this happens when the field $\gf$ is near the minimum of its potential. In this second case, we know the slow-roll approximation will also stop to be valid because $\ge$ is becoming large (similarly to single-field inflation). Hence, if this happens before the turn, as the slow-roll approximation is not valid anymore, we lose the analytical results for the Green's functions and $\fnl$. We have to check if the turn can occur before the first field reaches the minimum of its potential, or in simple terms, is it possible to have $\fnl$ of order unity without breaking the slow-roll approximation? To be able to make progress in answering that question, we will consider a specific class of two-field sum potentials, where both $U$ and $V$ are monomial plus a possible constant.

\subsection{Monomial potentials}
\label{Monomial potentials section}

Using the results of the previous section, we want to analytically study inflation between horizon-crossing and the beginning of the turn of the field trajectory. The idea is that the slow-roll approximation is broken when the dominating field $\gf$ gets close to the minimum of its potential, and we want to verify if the turn can occur before that time. This means that the form of the potential does not need to describe the end of inflation. 

We know that $V(\gs)$ has to be very flat around $\gs_*$, hence we can use an expansion in $\gs$ keeping only the largest term to write:
\be
V(\gs) = C + \gb (\gk\gs)^m,
\label{potential V}
\ee
where $C$, $m$ and $\gb$ are constants. Here $m > 1$, while $\gb$ can be either positive or negative. Because of the expansion in $\gs$, this potential is in fact quite general. Depending on the sign of $\gb$, the potential either corresponds to a ridge where $\gs_*$ is near the local maximum $C$ ($\gb<0$) or to a valley with $\gs_*$ near the minimum ($\gb>0$). For the potential $U$, there are many possibilities, we choose to focus on a monomial potential:
\be
U(\gf) = \ga (\gk\gf)^n,
\ee
with $\ga>0$ and $n>1$.

We redefine the fields as being dimensionless: $\tilde{\gf} = \gk \gf$ and $\tilde{\gs} = \gk \gs$ and we will omit the tildes in the redefined fields.
Using the expressions for the slow-roll parameters given at the end of the previous section \eqref{srparametersflat}, we have:
\be
\label{srparmonomial}
\begin{split}
\ge &= \frac{n^2}{2} \frac{1}{\gf^2}, \qquad
\getpa = - \frac{n(n-2)}{2}\frac{1}{\gf^2} = - \frac{n-2}{n} \ge,\\
\getpe &= - \frac{m\gb}{n \ga^2} \frac{\gs^{m-1}}{\gf^{2n-1}}
\lh n(n-1)\ga \gf^{n-2} - m(m-1)\gb \gs^{m-2} \rh\\ 
&= -\frac{m\gb}{n^{2n} \ga^2} 2^{n-\frac{1}{2}} \gs^{m-1} 
\lh n^{n-1}(n-1)2^{1-\frac{n}{2}}\ga \ge^{\frac{n}{2}+\frac{1}{2}}-m(m-1)\gb \gs^{m-2} \ge^{n-\frac{1}{2}}\rh,\\
\chi &= \frac{1}{\ga}\frac{1}{\gf^n} \lh n\ga \gf^{n-2} + m(m-1)\gb \gs^{m-2} \rh 
= \frac{2\ge}{n} + \frac{m(m-1)\gb}{n^{n}\ga}2^{n/2}\ge^{n/2}\gs^{m-2}.
\end{split}
\ee 
It is useful to express the slow-roll parameters as a function of $\ge$ instead of $\gf$ because $\ge$ increases after horizon-crossing, at least until the turn, and with $\ge$ we know exactly when the slow-roll approximation stops to be valid. $\ge$ and $\getpa$ are of the same order except in the case of $n=2$ where $\getpa$ is of order $\ge^2$ as can be checked with a second-order calculation.

The next step is to use the conditions that $\fnl$ should be of order unity and $\ns$ should be within the observational bounds to constrain the free parameters of this potential. With this form of $V$, we have the useful relation:
\be
\label{link12}
(m-1)V_\gs=\gs V_{\gs\gs}.
\ee
We know that $|e_{1\gs*}|\ll 1$ and substituting \eqref{link12} into the expression for $e_{1\gs}$ in \eqref{srparametersflat}, we can write:
\be
e_{1\gs} = -\frac{\gV_{\gs\gs}}{\sqrt{2\ge}}\frac{\gs}{m-1}.
\ee
Combining this with the contraints on the spectral index \eqref{nstwofield} which imply that $\ge_*$ and $\gV_{\gs\gs*}$ are both of order $10^{-2}$ at most, this imposes $\gs_*$ to be small compared to 1. Applying these constraints due to the observables to the potential gives:
\be
\label{explicitconstraints}
\frac{V_{\gs\gs*}}{W_*}=\frac{m(m-1)\gb \gs_*^{m-2}}{\ga \gf^n_* + C + \gb\gs_*^{m}} \sim \mathcal{O}(10^{-2}),\qquad \frac{V_{\gs\gs*}}{V_*}=\frac{m(m-1)\gb\gs_*^{m-2}}{C+\gb\gs_*^{m}}\sim \mathcal{O}(1).
\ee 
Within the limit $\gs_* \ll 1$, we learn from these equations that $\ga \gf_*^n \gg m(m-1)\gb \gs_*^{m-2} \sim C + \gb \gs_*^m$.

We also need to determine the slow-roll parameters at $t_*$, which requires to know $\gf_*$. One way to determine this is to know the amount of inflation due to each field between horizon-crossing and the end of inflation. We can start by solving the field equation:
\be
\dot{\gf} = - \frac{n}{\gf},
\label{eqphi(t)}
\ee
which integrates immediately to:
\be
\label{phi(t)}
\gf(t) = \gf_* \sqrt{1 - \frac{t}{N_\gf}},
\ee
with $N_\gf=\frac{\gf_*^2}{2n}$ the slow-roll approximation of the number of e-folds due to $\gf$ after horizon-crossing. 

The potential is known only before the turn of the field trajectory, especially for $V$ if it is an expansion of some more complicated function. This means that we do not know the value of $N_\gf$, however it is in the range of a few to 60 e-folds. We will test different values. Nevertheless, in the simplest cases $N_\gs$ (number of e-folds due to $\gs$) is small compared to $N_\gf$. As a simple argument here, we consider the case where $\gs$ falls off a ridge, so that $V_*\approx C$. If $V$ keeps the same form almost until the end of inflation, the minimum of the potential ($V=0$) corresponds approximately to $\gs_{e}=(-C/\gb)^{1/m} \sim [m(m-1)]^{1/m}\gs_*^{1-2/m}$, using the second part of \eqref{explicitconstraints}. For $m=2$, this is of order 1, for larger $m$ it becomes smaller (only $m$ close to 1 is problematic). In a pure monomial potential like $U$ without the constant term, having $\gf_*$ of order unity would imply that $N_\gf$ is itself of order unity. $V$ is a bit different because of the constant term, however once $\gs$ starts to fall at a non-negligible pace (the turn), it becomes quite similar and $\gs$ goes from $\gs_*$ negligible to $\gs_e$ of order unity. Hence this also corresponds to $N_\gs$ of order unity which can be neglected in the total number of e-folds compared to $N_\gf$. Note this is not a general proof, just a plausible argument to claim that $N_\gf$ is the dominant contribution. We can also see that $\gs_e$ becomes larger if $V_{\gs\gs*}/V_*$ in \eqref{explicitconstraints} becomes smaller. Hence the fact that $N_\gs$ is small is linked to having $\fnl$ of order unity or more. 

The parameter $\ge_*$ is related to the value of $N_\gf$, hence for these models where $N_\gs \ll N_\gf$, the value of $\ge_*$ is directly fixed by the total number of e-folds after horizon-crossing:
\be
\ge_*=\frac{n}{4N_\gf}.
\ee
When $\ge_*$ is fixed, we can use the spectral index formula \eqref{nstwofield} to constrain $\gV_{\gs\gs*}$:
\be
\gV_{\gs\gs*}=\frac{\ns-1}{2}+\ge_*.
\label{Wss spectral index}
\ee
Using $\ns = 0.968 \pm 0.006$ from the Planck data, table \ref{table:Vss} shows the constraints for integer values of $n$. Note that for $n\geq 5$, the second-order derivative has to be positive. 
\begin{table}
		\centering
		\begin{tabular}{|c|c|c|c|c|c|}
		\hline
		$n$ & 2 & 3 & 4 & 5 & 6\\
		\hline
		$10^3\,\gV_{\gs\gs*}$ & $ -7.7 \pm 3$ & $-3.5 \pm 3$ & $0.7 \pm 3$ & $4.8 \pm 3$ & $9 \pm 3$  \\
		\hline
	\end{tabular}
	\caption{Constraints from the spectral index on $\gV_{\gs\gs*}$ for different $n$ with $N_\gf=60$.}
\label{table:Vss}
\end{table}
 According to \eqref{fnlapp}, we also know that:
\be
\left|-\frac{6}{5}\fnl\right| < 0.65\left| \getpe_* - \frac{1}{2\getpe_*} (\ge_* + \getpa_* - \gc_*)(\gc_* - 2\ge_*)\right|, 
\label{fnllimit2}
\ee
which gave the estimation of $\getpe_*$ of order $10^{-4}$ to get $\fnl$ of order unity. We can neglect the first $\getpe_*$ which is already a few orders of magnitude smaller than the single-field slow-roll typical value of $\fnl$. Then we obtain:
\be
\left|-\frac{6}{5}\fnl \getpe_*\right| < 0.325\left|(\ge_* + \getpa_* - \gc_*)(\gc_* - 2\ge_*)\right|.
\label{etaperconstraint}
\ee
We can rewrite the right-hand side term:
\be
(\ge_* + \getpa_* - \gc_*)(\gc_* - 2\ge_*)= -\gV_{\gs\gs*}\frac{2(1-n)}{n}\ge_* - \gV_{\gs\gs*}^2.
\ee
This is largest for $\gV_{\gs\gs*} = {\textstyle\frac{n-1}{n}\ge_*}$, which corresponds to $\ns = 1-{\textstyle\frac{1}{2N_\gf}}\geq 0.992$ which is outside of the observed value. The maximum of the absolute value in \eqref{etaperconstraint} will then be given by the upper or the lower bound on $\ns$ (because in the interval of the observed value for $\ns$ it can change sign). Table \ref{table:getpe} gives the numerical constraints on $\left|-\frac{6}{5}\fnl \getpe_*\right|$ for integer values of $n$.
\begin{table}
		\centering
		\begin{tabular}{|c|c|c|c|c|c|}
		\hline
		$n$ & 2 & 3 & 4 & 5 & 6\\
		\hline
		$\left|-\frac{6}{5}\fnl \getpe_*\right|$ & $ 6.8 \times 10^{-5}$ & $5.0 \times 10^{-5}$ & $2.6\times 10^{-5}$ & $6.7 \times 10^{-5}$ & $1.2\times 10^{-4}$  \\
		\hline
	\end{tabular}
	\caption{Upper bounds from the spectral index on $\left|-\frac{6}{5}\fnl \getpe_*\right|$ for different $n$ with $N_\gf=60$.}
	\label{table:getpe}
\end{table}
We observe that the maximum value for $\getpe_*$ is two orders of magnitude smaller than $\ge_*$ for $\fnl$ of order unity. Moreover this limit is quite strong since the factor 0.325 \eqref{numfactor} is a limit which asks some fine tuning to be reached. This factor can easily be ten or a hundred times smaller. Hence, in most cases $\getpe_*$ will be a lot smaller than this limit.

To summarize, we know $\ge_*$ once we fix $N_\gf$. We then determine $\gV_{\gs\gs*}$ using $\ge_*$ and the observational constraints on $\ns$. This leads to an upper bound for $|\getpe_*|$ by imposing a value for $\fnl$. However, to see when the turn exactly happens, we need to know the full evolution of $\getpe$, not just its initial value. For this, some work needs to be done on the expression for $\getpe$ given in \eqref{srparmonomial}, where we can eliminate unknown quantities (like the parameters of the potential) by using the expressions for the slow-roll parameters at horizon crossing:
\be
\ge_*=\frac{n^2}{2}\frac{1}{\gf_*^2},\qquad \gV_{\gs\gs*}=\frac{m(m-1)\gb\gs_*^{m-2}}{\ga \gf_*^n}.
\ee
It is then straightforward to compute:
\be
\label{getpeevolution}
\begin{split}
&\gV_{\gs\gs}= \frac{V_{\gs\gs}}{\gk^2 U} =\gV_{\gs\gs*} \lh \frac{\gs}{\gs_*} \rh ^{m-2} \lh \frac{\ge}{\ge_*} \rh ^{n/2},\\
&\getpe = \getpe_* \lh \frac{\gs}{\gs_*} \rh ^{m-1} \lh \frac{\ge}{\ge_*} \rh ^{n/2} \frac{2\frac{n-1}{n}\ge^{1/2}-\gV_{\gs\gs}\ge^{-1/2}}{2\frac{n-1}{n}\ge_*^{1/2}-\gV_{\gs\gs*}\ge_*^{-1/2}}.
\end{split}
\ee
As already discussed, we want to express the time dependence in terms of $\ge$ which is directly related to $\gf$. However, the expression for $\getpe$ also depends on $\gs$, and while a bound for its initial value at horizon-crossing can be given using \eqref{srparmonomial} and the bounds on $\gV_{\gs\gs*}$ and $\getpe_*$, we need to know how it evolves with time. For this we solve the field equation:
\be
\dot{\gs} = - \frac{m\gb}{\ga} \frac{\gs^{m-1}}{\gf^n}.
\ee

Inserting the solution \eqref{phi(t)} for $\gf$ into the equation for $\gs$ we find
the following differential equation:
\be
\frac{\d\gs}{\gs^{m-1}} = - \frac{m\gb}{\ga} \frac{1}{\gf_*^n}
\frac{\d t}{(1-t/N_\gf)^{n/2}}.
\label{eqsig(t)}
\ee
We see that we need to consider the special cases $m=2$ and $n=2$ separately. We start with the most general cas $m\neq 2 $ and $n \neq 2$, where (with $\gs_*$ the initial value of $\gs$):
\be
\begin{split}
\gs &= \gs_* \left[1+\frac{m(2-m)}{n(2-n)}\frac{\gb}{\ga}\frac{\gs^{m-2}}{\gf_*^{n-2}}\lh \lh 1-\frac{t}{N_\gf}\rh^{1-n/2} - 1 \rh \right]^\frac{1}{2-m}\\
    &= \gs_* \left[1+ \frac{1}{2}\frac{m-2}{m-1}\frac{n}{n-2} \frac{\gV_{\gs\gs*}}{\ge_*^{n/2}} \lh \ge^{n/2-1}-\ge_*^{n/2-1}\rh\right]^{\frac{1}{2-m}}.
\end{split}
\ee
In the case $m\neq 2$ and $n=2$, we have:
\be
\begin{split}
\gs &= \gs_* \left[1+\frac{m(2-m)}{4\gs_*^{2-m}}\frac{\gb}{\ga}\ln{\lh 1-\frac{t}{N_\gf}\rh}\right]^{\frac{1}{2-m}}\\
    &= \gs_* \left[1+ \frac{1}{2} \frac{2-m}{m-1} \frac{\gV_{\gs\gs*}}{\ge_*} \ln{\lh\frac{\ge_*}{\ge}\rh}\right]^{\frac{1}{2-m}},
\end{split}
\ee
while for $m=2$ and $n \neq 2$:
\be
\begin{split}
\gs &= \gs_* \exp{\left[\frac{2\gb}{\ga}\frac{\gf_*^{2-n}}{n(2-n)} \lh \lh 1-\frac{t}{N_\gf}\rh^{1-n/2} - 1 \rh\right]}\\
    &= \gs_* \exp{\left[\frac{n}{2(2-n)}\frac{\gV_{\gs\gs*}}{\ge_*^{n/2}}\lh \ge^{n/2-1} - \ge_*^{n/2-1}\rh\right]}.
\end{split}
\ee
Inserting these expressions into \eqref{getpeevolution} gives the ratio $\getpe/\getpe_*$. In the last case $m=2$ and $n=2$, these equations take a nicer form:
\be
\gs =\gs_* \lh 1-\frac{t}{N_\gf}\rh^{\frac{\gb}{2\ga}}=\gs_* \lh 1-\frac{t}{N_\gf}\rh^{\frac{\gV_{\gs\gs*}}{2\ge_*}}= \gs_* \lh \frac{\ge}{\ge_*} \rh^{-\frac{\gV_{\gs\gs*}}{2\ge_*}},\qquad
\frac{\getpe}{\getpe_*}= \lh \frac{\ge}{\ge_*} \rh^{-\frac{\gV_{\gs\gs*}}{2\ge_*}+\frac{3}{2}}.
\ee

\subsection{Discussion}
\label{Mono Discussion section}

\begin{figure}
\centering
	\includegraphics[width=0.49\textwidth]{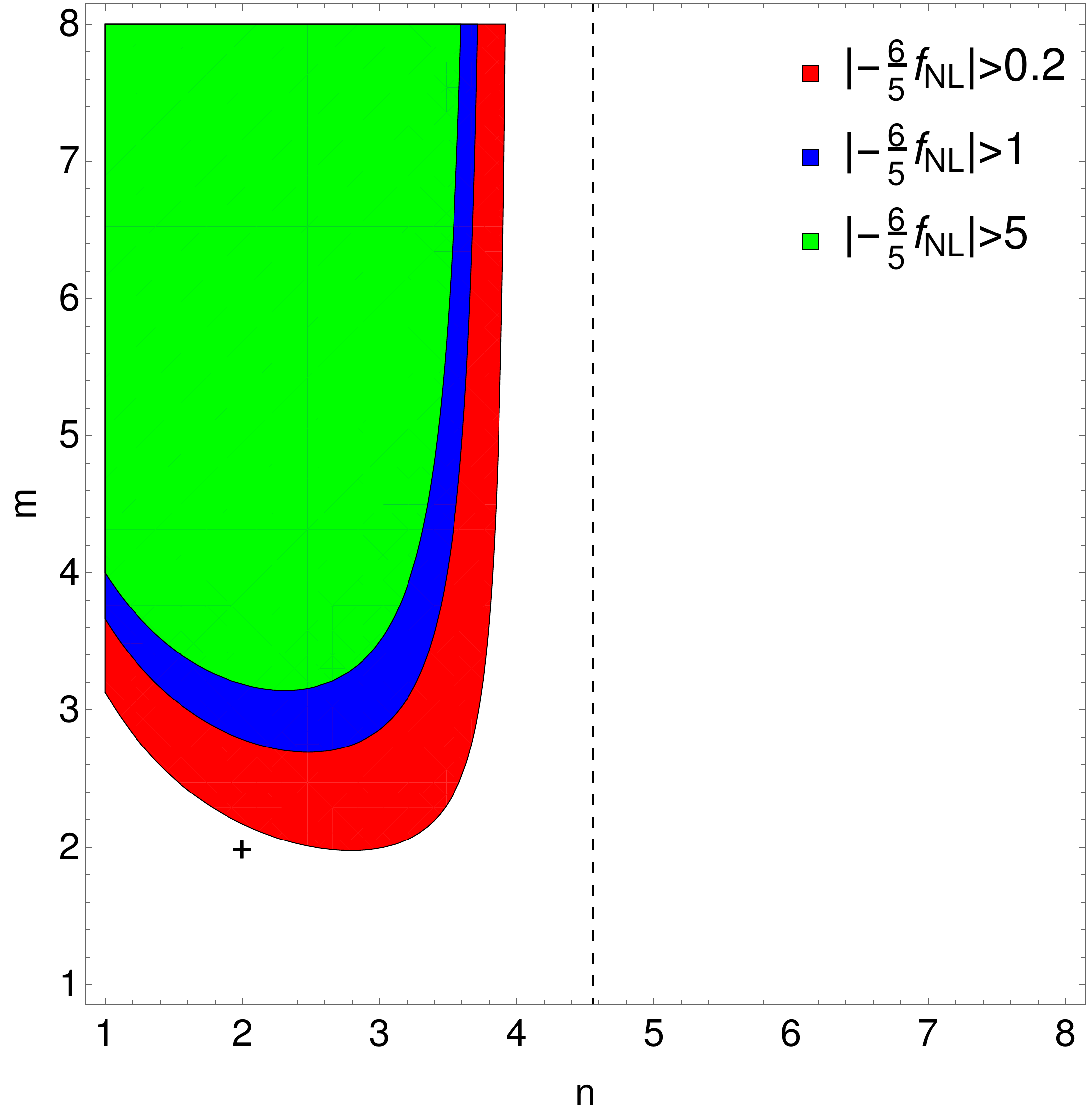}
	\includegraphics[width=0.49\textwidth]{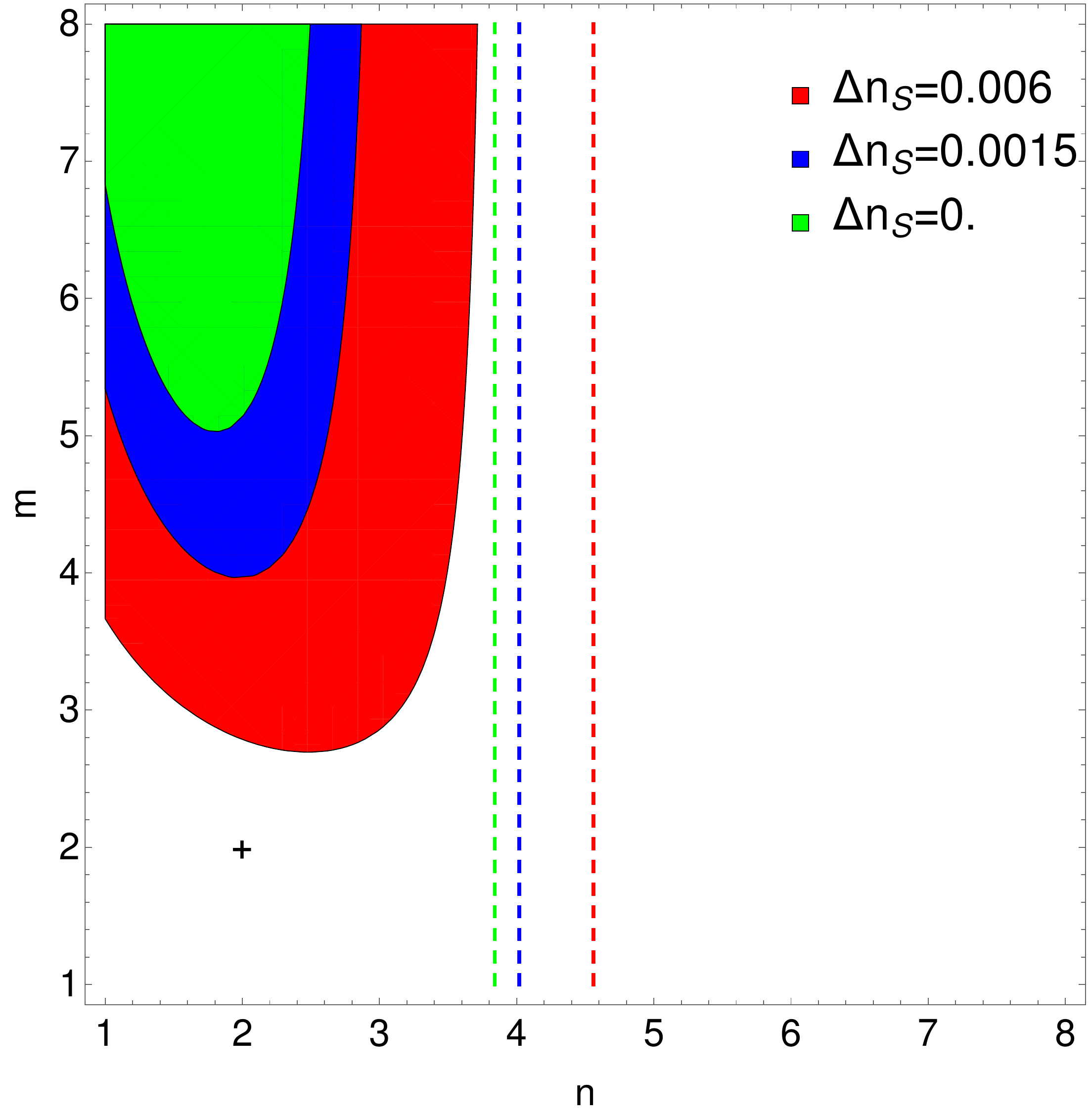}
	\includegraphics[width=0.49\textwidth]{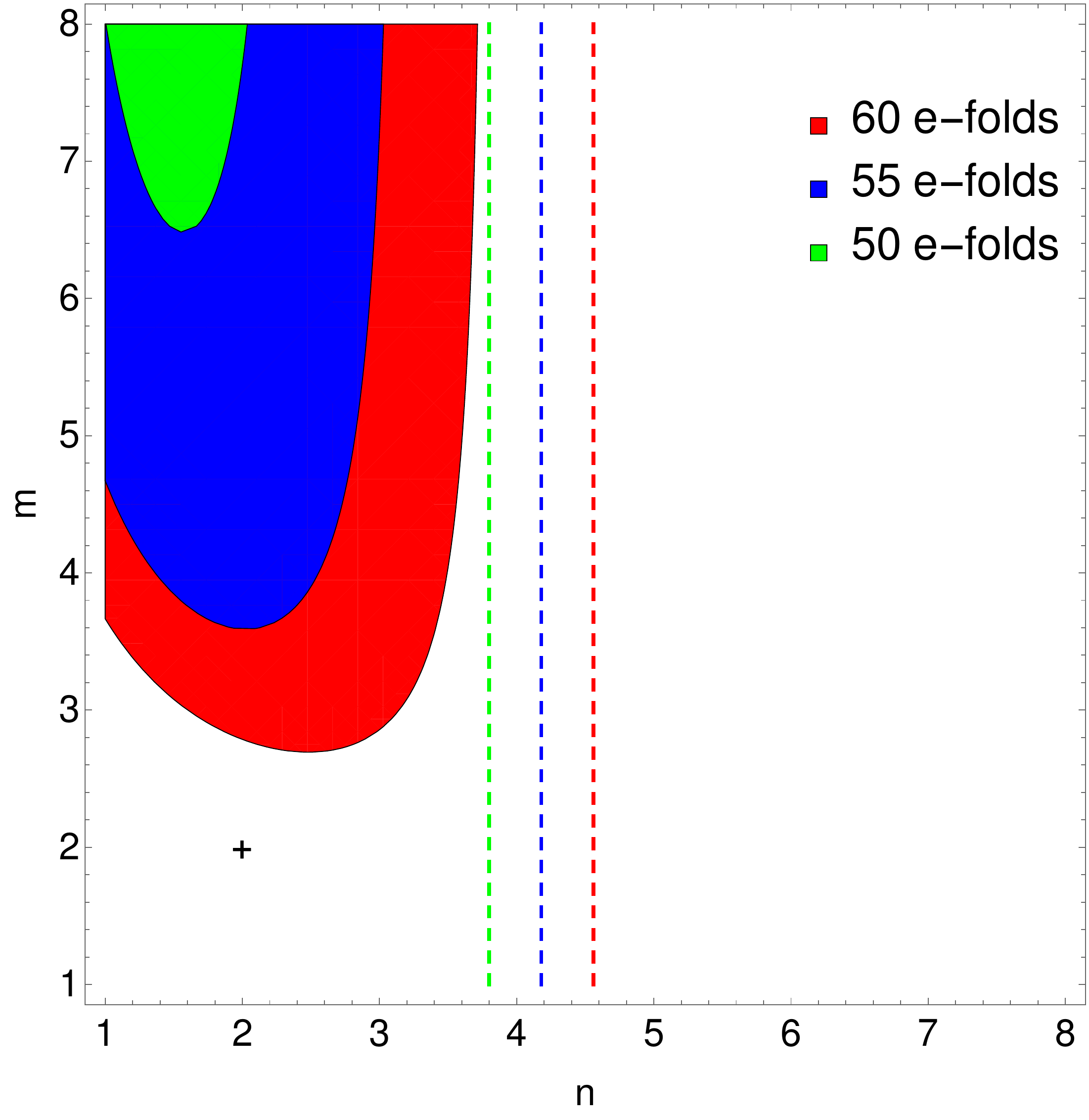}
	\includegraphics[width=0.49\textwidth]{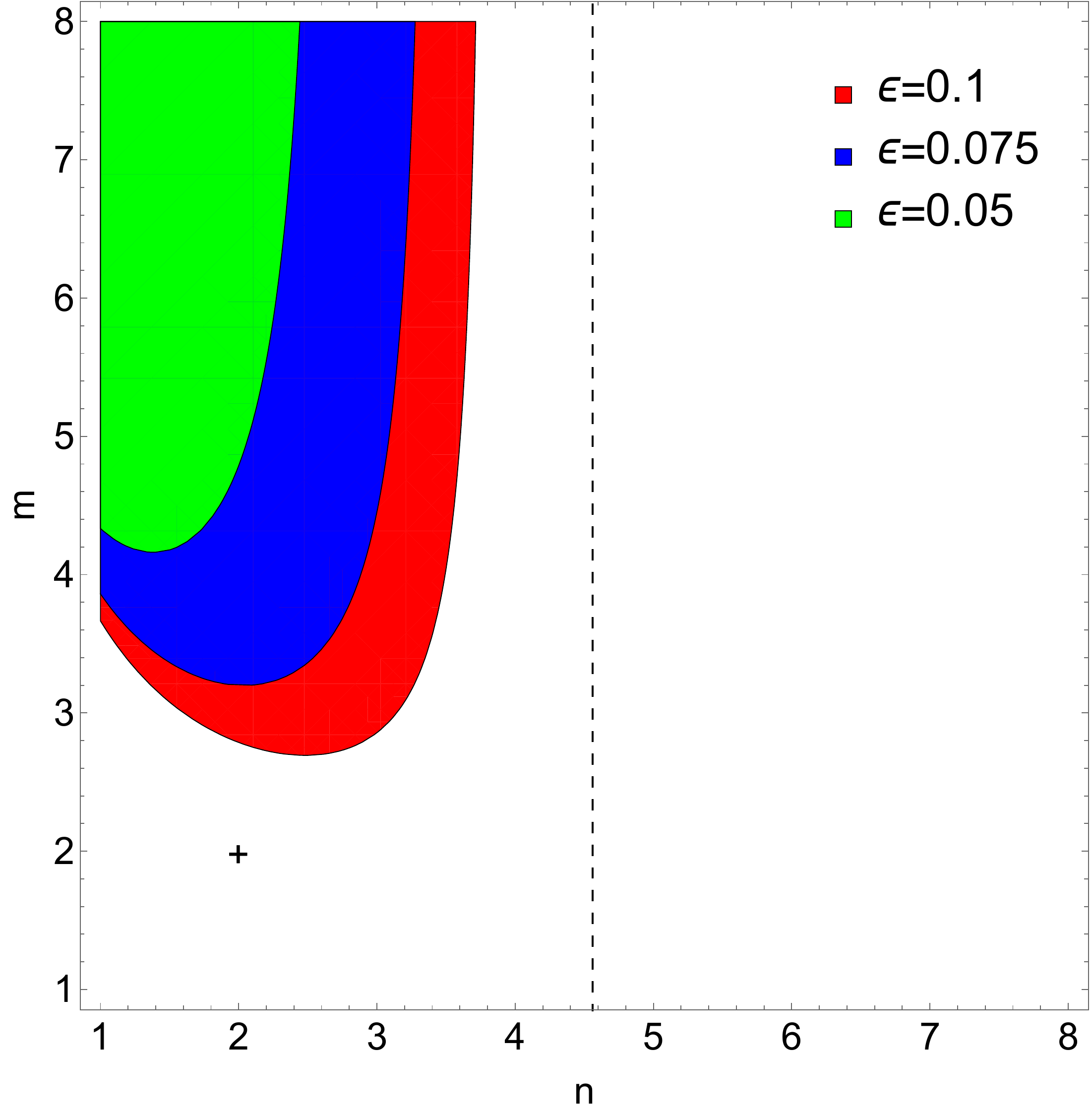}
	\caption{The regions of the parameter space of $m$ and $n$ where the turn of the field trajectory can occur before the end of the slow-roll regime. In the top left figure, these regions are determined for several values of $\fnl$: $|-\frac{6}{5}\fnl|$ larger than 5 (green), 1 (blue) or 0.2 (red). In the top right figure, we explore different error bars of the spectral index: the Planck constraint $\ns = 0.968 \pm 0.006$ (red), $\ns =0.968\pm 0.0015$ in blue (CORE-like experiment) and negligible error bars in green. On the bottom left, different values of $N_\gf$ are represented: $50$ (green), $55$ (blue) and $60$ (red). The last plot changes the constraint on $\ge$ to check when the turn can occur: $\ge=0.05$ (green), $0.075$ (blue) and $0.1$ (red). The dashed vertical lines indicate the change of sign of $\gV_{\gs\gs*}$, which depends on $n$, $N_\gf$ and $\ns$: it is necessarily positive on the right-hand side of this line. The small cross highlights that the double quadratic potential ($m=n=2$) is excluded in all plots.}
	\label{fig:etaper1}
\end{figure}

In figure~\ref{fig:etaper1}, we use the expressions of the previous section to determine the regions of the parameter space of $m$ and $n$ where a turn of the field trajectory might happen before the end of the slow-roll regime. For this we want to verify when multiple-field effects start to play a role or, in terms of slow-roll parameters, we want to find when $\getpe$ becomes of the same order as $\ge$. We choose $\ge$ and not $\getpa$ because $\getpa$ is of the same order as $\ge$ for most cases except if $n\approx 2$ when it is much smaller.

First, we choose the maximum value of $|\getpe_*|$ possible for $|-\frac{6}{5}\fnl|=1$ using the range of values for $\gV_{\gs\gs*}$ determined from the spectral index. Then we compute the maximum value of $|\getpe|$ when $\ge=0.1$. We choose this value of $\ge$ because this is already close to the end of inflation and the slow-roll approximation starts to break down after that point. Moreover, if the turn starts after this time, it is possible that there is not enough time for the isocurvature modes to decay. Finally, we plot the regions of the parameter space of $m$ and $n$ where $\getpe$ is at least as large as $\ge$ at that time, meaning there is a turn of the field trajectory. We also assume that $N_\gf = 60$. These are the default values for the parameters $\fnl$, $N_\gf$ and $\ge$. Next we vary them to test the validity of these choices. We also explore the effects of a future improvement of the spectral index measurements.

The main conclusion of figure~\ref{fig:etaper1} is that for most $m$ and $n$, the turn cannot happen before the end of the slow-roll regime, except in the top left part of the figures (small $n$ and large $m$). For example, the simple quadratic case $m=2$ and $n=2$ (indicated by a small cross) is excluded. 

The first figure shows that obviously the space of allowed parameters decreases if we want $\fnl$ to be larger. In fact, imposing a larger $\fnl$ is the same as imposing a smaller $\getpe_*$. This does not change the evolution of $\getpe$, only its initial condition, so that it will be harder to reach a final value of order $\ge$.

In the second figure, we explore the effects of an improvement of the measurements of the spectral index by comparing the Planck result $\ns=0.968 \pm 0.006$, with the accuracy expected with a CORE-like experiment where the error bar would be of order $\Delta\ns=0.0015$. We also add the case where the error bar becomes negligible. We see that the region where $\fnl$ is at least of order unity is strongly dependent on the spectral index. Decreasing the error bars on $\ns$ decreases the parameter region where $\fnl$ is of order unity. We will see later that in fact it is the lower bound of $\ns$ which matters. If a more accurate measurement would shift the central value of $\ns$, so that its lower bound would be slightly smaller than for Planck, then the size of the top-left region in this plot would increase. This is not indicated in the figure to keep the plot from being too busy, but $\ns = 0.94$ is sufficient to allow most of the parameter region in the figure ($m>2$ and $n<7$).

The third plot shows the effect of the parameter $N_\gf$. We do not know exactly the total duration of inflation; the usual value is between 50 and 60 e-folds. Moreover, we cannot be sure that $N_\gs$ can be neglected, which means that $N_\gf$ is not necessarily the full duration of inflation after horizon-crossing. In this figure, we observe that the surface of the top left region diminishes for smaller $N_\gf$. In fact, for $N_\gf$ smaller than 45 e-folds, it vanishes completely. The smaller $N_\gf$, the harder it will be to build a model where $\fnl$ is large.

The last figure is here to help to determine at what time the turn can occur. In the other figures, the only condition was before the end of the slow-roll regime. However, this regime is valid for most of the time after horizon-crossing. We can see that simply reducing $\ge$ by a factor two reduces a lot the allowed parameter region. This means that having a turn a few e-folds after horizon-crossing is extremely hard to have or even impossible. Most of the time the turn will happen near the end of slow-roll. 

To explain these different behaviours, we first need to discuss $\gV_{\gs\gs*}$. It is determined from the spectral index and $\ge_*$ using equation \eqref{Wss spectral index} which contains two terms: $ \half(\ns-1)$ which is negative and larger in absolute value for the lower bound on the spectral index, and $\ge_*$ which is positive and can be either smaller or larger than the first term. A small $\ge_*$ corresponds to small $n$ and/or large $N_\gf$. This means that in each of the four figures, the left (small $n$) corresponds to a negative $\gV_{\gs\gs*}$, while $\gV_{\gs\gs*}$ is positive on the right (large $n$). The transition happens between $n=4$ and $n=5$ for $N_\gf=60$ for example. If we decrease $N_\gf$, this value decreases and the transition is shifted to the left. The same happens if we increase the lower bound on the spectral index. In every figure this transition is indicated by a dashed vertical line. The sign of $\gV_{\gs\gs*}$ is important because this corresponds to the form of the potential $V$ at horizon-crossing. If it is positive we have a valley, while a negative value describes falling off a ridge.

Now that we have seen the role of the other parameters on $\gV_{\gs\gs}$, we have to explain the different regions by looking at the equations for the evolution of the ratio $\getpe/\getpe_*$ for the different cases. In the valley case ($\gV_{\gs\gs*}>0$), $\gs$ has to decrease to the minimum at $\gs=0$. However, because the potential has to be very flat at horizon-crossing, we start close to the minimum. Even if $\gs$ reaches its minimum before $\gf$, $\getpe$ does not have the time to become large because in $\getpe$, the decrease of $\gs$ is opposed by the increase of $\ge$. Hence, there is no allowed parameter region to the right of the dashed vertical line in the figures.

In the region of negative $\gV_{\gs\gs*}$, the situation is the opposite: $\gs$ increases to fall from the almost flat ridge where it started. Hence in $\getpe$ we have the effect of both $\ge$ and $\gs$ increasing. After inserting $\gs$ for the different cases into \eqref{getpeevolution}, the only dependence on $m$ appears in the ratio $(m-2)/(m-1)$ which tends to $1$ when $m$ increases. This explains the asymptotic behaviour which appears on the right-hand side of the allowed region.
 
Looking at the different expressions for $\gs$, we also see that the largest $\gV_{\gs\gs*}$ in absolute value makes $\gs$ increase the fastest. This implies that the lower bound on the spectral index is the most important to obtain $\gV_{\gs\gs*}$. When $n$ decreases, larger (in absolute value) $\gV_{\gs\gs*}$ are possible, which explains why smaller $m$ are allowed. But in $\gs$ there are also terms which decrease when $n$ becomes smaller and which compensate this effect, which is why for even smaller $n$ the minimum required value of $m$ starts to increase again.

At the end of section \ref{Sum Potential Section}, the difficulty, or at least the high level of fine-tuning, needed for a model where $\fnl$ is of order unity or more in slow-roll has been highlighted. Here, we showed explicitly that this is even impossible most of the time for simple monomial potentials. However, some examples exist, when $m>4$ and $n<4$ generally. We also showed that $N_\gf$ has to be close to the total number of e-folds after horizon-crossing which should be as large as possible given other constraints (around 60 e-folds), which implies that the turn of the field trajectory is quick. This also means that slow-roll parameters like $\ge_*$ and $\getpa_*$ are exactly the same as in the purely single-field case. However, the observables $\ns$ and $\fnl$ are different. Adding a second field which is responsible for the non-negligible $\fnl$ can help some single-field models which were not working well given the Planck constraints on $\ns$ to go back into the allowed range of parameters. However, this asks a lot of fine-tuning of the potential of the second field. For $\fnl$ to be of order unity or more, this asks even more fine-tuning as only the lowest spectral index values will work. This also means that the improvement of the spectral index measurements expected with a satellite like CORE would seriously constrain the possiblity of having a large $\fnl$, especially if the central value of the spectral index moves closer to the upper bound from Planck.

We have also seen that in the cases that do work, most of the time the turn is near the end of the slow-roll period. This means that $\ge$ and the other parameters are already of order $0.1$ at the start of the turn. Then parameters like $\getpa$ and $\getpe$ can easily become of order 1 or more during the turn when things are getting more violent. The slow-roll approximation is then broken anyway. If the turn happens a bit later, we can expect that isocurvature modes will not have enough time to vanish before the end of inflation (this does not exclude the existence of some cases where they vanish in time, but only a numerical study of such examples is possible). Finally, we can imagine a case where the turn has not started when $\gf$ reaches the minimum of its potential. If this happens, there is a period of large $\ge$ (which would be the end of inflation in the single-field case). Again, during this period the slow-roll approximation is no longer valid. Therefore, these different situations show the need to understand what happens if the very useful slow-roll approximation is not sufficient. This is the topic of the next section.

\section{Beyond the slow-roll regime} 
\label{Beyond slow-roll}

The previous section showed that it is difficult to have $\fnl$ not be slow-roll suppressed in the slow-roll regime. Is the situation the same if we leave this regime for a short period?
Here we discuss different cases where this can happen and we will show that like in the slow-roll situation, only the homogeneous part of the solution of \eqref{equadiff} is relevant once isocurvature modes have vanished. This means we will use the same quasi-single-field initial conditions at horizon-crossing as at the end of section \ref{Sum Potential Section}: $V_* \ll U_*$ and $|V_{\gs*}|\ll |U_{\gf*}|$ while $|V_{\gs\gs*} U_{\gf*}^2|\gg |U_{\gf\gf*}V_{\gs*}^2|$ and $|V_{\gs\gs*} V_{\gs*}^2|\ll |U_{\gf\gf*}U_{\gf*}^2|$.

\subsection{Two kinds of turns}
\label{two turns section}

\begin{figure}
	\centering
	\includegraphics[width=0.4\textwidth]{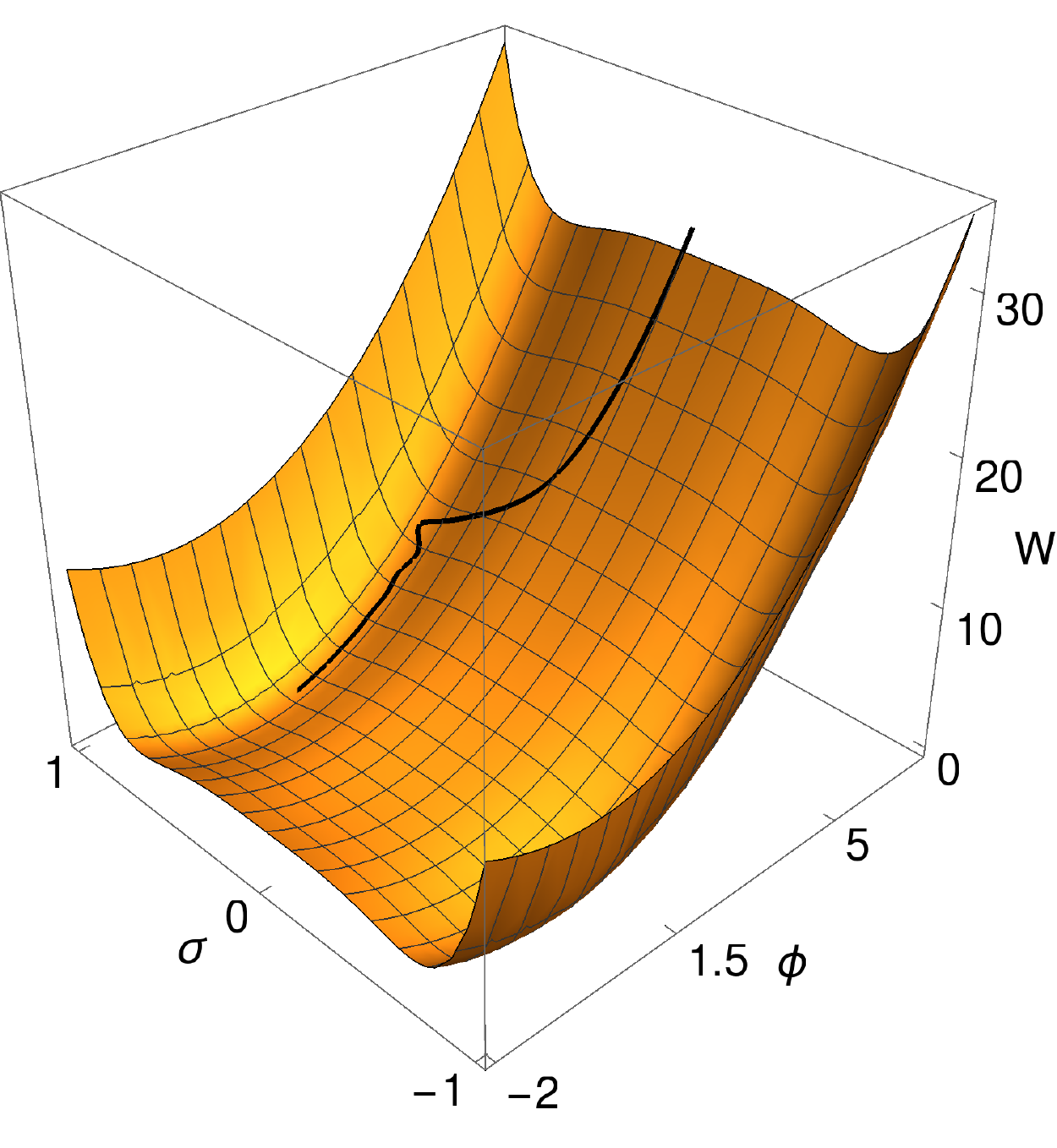}
	\includegraphics[width=0.55\textwidth]{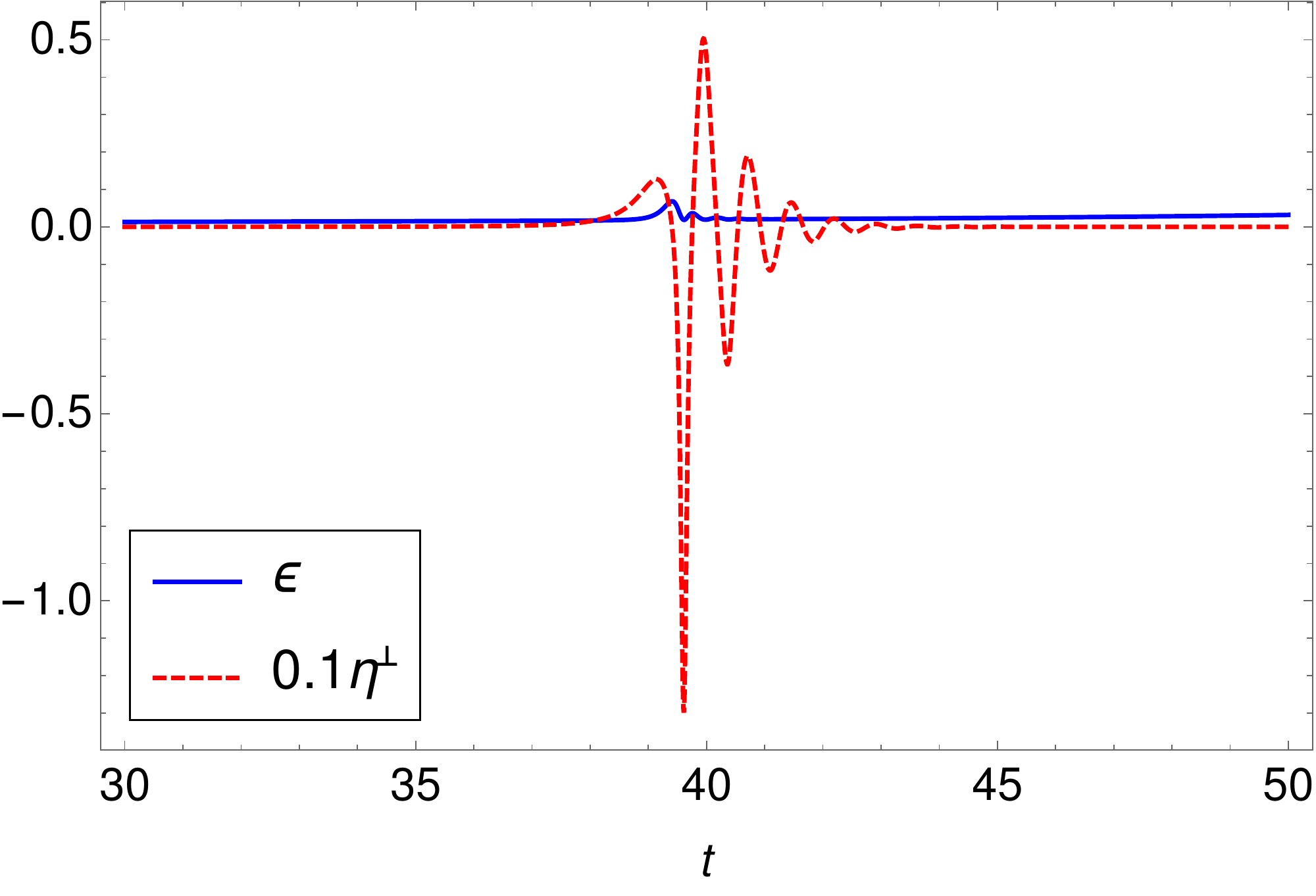}
	\includegraphics[width=0.4\textwidth]{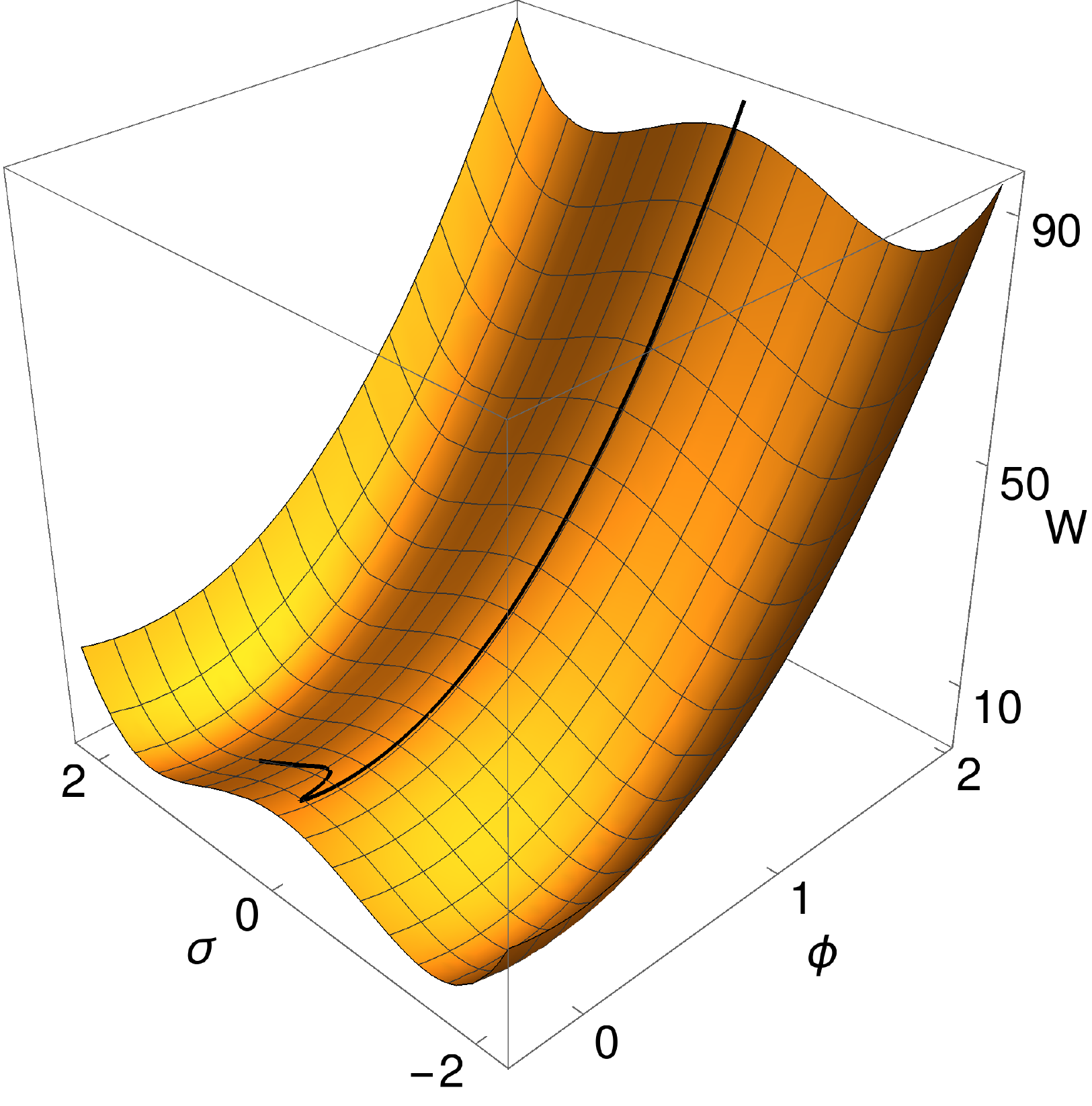}
	\includegraphics[width=0.55\textwidth]{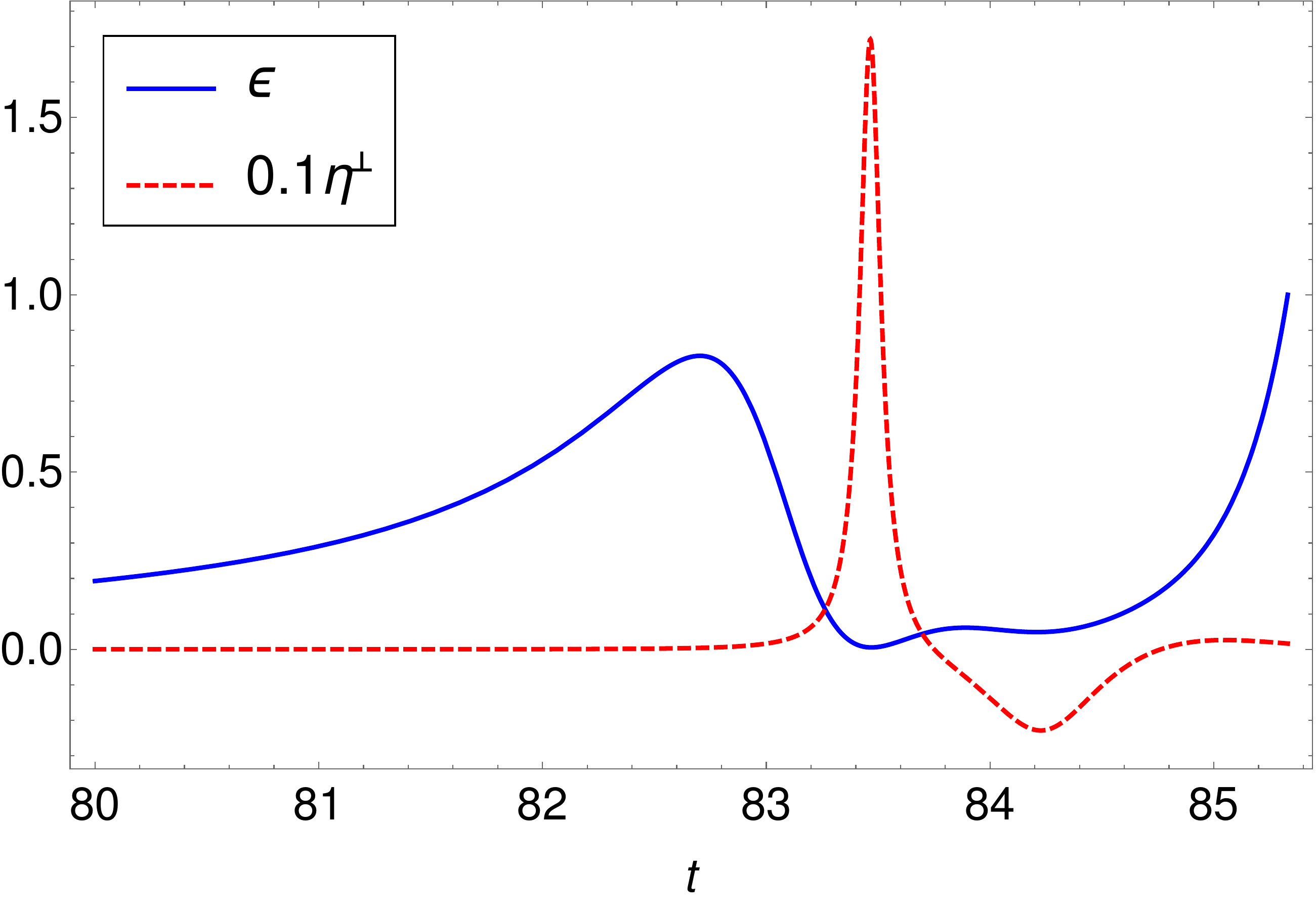}
	\caption{Illustrations of the two different types of turn where the slow-roll approximation is broken. On the left, the field trajectory is displayed in black on the potential while on the right the slow-roll parameters $\ge$ and $\getpe$ are shown for a typical example. The top correspond to what we call the first type (see sections \ref{n2m4 section} and \ref{Axion section} for examples), while the plots at the bottom show the second type (see sections \ref{Quad section}, \ref{n2m2 section} and \ref{last section} for examples).}
	\label{fig:slowrollbroken}
\end{figure}

We identified two different cases, illustrated in figure \ref{fig:slowrollbroken}, where the slow-roll approximation stops to be valid during the turn. 
In figure \ref{fig:slowrollbroken}, the main differences of the two situations are highlighted. With potentials of a quite similar form, we have the possibility for two different trajectories depending on the direction before and after the turn. In the previous section, the importance of the parameters $\ge$ and $\getpe$ to study the turn has been highlighted. Graphically they are useful to determine when the turn occurs and when the slow-roll regime is broken.

The first case is the one studied in the previous section. We determined that for a simple monomial potential, if the turn is possible before $\gf$ reaches the minimum of its potential, it is more likely to happen in the last few e-folds when slow-roll parameters are already of order $10^{-1}$, at the limit of the slow-roll approximation. Then, during the turn, $\eta$ parameters may become of order unity or more, which completely invalidates the idea of an expansion in terms of small slow-roll parameters. The turn is still early enough to have $\getpe$ small again at the end of inflation to make the isocurvature mode vanish. In this case the direction of the field trajectory is the same before and after the turn. This is compatible with a monomial potential where we established that $N_\gs$ has to be small compared to $N_\gf$ and the turn is then short. 

In the second case, perpendicular terms are still negligible when $\ge$ becomes of order $10^{-1}$. Then, like in single-field inflation, $\ge$ continues to grow. This is the end of the slow-roll regime. From \eqref{srderivatives} we see that this makes $\getpa$ also become large (in absolute value) and a maximum of $\ge$ is reached when $\getpa= -\ge$. A short time after that point, $\ge$ starts to decrease very fast as the $\getpa$ term dominates in $\dot{\ge}$. A large $\getpa$ also has an effect on the perpendicular parameter $\getpe$ which has been negligible until then. It is possible that $\getpe$ becomes large and that the turn will occur after a few e-folds at most if $\fnl$ is of order unity, see appendix \ref{appendix case 2}. Hence, it is possible to have the turn starting with $\ge \ll 1$. This is also motivated by the assumption of isocurvature modes vanishing before the end of inflation. Indeeed, this requires a turn not too close to the end of inflation ($\ge=1$) which is the case if $\ge$ is small compared to one during the turn. In this type of turn, the direction is not the same before and after. Before the turn $\gf$ is dominating but also near the minimum of its potential, while $\gs$ is still at a local maximum. Inflation ends when $\gf$ is still near its minimum but $\gs$ is also evolving towards its own minimum.

In both theses cases, we established that the slow-roll approximation can be broken. We know that solving the equations without any approximation is not possible, even in the simple case of a sum potential. However, we have also seen that $\ge$ is small at the start of the turn simply because of the assumption of vanishing isocurvature modes. Moreover, in $\dot{\ge}$ \eqref{srderivatives}, there is a factor $\ge$ in front. This means that when $\ge$ is small, $\ge$ cannot evolve very fast and will stay small during a short period like the turn, unless the turn is very sharp with $\eta$ parameters becoming very large. Hence during the turn, except in the most extreme cases we do not treat, we still have that $\ge$ is small compared to one which will play an important role in this section.

In the first type of turn, this hypothesis of small $\ge$ has the important consequence that the slow-roll approximation is in fact broken only for the field $\gs$. Indeed, in the field equation \eqref{fieldeq}, each field can only affect the other through $H$ which evolves slowly if $\ge \ll 1$. Hence, even if $\gs$ starts to evolve fast, it is only a small perturbation for $\gf$ which continues to evolve slowly during and after the turn until near the end of inflation when $\ge \approx \gk^2 \dot{\gf}^2 /2$ becomes of order unity. Hence, the derivatives of $\gf$ of order two and more are negligible. This can be used to simplify the slow-roll parameter expressions from \eqref{defeta}, keeping only the terms which are larger than order slow-roll:
\be
\getpa = \frac{\ddot{\gs}\dot{\gs}}{\dot{\gf}^2+\dot{\gs}^2}, \qquad \getpe = -\frac{\ddot{\gs}\dot{\gf}}{\dot{\gf}^2+\dot{\gs}^2}, \qquad \gxpa = \frac{\dddot{\gs}\dot{\gs}}{\dot{\gf}^2+\dot{\gs}^2} \qquad \text{and}\qquad \gxpe = -\frac{\dddot{\gs}\dot{\gf}}{\dot{\gf}^2+\dot{\gs}^2}.
\ee
Using this, a direct computation gives useful relations between the parallel and perpendicular parameters of the same order:
\be 
\label{first case}
e_{1\gf}\getpa = -e_{1\gs}\getpe \qquad\text{and}\qquad e_{1\gf}\gxpa = -e_{1\gs}\gxpe.
\ee

In the second type of turn, the slow-roll approximation is broken for the two fields, so that these relations are then not valid. However, there is also an important approximation we can make in this case. Before the turn, the slow-roll approximation is broken during the period of large $\ge$. Having $\ge$ large for some time also means that $H$ decreases a lot during that period. This means that during the turn, we have:
\be
H^2 \ll H_*^2.
\label{second case}
\ee
A brief remark about the end of inflation is necessary. We use the common definition that the period of inflation finishes when $\ge=1$. However, in the second type of turn, $\ge$ can be larger than 1 for a very small number of e-folds before the turn. A more complete definition of the end of inflation is then that $\ge=1$ with $U\ll U_*$ and $V\ll V_*$, which ensures that the second field as well had time to evolve.

The main tool in this section is the differential equation \eqref{equadiff} which we will  call the $\gint$ equation. We have already solved it during the period of slow-roll which goes from horizon-crossing to the turn or to $\ge$ of order 1. We also know the exact homogeneous solution of the full equation. The only remaining work is to understand what happens to the particular solution beyond the slow-roll approximation. We will each time follow the same method. First we discuss each equation in the more general case, only supposing that $\getpa$ and $\getpe$ are large while $\ge \ll 1$. Then, when needed to go further, we will study separately each case using \eqref{first case} or \eqref{second case} depending on the type of turn considered.

\subsection{Green's functions}
\label{Green's functions section}

Beyond the slow-roll regime, we have to solve the second-order differential equation \eqref{G22eq} to compute the Green's functions (recalling that $\gv_{22}(t)$ and $G_{22}(t,t_*)$ obey the same equation). We assume that the solution has the form $\gv_{22} \propto f e_{1\gf}e_{1\gs}$, similar to the slow-roll case (\ref{Green srsolution}). One motivation is that, during the turn, the dominant term will be $(\getpe)^2$ and this is canceled by this form of solution. Substituting this into \eqref{G22eq}, we find a differential equation for the function $f$:
\be
\label{feq}
\begin{split}
e_{1\gf}e_{1\gs}\,\ddot{f}&+\left[2\getpe(e_{1\gs} ^2 - e_{1\gf} ^2) + (3 + \ge + 2\getpa)e_{1\gf}e_{1\gs}\right]\dot{f} \\ &+\left[2\ge\getpe(e_{1\gs} ^2 - e_{1\gf} ^2)+(6\ge+2\ge^{2}+4\ge\getpa)e_{1\gf}e_{1\gs}\right]f=0.
\end{split}
\ee
In the slow-roll regime, a first-order expansion of this equation gives
\be
\label{feqsumsr}
e_{1\gf}e_{1\gs}\,\dot{f} +2\ge\, e_{1\gf}e_{1\gs}f=0,
\ee
and then it is easy to show that $f = H^2$ to find the slow-roll result (\ref{Green srsolution}). During this initial period of slow roll, having a first-order equation as a very good approximation means that the second mode needed to solve the full equation rapidly becomes negligible. Once slow roll is broken, we only need to study how the remaining mode evolves.

In the general case, an analytical solution cannot be found. However, if we take a solution of the form $f=H^\ga$ by inspiration from the slow-roll solution (because that is the form of the solution until the moment when the slow-roll regime is broken), \eqref{feq} becomes:
\be
\begin{split}
&e_{1\gf}e_{1\gs}\left[\ddot{\ga}\ln H + \dot{\ga}^{2} (\ln H)^2-\ga\dot{\ga}\ln H + \dot{\ga}\lh -2\ge+(3+\ge+2\getpa)\ln H\rh \right. \\
&\left. + (\ga-2)(\ga-1)\ge^2 -4(\ga-1)\ge\getpa-3(\ga-2)\ge\right] + 2\getpe(e_{1\gs} ^2 - e_{1\gf} ^2) \left[ \dot{\ga}\ln H +(1-\ga)\ge \right] = 0.
\end{split}
\ee
There are two interesting values for $\ga$ which are 1 and 2. They can be linked to the two regimes already discussed previously where the slow-roll approximation is not valid.

We can see directly that the lowest order term in slow-roll is canceled by $\ga = 2$ as expected. Moreover, the $\ge^2$ term also vanishes with this value. This means that when $\ge$ becomes larger while the other parameters are still small compared to 1, $f=H^2$ is still a good approximation. This is exactly what happens at the end of the slow-roll regime just before the second type of turn, when the first field is near the minimum of its potential. Then, the complete solutions for the Green's functions are:
\be
\gv_{22} = \frac{H^2 e_{1\gf} e_{1\gs}}{H^2_* e_{1\gf*}e_{1\gs *}},\qquad
\gv_{32} =  \frac{H^2}{H^2_* e_{1\gf*}e_{1\gs *}} \lh -2\ge e_{1\gf}e_{1\gs} +  \getpe (e_{1\gs}^2 - e_{1\gf}^2)\rh.
\ee
The same integration as in the slow-roll case works to compute $\gv_{12}$:

\be
\gv_{12} = \frac{Z-Z_*}{W_* e_{1\gf*}e_{1\gs*}},
\ee
with $Z$ previously introduced in \eqref{Z}.

The other interesting value $\ga=1$ cancels every second-order term in the equation. Hence, this is a good solution when $\getpe$ and $\getpa$ are large but $\ge$ is small compared to 1, hence during the turn. The solutions are then, 

\be
\label{v22turn}
\gv_{22} = \frac{H}{\cal{N}}e_{1\gf}e_{1\gs}, \qquad
\gv_{32} =  \frac{H}{\cal{N}} \lh -\ge e_{1\gf}e_{1\gs} +  \getpe (e_{1\gs}^2 - e_{1\gf}^2)\rh,
\ee
where $\cal{N}$ is a constant used to satisfy the continuity of $\gv_{22}$. If we call the time when this solution becomes better than the previous one $t_0$, we have $\cal{N}$ $=\frac{H^2_*}{H(t_0)} e_{1\gf*}e_{1\gs*}$. 

We cannot directly compute $\gv_{12}$ in this regime. However, $\ge$ is supposed to be very small compared to 1 which means that $H$ is almost a constant. We can then write $H(t) = H_0 + \delta H(t)$ where $\delta H(t)$ is only a small correction. Taking the square of this expression and doing a first-order slow-roll expansion gives $\delta H(t)=\half \textstyle\frac{ H^2-H_0^2}{H_0}$. Then it is easy to deduce $H(t) = \half \textstyle\frac{ H^2+H_0^2}{H_0}$. Substituting this into \eqref{v22turn}, we can perform the integration and we get:
\be
\label{v_12}
\gv_{12} = - \frac{H_0^2}{H_*^2} \frac{S-S_0}{4 e_{1\gf*}e_{_1\gs*}} + \frac{Z/2 + Z_0/2 - Z_*}{W_* e_{1\gf*} e_{_1\gs*}},
\ee
with $S\equiv e_{1\gf}^2- e_{1\gs}^2$.

\subsection{The $\gint$ equation during the turn}

A first use of the Green's functions during the turn computed in the previous section is to insert them into \eqref{equadiff} to simplify the right-hand side of the equation: r.h.s.~$\equiv K_{22} (\gv_{22})^2 + K_{23} \gv_{22}\gv_{32} + K_{33}(\gv_{32})^2$. After this step, every term of r.h.s.\  has one factor depending on the basis components: $e_{1\gf}^2 e_{1\gs}^2$, $e_{1\gf}e_{1\gs}(e_{1\gs}^2-e_{1\gf}^2)$ or $(e_{1\gs}^2-e_{1\gf}^2)^2$. We use the relation $(e_{1\gs}^2-e_{1\gf}^2)^2=1-4e_{1\gf}^2 e_{1\gs}^2$ coming from the normalization of the basis to eliminate one of the factors. Having terms with these factors permits us to use equations \eqref{W11eq2} and \eqref{sumequation} to eliminate the slow-roll parameters $\gc$, $\gw_{221}$ and $\gw_{222}$. Finally, we obtain:
\be
\begin{split}
\mathrm{r.h.s.}= &\ge \lh\frac{H}{\cal{N}}\rh ^2 \left\{ 2(\getpe)^4(\ge+3\getpa) + e_{1\gf}^2 e_{1\gs}^2 \left[ (\getpe)^4 \lh -18-14\ge-36\getpa\rh - 2(\getpe)^3 \gxpe \right.\right. \\
&+ (\getpe)^2 \lh-3\gw_{111} -18\ge - 6\ge^{2} -24\ge\getpa + 18(\getpa)^2 + 6\gxpa + 2 \ge^2 \getpa + 10 \ge (\getpa)^2 + 12 (\getpa)^3  \right. \\
&\left.\left. + 2\ge \gxpa + 12 \getpa \gxpa\rh + \getpe\gxpe \lh 6\ge-6\getpa-2\ge\getpa-10(\getpa)^2 -2\gxpa\rh + 2\getpa (\gxpe)^2 + 3\getpe \getpa \gw_{211}  \right]\\
&+ e_{1\gf}e_{1\gs} (e_{1\gs}^2-e_{1\gf}^2)  \left[ -6(\getpe)^5 + (\getpe)^3 \lh- 6\ge + 18\getpa + 2\ge^2 + 12 \ge \getpa + 18 (\getpa)^2 + 6 \gxpa \rh\right.\\
&\left.\left. -4\getpa(\getpe)^2\gxpe \right]\right\}. 
\label{rhs}
\end{split}
\ee 
At first sight, this expression does not look simpler than the original one. However, it has an important new feature which is the $\ge$ factor in front of the whole expression. In fact, in the computation every term without $\ge$ cancels. Recalling that the main assumption we made is that $\ge$ is small during the turn, this indicates that r.h.s.\  might be negligible during the turn, which means that only the homogeneous solution (which is known) is needed. In the rest of this section we will show that this is indeed the case. 

First we have to figure out compared to what r.h.s.\  has to be negligible. One way to answer this question is to use what we already know about the solution: the slow-roll expression given in \eqref{gint} which we write as $\dot{g}_{\mathrm{int}} = P_{\mathrm{sr}} + h_{\mathrm{sr}}$ with:
\be
P_{\mathrm{sr}} = 2\ge(\ge+\getpa)(\gv_{22})^2 + 2\ge\gv_{22}\gv_{32},\qquad
h_{\mathrm{sr}} = - \frac{e_{1\gf*}^2\gV_{\gs\gs*} - e_{1\gs*}^2\gU_{\gf\gf*}}{e_{1\gf*}e_{1\gs*}} \getpe\gv_{22}.
\label{dotgint}
\ee
Here we used that $\gv_{32}=-\gc \gv_{22}$ in the slow-roll regime. $P_{\mathrm{sr}}$ corresponds to the particular solution while $h_{\mathrm{sr}}$ is the homogeneous part. We will study these two parts of the solution in the two next sections to see how they evolve beyond the slow-roll regime. In section \ref{solution section}, we will discuss why they are sufficient to solve the $\gint$ equation even beyond the slow-roll approximation. We start by focusing on this homogeneous solution.

\subsection{Fate of the slow-roll homogeneous solution}
\label{homogeneous solution section}

As already discussed at the end of section \ref{Definitions}, the homogeneous slow-roll solution is also a homogeneous solution of the full second-order equation. Hence, we can use it and substitute it into \eqref{equadiff}. Then we look at each term (order1 $\propto \dot{h}_{\mathrm{sr}}$, order2 $\propto \ddot{h}_{\mathrm{sr}}$ and order3 $\propto \dddot{h}_{\mathrm{sr}}$) individually and not at the total sum because that is obviously zero. We want to show that these terms are large compared to r.h.s., so that, during the turn, r.h.s.\  is only a small correction which can be neglected to get a good approximation of $\gint$. To compute the three left-hand side terms, we use the same steps as in deriving (\ref{rhs}) to get:
\be
\begin{split}
\mathrm{order1} = & -\frac{H}{\cal{N}} \frac{e_{1\gf*}^2\gV_{\gs\gs*} - e_{1\gs*}^2\gU_{\gf\gf*}}{e_{1\gf*}e_{1\gs*}} \left\{  (e_{1\gs}^2-e_{1\gf}^2) \left[-6(\getpe)^4 -2(\getpe)^3\gxpe \right] + e_{1\gf}e_{1\gs} \right. \\
&\left. \times\left[ 6(\getpe)^5 + (\getpe)^3 \lh 6(\getpa)^2 +2\gxpa\rh  -8\getpa\gxpe(\getpe)^2  + 2\getpe (\gxpe)^2 + 3 (\getpe)^2\gw_{211} \right] \right\} \\
\mathrm{order2} = & \frac{H}{\cal{N}} \frac{e_{1\gf*}^2\gV_{\gs\gs*} - e_{1\gs*}^2\gU_{\gf\gf*}}{e_{1\gf*}e_{1\gs*}} \left\{ (e_{1\gs}^2-e_{1\gf}^2) \left[ (\getpe)^4 \lh-3+\ge-6\getpa\rh + 2(\getpe)^3 \gxpe\right] + e_{1\gf}e_{1\gs} \right. \\
& \left. \times \left[ (\getpe)^3 \lh 6\getpa-2\ge\getpa+12 (\getpa)^2\rh + (\getpe)^2 \gxpe \lh -3 +\ge -10 \getpa\rh + 2\getpe (\gxpe)^2 \right]\right\}\\
\mathrm{order3} = & \frac{H}{\cal{N}} \frac{e_{1\gf*}^2\gV_{\gs\gs*} - e_{1\gs*}^2\gU_{\gf\gf*}}{e_{1\gf*}e_{1\gs*}} \left\{ (e_{1\gs}^2-e_{1\gf}^2) \left[ (\getpe)^4 \lh-3-\ge+6\getpa\rh - 4 (\getpe)^3 \gxpe\right] + e_{1\gf}e_{1\gs}\right. \\ 
&\hspace{-1.5cm} \left. \times \left[ 6(\getpe)^5 + (\getpe)^3 \lh -6\getpa +2\ge\getpa - 6 (\getpa)^2 +2 \gxpa\rh + (\getpe)^2 \gxpe \lh 3 -\ge + 2 \getpa \rh +3 (\getpe)^2 \gw_{211}\right]\right\}.
\label{order}
\end{split}
\ee
We separate our equations into parts easier to compare. We start by comparing the factors in front of the braces of each expression in \eqref{rhs} and \eqref{order} which are:
\be
\label{equadiff_factors}
\frac{H}{\cal{N}} \frac{e_{1\gf*}^2\gV_{\gs\gs*} - e_{1\gs*}^2\gU_{\gf\gf*}}{e_{1\gf*}e_{1\gs*}} \qquad \text{and}\qquad  \ge(\frac{H}{\cal{N}})^2.
\ee
After simplifying the common factor $H/$ $\cal{N}$ and inserting $\cal{N}$ $= H^2_* e_{1\gf *}e_{1\gs *}/H_0$ from \eqref{v22turn}, we use the quasi single-field initial conditions at horizon-crossing to write \eqref{equadiff_factors} as
\be
\gV_{\gs\gs*}\frac{H_*^2}{H^2} \qquad\text{and}\qquad \ge.
\ee
The discussion about the spectral index from section \ref{Sum Potential Section} is still valid, because the only difference from the slow-roll regime is the value of $\gv_{12e}$, but for a large enough value (larger than four) the dependence on $\gv_{12e}$ in \eqref{ns} disappears and \eqref{nstwofield} can be used. Hence, $\gV_{\gs\gs*}$ is typically of order $10^{-2}$, or at least not hugely smaller.

As for the size of $\ge$ and $H_*^2/H^2$, this depends on the type of turn. For the first type, $\ge$ is still of order slow-roll but it can be easily larger than $\gV_{\gs\gs*}$ by an order of magnitude. However $H_*^2 /H^2$ is also larger than one. Moreover, if $\ge$ had enough time to increase since horizon-crossing, the situation is the same for $H_*^2 /H^2$ because $H$ decreases faster if $\ge$ is larger. During a few dozens of e-folds with $\ge$ of order slow-roll, it can also increase by an order of magnitude. This means that both terms will be of the same order during the turn in this case, or at least that neither of them is hugely smaller or larger than the other. For the second type of turn, the situation is different. During the turn, $\ge$ is again of order slow-roll so it is not hugely larger than $\gV_{\gs\gs*}$. However, because of the period of large $\ge$, we know that $H_*^2/H^2 \gg 1$ from \eqref{second case}. Hence the factor in front of order1, order2 and order3 is large compared to the one in r.h.s.\  in this case.

Next we focus on the second part of each expression, which is the part inside the braces and which is a complicated expression depending on basis components and slow-roll parameters. We start with some comments on the factors $e_{1\gf}e_{1\gs}$ and $e_{1\gs}^2 -e_{1\gf}^2$. By definition of the basis, $e_{1\gf}e_{1\gs}$ goes from $-\half$ to $\half$ and $e_{1\gs}^2 - e_{1\gf}^2$ from $-1$ to $1$ and when one is at an extremum, the other one vanishes. When one vanishes, the leftover slow-roll parameter terms are similar in the different expressions. It is also not possible to have both of them small compared to one at the same time, hence the term in r.h.s.\  without a factor depending on the basis is not an issue. Hence, we can forget about these basis component factors which cannot change the conclusion.

The different expressions depend on all the first and second-order slow-roll parameters, except $\gc$, $\gw_{221}$ and $\gw_{222}$ which have been eliminated using the relations specific to sum potentials \eqref{W11eq2} and \eqref{sumequation}. The first step is to study the cancellations of the left-hand side terms. An obvious one is when $\getpe$ vanishes because it multiplies every term in \eqref{order}; the homogeneous solution vanishes in that case. It also multiplies every term in r.h.s.\  except the one term $2\getpa(\gxpe)^2$. However $\gxpe$ is also small when $\getpe$ becomes small. During the turn of the field trajectory, it is usual that the slow-roll parameters oscillate, hence $\getpe$ can vanish several times. At those times our hypothesis that r.h.s.\  is much smaller than the other terms is not valid and we cannot neglect the particular solution. However, we will show in section \ref{solution section} that we have a way of dealing with this. Apart from this vanishing of $\getpe$, there is no other possibility to cancel order1, order2 and order3 simultaneously. Indeed the expressions contain similar terms, but with opposite signs or different numerical constants.

Once we know there are no cancellations in the left-hand side terms (apart from the  moments when $\getpe=0$), we can compare their expressions to r.h.s.\  and verify they are of the same order. As the expressions contain terms up to order five in slow-roll parameters, two cases have to be differentiated. First, the slow-roll parameters can be of order unity. Then the powers do not matter and most of the terms have to be taken into account. We remark that the terms are similar on each side of the equation, and that the numerical constants are also of the same order, so that r.h.s.\  cannot be very large compared to the other expressions in this case. However, the slow-roll parameters can also become larger than order unity and this situation requires more discussion. An important remark is that when the slow-roll approximation is broken, the slow-roll cancellations in \eqref{srpareq} disappear which means that $\gxpa$ and $\gxpe$ are of order a few times $\getpa$ and $\getpe$ respectively, and not of order $(\getpa)^2$ and $(\getpe)^2$. Using the expressions for $\dot{\eta}^\parallel$ and $\dot{\eta}^\perp$ in \eqref{srderivatives}, we can see that when $|\getpa|$ is at a maximum, $|\getpe|$ has to be of the same order because the only possibility to cancel the largest term $(\getpa)^2$ in the derivative expression is to have $(\getpe)^2$ of the same order. However, when $|\getpe|$ is at a maximum, we can see in a similar way that $|\getpa|$ must be of the order of a few at most.

Then we can study what happens if the perpendicular parameters are the largest (near the maximum of $|\getpe|$). If $\getpa$ is only a few, the dominant terms in r.h.s.\  and the order1,2,3 are the ones in $(\getpe)^5$ and $(\getpe)^4$ (or the equivalent $(\getpe)^3 \gxpe$). The same terms exist in all the different expressions meaning the part inside the braces has to be of the same order in general. If, on the other hand, the parallel parameters are the largest, there is a term in $(\getpe)^2(\getpa)^3$ in r.h.s.\  which does not exist in the other expressions. However, as discussed a few lines earlier, $\getpe$ is also of the same order as $\getpa$ at that time. Using this, the dominant terms are actually of order $(\getpe)^5$. Again we find similar terms inside the braces for the different expressions which have to be of the same order. Finally, the only term in r.h.s.\  that has no equivalent in the other expressions is $(\getpe)^2 \gw_{111}$. This term, which is only of order three, can never be dominant because $\gw_{111}$ cannot be large enough to make this term a lot larger than the order five ones because this parameter is also in the derivative of $\gxpa$ (see \eqref{srderivatives}).

Hence, we have established that the terms inside the braces are of the same order in the general case for each expression in \eqref{rhs} and \eqref{order}. This is exactly the situation for the second type of turn where the only hypothesis not used \eqref{second case} has no consequence for the terms inside the braces. However, for the first type of turn, the relations \eqref{first case} between the parallel and perpendicular slow-roll parameters of the same order can change the result. To verify this, we substitute them into \eqref{rhs} and \eqref{order}. We also introduce the notation with $\{\}$ in subscript, meaning we consider only the terms inside the braces. The computation gives:
\be
\begin{split}
\mathrm{r.h.s.}_{\{\}}&= -e_{1\gf}^2 e_{1\gs}^2 \left[(\getpe)^2 \lh 3 \gw_{111}+6 \ge^2+18 \ge\rh -6 \getpe \gxpe \ge\right] \\
&\!\!\! - e_{1\gf} e_{1\gs}^3 \left[ 3 (\getpe)^2 \gw_{211}+(\getpe)^3 \lh -2 \ge^2-12 \ge\rh \right]
- e_{1\gf} e_{1\gs} (\getpe)^3 \lh 2 \ge^2-6 \ge\rh +2 e_{1\gf}^2 (\getpe)^4 \ge,\\
\mathrm{order1}_{\{\}}&= \frac{e_{1\gs}}{e_{1\gf}}  \left[ 4 \getpe (\gxpe)^2 + 12(\getpe)^5 + 6(\getpe)^2 \gw_{211}\right]  + \frac{e_{1\gs}^3}{e_{1\gf}} \left[ -4 \getpe (\gxpe)^2-6 (\getpe)^2 \gw_{211}\right] \\
&~~~ + e_{1\gs}^2 \left[ 4 (\getpe)^3 \gxpe-24 (\getpe)^4\right] + 4 (\getpe)^3 \gxpe + 12 (\getpe)^4 ,\\
\mathrm{order2}_{\{\}}& = 12\frac{ e_{1\gs} }{e_{1\gf}}(\getpe)^5 - 48\frac{ e_{1\gs}^3}{e_{1\gf}}(\getpe)^5 - 2 e_{1\gf} e_{1\gs} \left[ 2 \getpe (\gxpe)^2+(\getpe)^2 \gxpe (\ge-3)\right] \\
&~~~-2 e_{1\gs}^2 \left[ 14 (\getpe)^3 \gxpe+(\getpe)^4 (4 \ge-12)\right] +4 (\getpe)^3 \gxpe-6 (\getpe)^4+2 (\getpe)^4 \ge ,\\
\mathrm{order3}_{\{\}}& = \frac{e_{1\gs} }{e_{1\gf}} \left[ (\getpe)^2 (\gxpe (2 \ge-6)-6 \gw_{211})-24 (\getpe)^5\right] \\
&~~~ + \frac{e_{1\gs}^3 }{e_{1\gf}} \left[ 48 (\getpe)^5+(\getpe)^2 (6 \gw_{211}+\gxpe (6-2 \ge))\right] \\
&~~~ + e_{1\gs}^2 \left[ 24 (\getpe)^3 \gxpe+8 (\getpe)^4 \ge\right] -8 (\getpe)^3 \gxpe+(\getpe)^4 (-2 \ge-6).
\end{split}
\label{rhs first case}
\ee
We can directly see that the higher order terms in r.h.s.$_{\{\}}$ have disappeared but are still present in the left-hand side terms. Moreover, most of the remaining terms in r.h.s.$_{\{\}}$ are now proportional to $\ge$, which makes them even smaller. Finally, the divisions by the basis components $e_{1\gf}$ and $e_{1\gs}$ which are smaller in absolute value than one only appear in order1$_{\{\}}$, order2$_{\{\}}$ and order3$_{\{\}}$. All these observations leads to the conclusion that r.h.s.$_{\{\}}$ is in fact small compared to left-hand side terms for the first type of turn.

\begin{figure}
	\centering
	\includegraphics[width=0.49\textwidth]{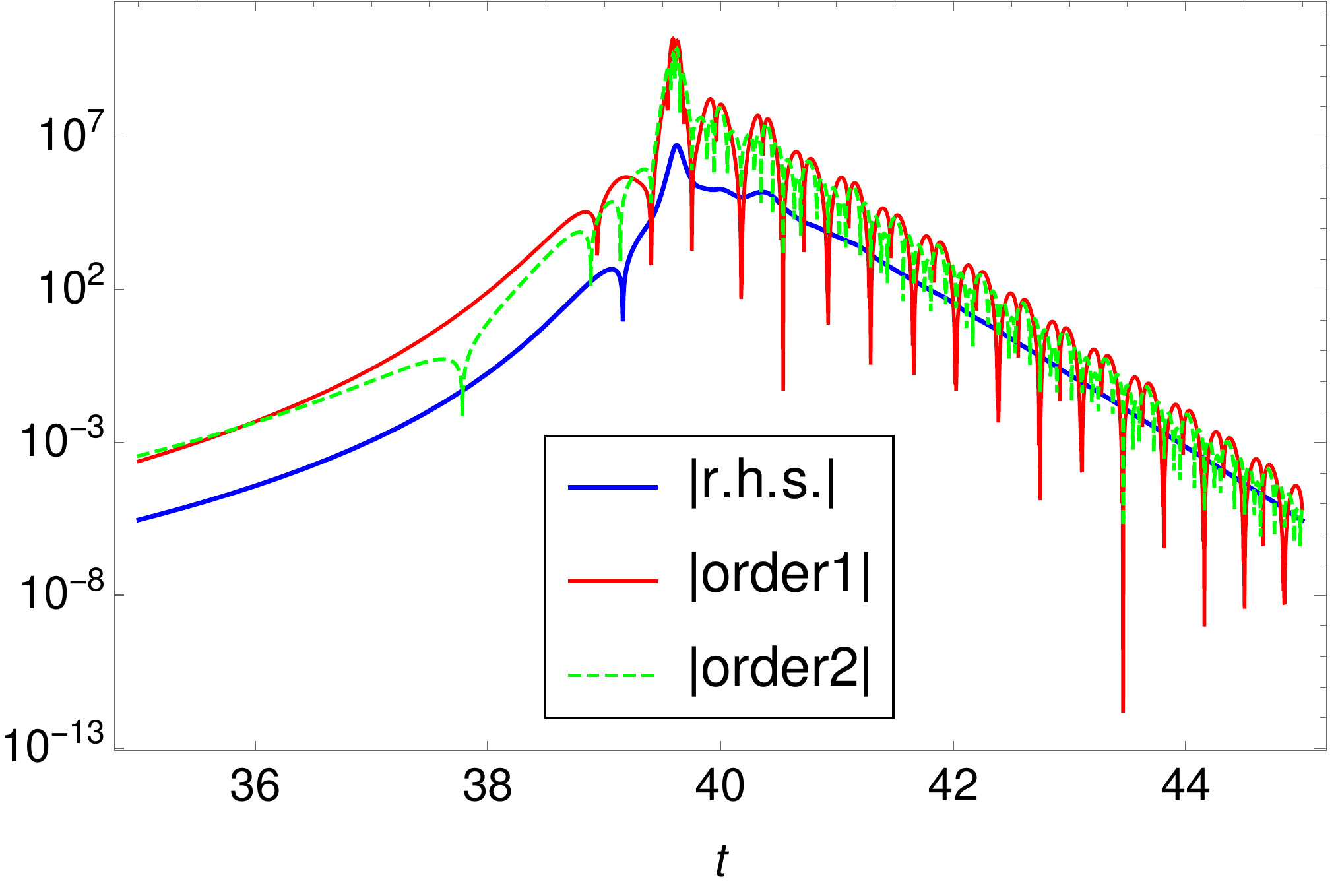}
	\includegraphics[width=0.485\textwidth]{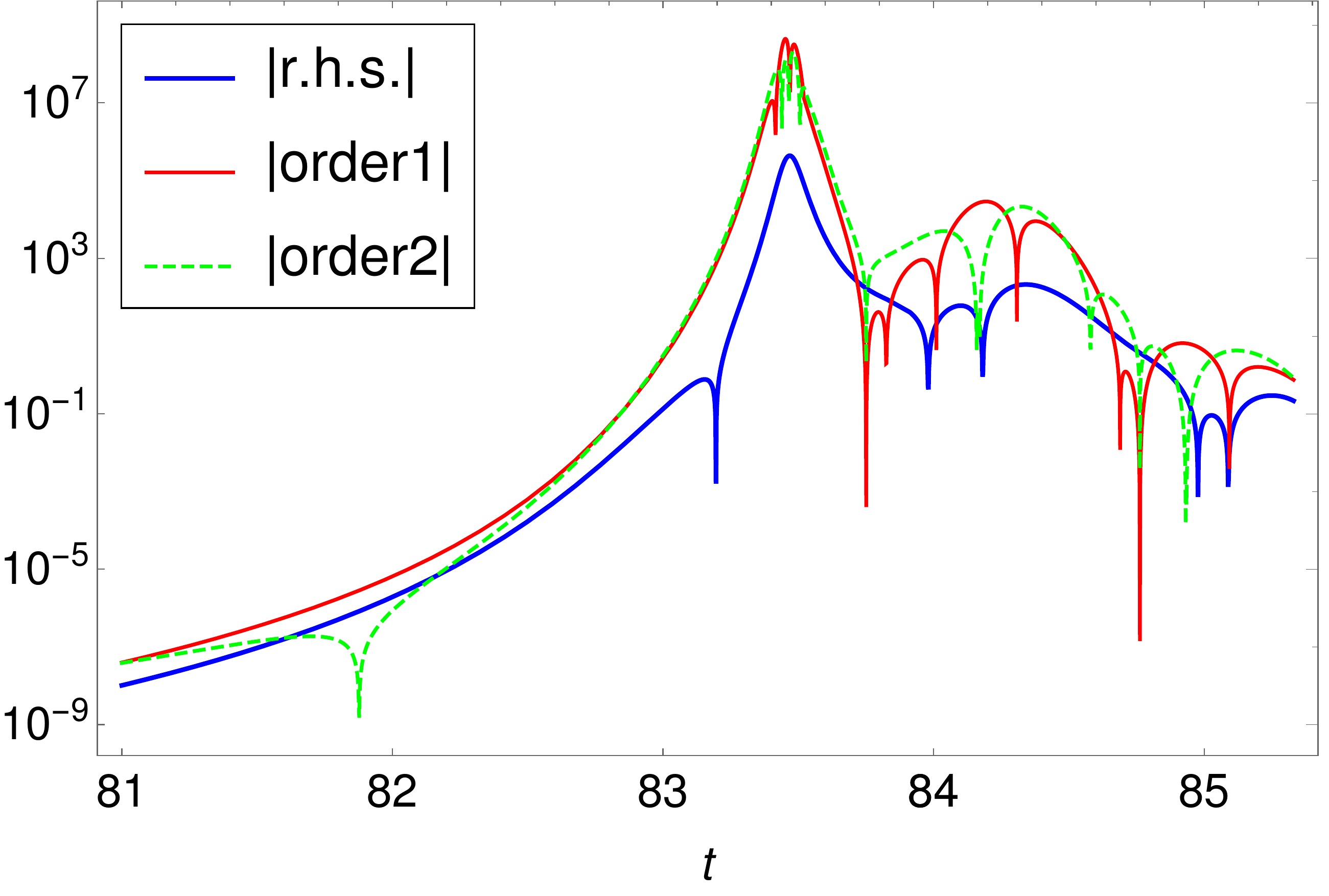}
	\caption{This plot displays $|$r.h.s.$|$ in blue (thick line), $|$order1$|$ in red and $|$order2$|$ in green (dashed) during the turn of the field trajectory for the potentials studied in sections \ref{n2m4 section} (on the left) and \ref{n2m2 section} (on the right), which correspond to the first and second types of turn, respectively. Note the logarithmic scale.}
	\label{fig:rhs}
\end{figure}

To summarize the results of the section, we have established that r.h.s.\  is negligible compared to order1, order2 and order3. With the first type of turn, this is due to the cancellations of the dominant terms in r.h.s.\  due to the relations between the parallel and the perpendicular parameters which exist in that case. For the second type of turn, this is simply due to the factor in front of r.h.s.\  which is smaller than the one in order1,2,3 because $H^2 \ll H_*^2$. This means that even if the slow-roll approximation is broken, if the initial condition of that period is the slow-roll homogeneous solution, then the right-hand side of \eqref{equadiff} can be neglected. This is illustrated in figure \ref{fig:rhs} which displays $|$r.h.s.$|$, $|$order1$|$ and $|$order2$|$ (obviously order3 is not needed because it is minus the sum of the two others) for the potentials of each type of turn that are studied in section \ref{numerical section}. This figure (with a logarithmic scale) shows that r.h.s.\  is always several orders of magnitude smaller than the others during the turn (except at the times where $\getpe$ crosses zero, which will be discussed in section \ref{solution section}).

From this section we learn that the homogeneous solution, which is known, is sufficient to solve \eqref{equadiff} during the turn when the slow-roll approximation is broken (large $\getpa$ and $\getpe$) as long as $\ge$ remains small, since the particular solution is negligible. 

\subsection{Fate of the slow-roll particular solution}
\label{particular solution section}

In the previous section, we showed that we only need the homogeneous solution of the $\gint$ equation during the turn when the slow-roll approximation is broken. However, this does not mean that we can forget about the particular solution completely. It is still required  during the slow-roll evolution before and after the turn as we will show explicitly in this section (and potentially during the turn when $\getpe$ crosses zero, see next section) and hence plays a role in principle in the determination of the integration constants in the various regions. In fact, to avoid having to perform an explicit matching at every transition it would be very convenient if we could just add the slow-roll particular solution to the homogeneous solution everywhere. We will come back to this point in the next section. As a preliminary we will in this section investigate the behaviour of the slow-roll particular solution $P_{\mathrm{sr}}$ before and during the turn. We start by comparing $P_{\mathrm{sr}}$ to the homogeneous solution in the different regimes.

First, we focus on the slow-roll regime using the Green's functions determined in \eqref{Green srsolution} when the slow-roll particular solution can be written as:
\be
P_{\mathrm{sr}} = -2\ge\lh e_{1\gf}^2 \gV_{\gs\gs} + e_{1\gs}^2 \gU_{\gf\gf}\rh \frac{H^4 e_{1\gf}^2 e_{1\gs}^2}{H_*^4 e_{1\gf*}^2 e_{1\gs*}^2}
=\frac{2}{3}\ge \lh e_{1\gf}^2 V_{\gs\gs} + e_{1\gs}^2 U_{\gf\gf}\rh e_{1\gf} e_{1\gs} \frac{H^2 e_{1\gf} e_{1\gs}}{H_*^4 e_{1\gf*}^2 e_{1\gs*}^2}.
\ee
Doing the same for the homogeneous part using the quasi single-field initial conditions discussed at the end of section \ref{Sum Potential Section} and recalled at the beginning of this one, as well as \eqref{srpareq}, we get:
\be
h_{\mathrm{sr}}= -e_{1\gf*}^2\gV_{\gs\gs*} \getpe \frac{H^2 e_{1\gf} e_{1\gs}}{H_*^2 e_{1\gf*}^2 e_{1\gs*}^2}=-e_{1\gf_*}^2 \gV_{\gs\gs*} \lh e_{1\gf}e_{1\gs}(\gV_{\gs\gs}-\gU_{\gf\gf}) -\frac{1}{3}\gxpe \rh H_*^2 \frac{H^2 e_{1\gf} e_{1\gs}}{H_*^4 e_{1\gf*}^2 e_{1\gs*}^2}.
\ee 
In the slow-roll approximation (neglecting the higher-order term $\gxpe$ in $h_{\mathrm{sr}}$), we end up with $\ge \, e_{1\gf} e_{1\gs} \lh e_{1\gf}^2 V_{\gs\gs} + e_{1\gs}^2 U_{\gf\gf}\rh$ and $\gV_{\gs\gs*} e_{1\gf}e_{1\gs}\lh V_{\gs\gs}-U_{\gf\gf}\rh\frac{H_*^2}{H^2}$ to compare, because $e_{1\gf*}^2 \approx 1$ and by definition $\gV_{\gs\gs}= V_{\gs\gs}/(3H^2)$ and $\gU_{\gf\gf}= U_{\gf\gf}/(3H^2)$. As a reminder, we want to see if $P_\mathrm{sr}$ can be negligible compared to $h_\mathrm{sr}$ during the slow-roll regime. First, we look at the terms inside the parentheses which both contain second-order derivatives of the potential. Then, for our models where neither of the derivatives is negligible compared to the other at horizon crossing, we can expect that in general this remains true later, at least up to the turn (it can change during the turn, but at that time the slow-roll approximation is broken and these expressions are not valid as we will discuss later in this section). So we conclude that the terms between parentheses in the two expressions are in general of a comparable order (the basis components in $P_{\mathrm{sr}}$ can make it smaller, but not a lot smaller). If there is a difference between the two expressions, it has to come from the remaining factors, which means we have to compare $\ge$ to $\gV_{\gs\gs*}\frac{H_*^2}{H^2}$ like in the previous section. As discussed there, these have to be of the same order because in the slow-roll regime $H$ is still of the same order as $H_*$. There is one exception which corresponds to models where $\ge$ is extremely small compared to $\getpa$ even in the slow-roll regime (Starobinsky-like inflation for example), so that $\ge$ is also small compared to $\gV_{\gs\gs*}$ (in that case, there would be a similarity with the beyond-slow-roll situation studied in this section where $\ge\ll \getpa,~\getpe$ as well). But apart from those specific models, this leads to the conclusion that in general both the particular solution and the homogeneous solution have to be taken into account during the slow-roll regime.

As shown in section \ref{Green's functions section}, the slow-roll expressions for the Green's functions are also valid in a region of large $\ge$, which occurs just before a turn of the second type. The same expressions as in the previous paragraph can be used, however $\gxpe$ can no longer be neglected in $h_{\mathrm{sr}}$. On the other hand, there is no reason for $\gxpe$ to become much larger than the other term between the parentheses (which is $\getpe$) either, given that we are still before the turn, so that in the end the conclusion about the terms between parentheses from the previous paragraph still holds. As for the other factors, both the homogeneous and the particular solutions will grow because $\ge$ becomes of order unity, which makes $H_*^2/H^2$ large compared to 1. However, at the end of this period $\ge$ will decrease and becomes of order slow roll again, but the ratio $H_*^2/H^2$ will stay large. This means that the slow-roll particular solution finishes the period of large $\ge$ being small compared to the slow-roll homogeneous solution. We will show below that this is fully consistent with the result for the second type of turn (the type that has a period of large $\ge$ right before the turn) that the slow-roll particular solution is negligible during the turn.

We continue by considering the behaviour of the slow-roll particular solution during the different types of turn. Obviously, it is not an actual solution at that time, but we want to know if it would cause any problems if we were to simply add it to the solution. Again, we follow the same method using the Green's function expressions given in \eqref{v22turn} to write:
\be
P_{\mathrm{sr}} = \ge \frac{H^2 H_0 ^2 e_{1\gf}e_{1\gs}}{H_*^4 e_{1\gf*}^2 e_{1\gs*}^2} \lh \getpa e_{1\gf}e_{1\gs} + \getpe (e_{1\gs}^2-e_{1\gf}^2) \rh , \qquad
h_{\mathrm{sr}} = -e_{1\gf*}^2\gV_{\gs\gs*} \getpe \frac{H H_0 e_{1\gf} e_{1\gs}}{H_*^2 e_{1\gf*}^2 e_{1\gs*}^2}.
\label{srsolutionturn}
\ee
This time we end up with ${\textstyle\frac{H H_0}{H_*^2}}\ge\lh \getpa e_{1\gf}e_{1\gs} + \getpe (e_{1\gs}^2-e_{1\gf}^2) \rh$ and $\gV_{\gs\gs*}\getpe$ to compare. Again the two expressions have a similar form, excluding the factor $H H_0/H_*^2$. As discussed in the previous section, $\gV_{\gs\gs*}$ is typically of order $10^{-2}$ and hence cannot be much smaller than $\ge$ which is of order slow-roll. During the turn, the terms depending on the $\eta$ parameters are also of the same order, except in the rare case when $\getpe$ vanishes. Finally, the only large difference can come from the factor in front in the slow-roll particular solution. The two types of turn described in section~\ref{two turns section} give different results. In the first type where $\ge$ is of order slow-roll since horizon-crossing, $H$ and $H_0$ are not much smaller than $H_*$. Then the factor is not much smaller than one. Moreover, the cases where it is the smallest are also the cases where $\ge$ has increased the most (and can then be larger than $\gV_{\gs\gs*}$ by an order of magnitude), so that these two effects compensate each other. Hence, the slow-roll particular solution is then typically of the same order as the homogeneous solution during the turn. In the second type of turn, the situation is different, indeed $H$ and $H_0$ are of the same order and we know that $H^2 \ll H_*^2$. This means that this time $P_{\mathrm{sr}}$ is small and negligible during the turn compared to $h_{\mathrm{sr}}$, fully consistent with the result that $P_{\mathrm{sr}}$ has become very small during the period of large $\ge$ just before the turn, as shown above.

More must be said about the slow-roll particular solution during a turn of the first type and we will now show that it becomes in fact proportional to the homogeneous solution of \eqref{equadiff}. To show this, we substitute $P_{\mathrm{sr}}$ in the left-hand side of \eqref{equadiff}, using the Green's function expressions from \eqref{v22turn}, the sum potential relations from \eqref{W11eq2} and \eqref{sumequation} to eliminate $\gc$, $\gw_{221}$ and $\gw_{222}$, and also the relations between parallel and perpendicular parameters \eqref{first case}. We then compare the three terms of the equation corresponding to the three different orders of derivative (called term1, term2 and term3)\footnote{These are the same terms we called order1,2,3 before, however now with the particular solution substituted and not the homogeneous one.} to their sum (called l.h.s.):
\be
\begin{split}
\mathrm{l.h.s.}_{\{\}} & = e_{1\gf}^2 e_{1\gs}^2 \left[ (\getpe)^2 \lh -6\gw_{111} - 12 \ge^2 \rh +12 \ge \getpe \gxpe \right]  \\
&~~~ + e_{1\gf} e_{1\gs}^3 \left[ (\getpe)^3 \lh 4 \ge^2+24 \ge\rh -6 (\getpe)^2 \gw_{211}\right]+e_{1\gf} e_{1\gs} (\getpe)^3 \lh 12 \ge-4 \ge^2\rh +4 e_{1\gf}^2 (\getpe)^4 \ge ,\\
\mathrm{term1}_{\{\}} & = -12 e_{1\gf} e_{1\gs} (\getpe)^5 + e_{1\gf}^3 e_{1\gs} \left[ -4 \getpe (\gxpe)^2-6 (\getpe)^2 \gw_{211}\right] +e_{1\gs}^4 \left[ 4 (\getpe)^3 \gxpe-24 (\getpe)^4\right] \\
&~~~ +36 e_{1\gs}^2 (\getpe)^4 -4 (\getpe)^3 \gxpe-12 (\getpe)^4,\\
\mathrm{term2}_{\{\}} & = -12 e_{1\gf} e_{1\gs} (\getpe)^5 +e_{1\gf}^2 \left[ (\getpe)^4 (6-2 \ge)-4 (\getpe)^3 \gxpe\right] + e_{1\gf} e_{1\gs}^3 \left[ 48 (\getpe)^5+36 (\getpe)^3 \ge\right] \\
&~~~ + e_{1\gf}^2 e_{1\gs}^2 \left[ 28 (\getpe)^3 \gxpe + 12 \getpe \gxpe \ge+(\getpe)^4 (20\ge-24)+(\getpe)^2 \lh 6\ge^2-18 \ge\rh \right]\\
&~~~ +e_{1\gf}^3 e_{1\gs} \left[ 4 \getpe (\gxpe)^2+(\getpe)^2 \gxpe (6 \ge-6)+(\getpe)^3 \lh 2 \ge^2-6 \ge \rh \right],\\
\mathrm{term3}_{\{\}} & = e_{1\gf} e_{1\gs}^3 \left[ -48 (\getpe)^5+(\getpe)^2 (\gxpe (6 \ge-6)-12 \gw_{211})+(\getpe)^3 \lh 6 \ge^2 -18 \ge\rh \right]\\
&~~~ + e_{1\gf} e_{1\gs} \left[ 24 (\getpe)^5 + (\getpe)^2 (6 \gw_{211}+\gxpe (6-6 \ge))+(\getpe)^3 \lh 18 \ge-6 \ge^2\rh \right] \\
&~~~+ e_{1\gs}^2 \left[ -32 (\getpe)^3 \gxpe+ (\getpe)^2 \lh -6 \gw_{111}-18 \ge^2 +18 \ge\rh +(\getpe)^4 (-26 \ge-6) \right]\\
&~~~ +e_{1\gs}^4 \left[ 24 (\getpe)^3 \gxpe+(\getpe)^2 \lh 6 \gw_{111}+18 \ge^2 - 18 \ge\rh +20 (\getpe)^4 \ge\right] \\
&~~~ + 8 (\getpe)^3 \gxpe +(\getpe)^4 (6 \ge+6).
\end{split}
\ee
The discussion of these expressions is very similar to the one for \eqref{rhs} and \eqref{order}. We use again the subscript $_{\{\}}$ to indicate that we have left out an overall factor (cf.\ (\ref{rhs}) and (\ref{order})), which is here
the same for all four expressions.
We can see that in l.h.s.$_{\{\}}$ the higher order terms like $(\getpe)^5$ have disappeared. Moreover, most of the terms in l.h.s.\  have an extra factor of $\ge$, which is not the case for the other expressions. This implies that the sum of the three terms is much smaller than the individual terms of \eqref{equadiff} with the slow-roll particular solution. Hence this function is in fact an approximated solution of the homogeneous equation during a turn of the first type when the slow-roll approximation is broken.

If $P_{\mathrm{sr}}$ becomes a homogeneous solution it means that it has to be proportional to a linear combination of the two previously determined exact independent homogeneous solutions $\getpe \gv_{22}$ and $\getpe G_{22*}$. However, using \eqref{proportionality}, these independent solutions have in fact become proportional before the turn. Hence, we simply have that $P_{\mathrm{sr}}$ and $h_{\mathrm{sr}}$ are proportional. Using \eqref{srsolutionturn} and \eqref{first case}, we rewrite the particular solution as:
\be
P_{\mathrm{sr}}= -\ge\frac{H^2 H_0^2 e_{1\gf} e_{1\gs}}{H_*^4 e_{1\gf*}^2 e_{1\gs*}^2}\getpe e_{1\gf}^2.
\ee
We find the same factor $\getpe$ as in the homogeneous solution (\ref{srsolutionturn}), but also another factor $\ge H e_{1\gf}^2$. Hence, the proportionality is true only if $ \ge H e_{1\gf}^2$ is constant during the turn. This happens if $e_{1\gf}^2 \approx 1$, in that case $\gf$ is dominating meaning that $\ge$ and $H$ are purely slow-roll and are almost constant during a short turn. At first, the idea of $\gf$ dominating during the turn might seem odd. However, we recall that this does not have to be during the whole turn, but only when $\getpe$ and $\getpa$ are large enough to break the slow-roll approximation. Looking at the form of trajectory in the top left plot of figure~\ref{fig:slowrollbroken}, the only period when $\gf$ dominates is in fact at the end of the turn when $\gs$ is oscillating around its minimum. This can also be verified with the explicit examples of the next section (see figures \ref{fig:n2m4} and \ref{fig:axion}). Here, we can observe that $\getpe$ becomes large only after the period when $e_{1\gs}$ was not negligible (the turn).

Different behaviours of the slow-roll particular solution depending on the type of turn have been highlighted in this section. In the next section we will discuss how these results can be used to solve the differential equation \eqref{equadiff} beyond the slow-roll regime.

\subsection{Solution of the $\gint$ equation}
\label{solution section}

As usual, we will discuss separately the two types of turn, but we start by reminding the reader about the main result of the previous sections. The solution of \eqref{equadiff} is known until the end of the slow-roll regime and it is composed of a homogeneous solution and a particular solution that both have to be taken into account. When $\getpa$ and $\getpe$ become large, during the turn, only the homogeneous solution (which is exact and does not depend on any slow-roll approximation) is needed to solve the equation. The difficulty is then to ensure the continuity of the solution at the transition between the two regimes. In fact, after the turn, there may also be another period of slow-roll before the end of inflation, and during the turn the slow-roll parameters can oscillate and vanish for a short time, which could lead to a very brief restoration of the slow-roll conditions. So in the end there might be many transitions and it would be very inconvenient if we had to perform an explicit matching of the solutions at each of them. Fortunately, there is another option as we will now show. Finally, we also recall that the slow-roll particular solution evolves differently depending on the type of turn. In the first type, it becomes proportional to the homogeneous solution of \eqref{equadiff}, while in the second type it becomes negligible compared to the homogeneous solution.

It is then easy to see that the case of the first type of turn is most simply treated by keeping the full slow-roll solution at all times. Indeed, at the moment when the slow-roll regime ends and the turn starts, the solution should become only homogeneous, and that is exactly the case because the slow-roll particular solution becomes a homogeneous solution at that time. Continuity at the transition is then automatic, without the need for any explicit matching. Then, if later during the turn or at the end of the turn the slow-roll approximation is re-established, continuity is also ensured since the same solution works on both sides of the transition. Note that if $\getpe$ vanishes, from \eqref{first case}, $\getpa$ has to be of order slow-roll, meaning that the slow-roll approximation is indeed restored during these brief moments.

The second type of turn deserves a longer discussion. Indeed, we do not know the full particular solution during the period of large $\ge$ just before the turn but we know two things: the slow-roll particular solution vanishes (but it is not an exact particular solution at that time) and the right-hand side of \eqref{equadiff} can be neglected once this period has finished, because r.h.s.\  is negligible at the start of the turn as shown in section \ref{homogeneous solution section}. These two ingredients are sufficient to prove that the particular solution during the period of large $\ge$ vanishes, even without having its explicit form. To stay general, we write the particular solution $P$ as $P=P_{\mathrm{sr}} + A h + P_{\perp h}$, where $A$ is a constant, $h$ the homogeneous solution, and $P_{\mathrm{sr}}$ the slow-roll particular solution. $P_{\perp h}$ is the function that, when inserted into \eqref{equadiff}, gives those right-hand side terms that are not given by $P_{\mathrm{sr}}$, and which is zero when these terms vanish (in other words, it does not contain a homogeneous solution). We know that over the course of the period of large $\ge$, $P_{\mathrm{sr}}$ vanishes (see \eqref{srsolutionturn}). The right-hand side of \eqref{equadiff} vanishes during that period too, which means that $P_{\perp h}$ has to vanish by definition. The only remaining term could then be the one proportional to the homogeneous solution, but it has to be zero because of the matching conditions at the start of the period of large $\ge$. Indeed at the end of the slow-roll regime, the particular solution is simply $P_{\mathrm{sr}}$ while $P_{\perp h}$ has to be zero, because the terms of higher order in slow-roll are still negligible and will grow only later during that period of large $\ge$. The function $h$ is not zero at the transition, hence $A$ has to be. Without knowing the exact formula for $P$, we can conclude that it vanishes during that period of large $\ge$. Hence, at the start of the turn, the solution is simply the slow-roll homogeneous solution. 

During the second type of turn, keeping the slow-roll particular solution, even if it is not a particular solution of the exact equation at that time, only induces a negligible error, but it solves any potential issues with matching to later slow-roll periods. When $\getpe$ vanishes, $\getpa$ can be larger than order slow-roll in this type of turn. This is not an issue because then the parameters evolve very fast, meaning that a very short time before $\getpe$ vanishes, the particular solution is still negligible compared to the homogeneous solution, and the same a very short time after. Moreoever, one can verify that at the exact time when $\getpe=0$, the particular solution is $\half P_{\mathrm{sr}}$ and we know that this function is negligible during the rest of the turn. Then it is possible to add this particular solution to the full solution only for these very short periods (without using matching conditions, because at the time of the matchings it it is negligible). It is also important to remember that in the end we are interested in the integrated $\gint$, and when $\getpe$ vanishes, the right-hand side of \eqref{equadiff} is also very small compared to its value a short time before or after (because every term contains $\getpe$ except one which also becomes small), meaning this particular solution is also small at that time compared to its usual value during the turn. In the integral it is then negligible. In fact, when $\getpe$ vanishes, the only thing that happens is that the whole solution almost vanishes (but the particular solution does not vanish at that exact same time), but because the homogeneous solution is zero, it cannot be large compared to the particular solution for once. 

To summarize, we have shown that for both types of turn, the slow-roll solution of \eqref{equadiff} is sufficient to solve this equation even beyond the slow-roll regime, under the condition that $\ge$ stays of order slow-roll during the turn. Of course, knowing the solution $\dot{g}_{\mathrm{int}}$ which is given in \eqref{dotgint} is not sufficient, we also have to integrate it. But the computation is exactly the same as in the slow-roll case even if the slow-roll approximation is not valid, meaning that $\gint$ has again the same form:
\be
\gint = \ge \gv_{22}^2 - \ge_* - \frac{e_{1\gf*}^2\gV_{\gs\gs*} - e_{1\gs*}^2\gU_{\gf\gf*}}{2e_{1\gf*}e_{1\gs*}} \gv_{12}.
\label{gintbsr}
\ee

\subsection{End of inflation and $\fnl$}
\label{end of inflation section}

Once the form of $\gint$ is known, it is possible to compute $\fnl$ at the end of inflation: 
\be
-\frac{6}{5}\fnl =  \frac{e_{1\gf*}^2\gV_{\gs\gs*} - e_{1\gs*}^2\gU_{\gf\gf*}}{e_{1\gf*}e_{1\gs*}} \frac{(\gv_{12e})^3}{\lh 1+(\gv_{12e})^2\rh ^2} + \mathcal{O}(10^{-2}).
\ee
This expression has the same form as the slow-roll one \eqref{fnlapp}, the difference is hidden in the Green's functions which have been computed in section \ref{Green's functions section}. The same discussion of this expression as in section \ref{Sum Potential Section} holds and the conclusions are the same, see \eqref{fnlsrfactor}. Like in that section, we use the limit \eqref{v12limit} which is a good approximation when $|\gv_{12e}|>4$. Then the only remaining step is to study the value of $\gv_{12}$ at the end of the turn using \eqref{v_12}, when the slow-roll approximation is valid again, which is equal to $\gv_{12e}$.

As usual, we need to distinguish the two types of turn because they have different initial and final conditions. In the first case, the turn occurs early which means that $U_0\gg V_0$ (as defined before, the subscript 0 indicates that the function is evaluated at $t_0$ when the slow-roll approximation stops to be valid). However because there is a turn, we cannot neglect $e_{1\gs0}$ anymore. We can then write $S_0=e_{1 \gf0}^2 -e_{1\gs0}^2 = 1-2e_{1\gs0}^2$ and $Z_0\approx -U_0 e_{1\gs0}^2$. Moreover, before the turn we are still in slow-roll, meaning that $\ge_0\ll 1$ and we can use the slow-roll expression $H_0^2=\gk^2 U_0/3$. At the end of the turn, the situation is similar to single-field inflation in the direction $\gf$ meaning that $Z\approx 0$ and $S \approx 1$. Inserting this into \eqref{v_12}, we obtain:
\be
\gv_{12e}=\frac{U_0}{W_*}\frac{2e_{1\gs0}^2}{4e_{1\gf*}e_{1\gs*}} +\frac{-U_0 e_{1\gs_0}^2}{2W_*e_{1\gf*}e_{1\gs*}}+\frac{-V_*}{W_*e_{1\gf*}e_{1\gs*}}=\frac{-V_*}{W_*e_{1\gf*}e_{1\gs*}}.
\ee
This is exactly the same limit as in the slow-roll situation. Hence for this first type of turn, we get the same result:
\be 
-\frac{6}{5}\fnl  = -\frac{V_{\gs\gs*}}{\gk^2 V_*}.
\ee
The implications of this result were already discussed in section \ref{Sum Potential Section}.

In the second type of turn, the situation is slightly different. Firstly, the slow-roll approximation is not valid at the time $t_0$, at the end of the period of large $\ge$. Moreover, at that time we are still in a single-field case ($\gf$ dominates), hence $S_0\approx 1$ and $Z_0 \approx -V_0\approx V_*$ (because even if $U_0$ is not zero, it cannot be large compared to $V_0$ because we are near the moment when $\gf$ reaches the minimum of $U$). After the turn, the single-field situation is now in the $\gs$ direction, hence $S \approx -1$. At the end of inflation, the situation is:
\be
\gv_{12e}= -\frac{H_0^2}{2 H_*^2 e_{1\gf*}e_{_1\gs*}} + \frac{- Z_*}{2W_* e_{1\gf*} e_{_1\gs*}}=  \frac{-\frac{3}{\gk^2} H_0^2 - Z_*}{2W_* e_{1\gf*} e_{_1\gs*}}.
\ee
Substituting this into $\fnl$, we obtain:
\be
-\frac{6}{5}\fnl  = -\frac{2V_{\gs\gs*}}{3 H_0^2 + \gk^2 V_*}.
\ee
However, we can add that $H_0^2 >\gk^2 W_0/3$ because $\ge$ is not negligible (equality in the slow-roll case). Moreover, $W_0=U_0 + V_0 \approx U_0 + V_* > V_*$. We can then write:
\be
|\gv_{12e}|> \left|\frac{-V_*}{W_*e_{1\gf*}e_{1\gs*}}\right|,
\ee
which has an immediate consequence for $\fnl$:
\be 
\left|-\frac{6}{5}\fnl\right|  < \left|-\frac{V_{\gs\gs*}}{\gk^2 V_*}\right|.
\ee
In this case, the value of $\fnl$ is smaller than the slow-roll result. However, it is easily of the same order because $U_0$ and $V_0$ are of the same order while even if $\ge_0=1$, it only changes the factor between $H_0^2$ and $\gk^2 W_0$ from $1/3$ to $1/2$.

So in the end we have derived the rather surprising result that in the class of models considered (two-field sum potentials), the slow-roll expression for $\fnl$ gives a very good approximation of the exact result, even in the case where the slow-roll approximation breaks down during the turn. Allowing for the break-down of slow-roll does however increase the region of the parameter space where large non-Gaussianity can occur compared to the results shown in figure~\ref{fig:etaper1}, because we no longer have the constraint that the turn has to happen before the end of the slow-roll regime.

\section{Numerical examples}
\label{numerical section}

Here, we provide several explicit examples to illustrate the different results of the previous sections. We also show how to explicitly construct a model that produces $\fnl$ of order unity while satisfying all observational constraints.

\subsection{Double quadratic potential}
\label{Quad section}

The double quadratic potential has the form:
\be
W({\gf,\gs})=\frac{1}{2}m_\gf^2 \gf^2 + \frac{1}{2}m_\gs^2 \gs^2.
\label{pot quad}
\ee 
It has been studied and discussed in many papers, see e.g.~\cite{Vernizzi:2006ve,Rigopoulos:2005us,Tzavara:2010ge}. However, it is always a good introductory example.

Without taking into account the exact constraints of the monomial potential yet, we keep the main idea that the second field has a negligible effect at the time of horizon-crossing. This can be achieved by taking $m_\gf^2 \gg m_\gs^2$ and we will use the same values as in \cite{Tzavara:2010ge}: $m_\gf = 20 m_\gs$ and $m_\gs = 10^{-5}\gk^{-1}$. As initial conditions, we use $\gf_i=13\gk^{-1}$ and $\gs_i=13\gk^{-1}$, while their derivatives $\dot{\gf}_i$ and $\dot{\gs}_i$ are determined by the slow-roll approximation.
\begin{figure}
	\centering
	\includegraphics[width=0.49\textwidth]{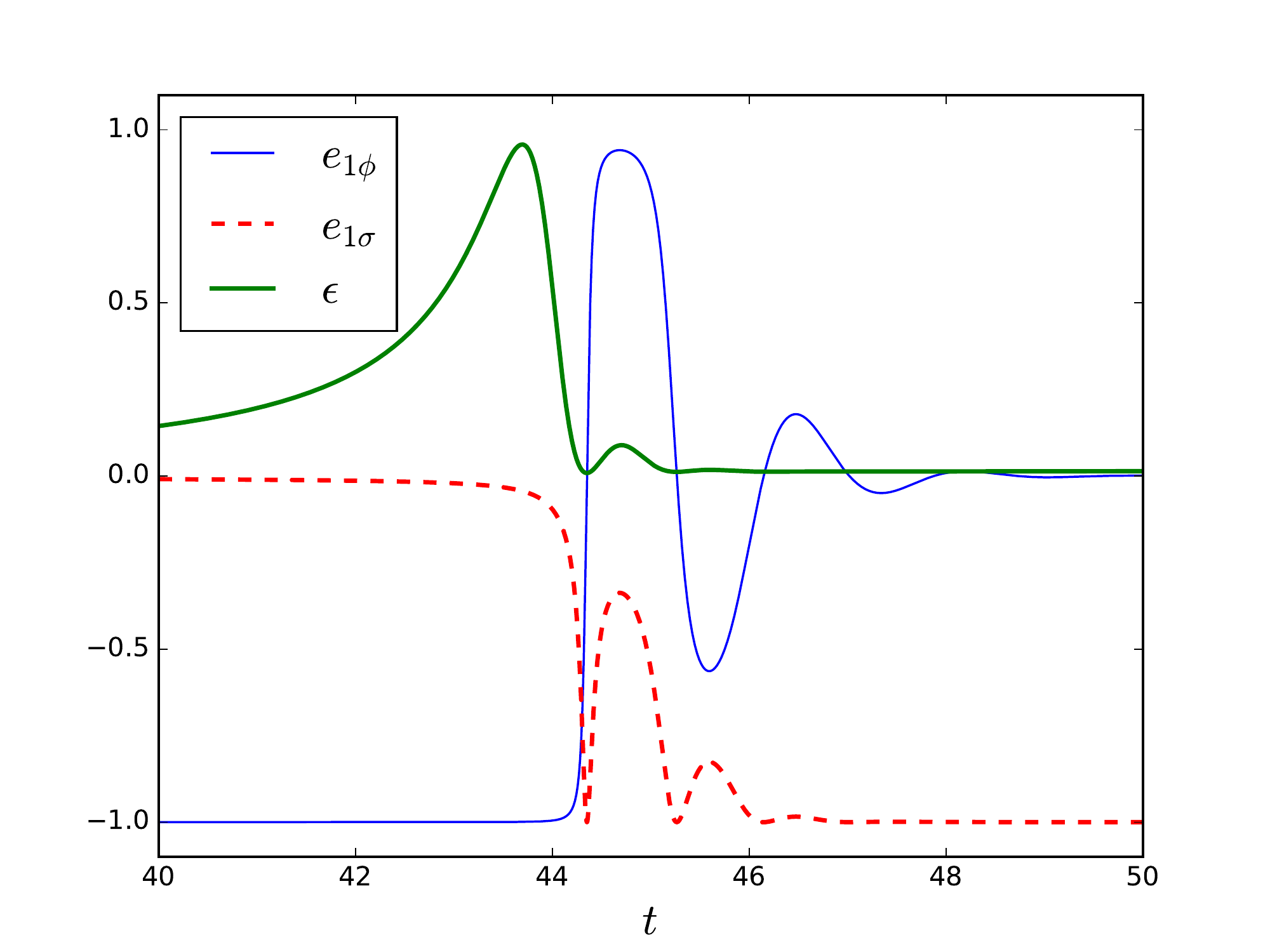}
	\includegraphics[width=0.49\textwidth]{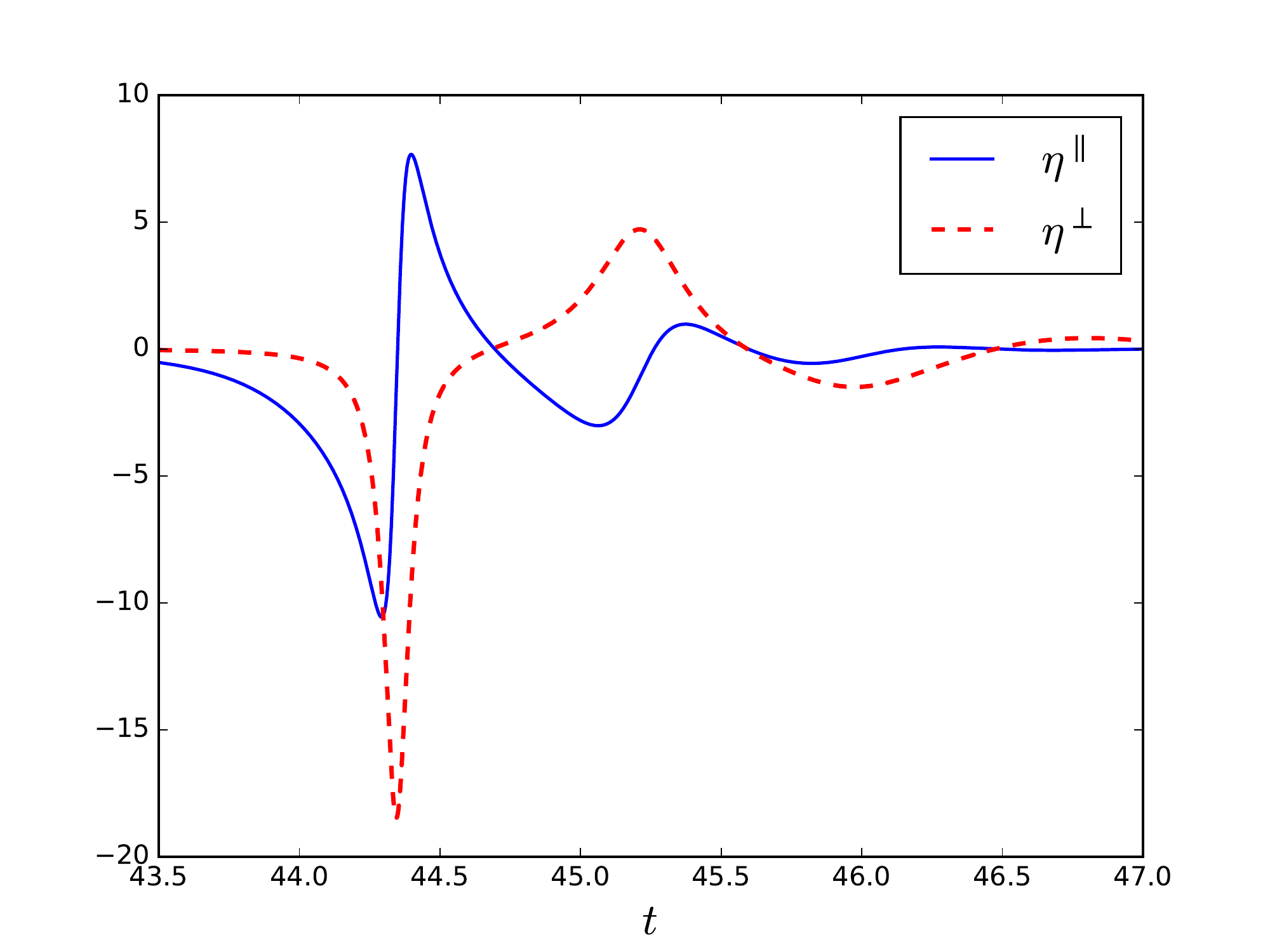}
	\includegraphics[width=0.49\textwidth]{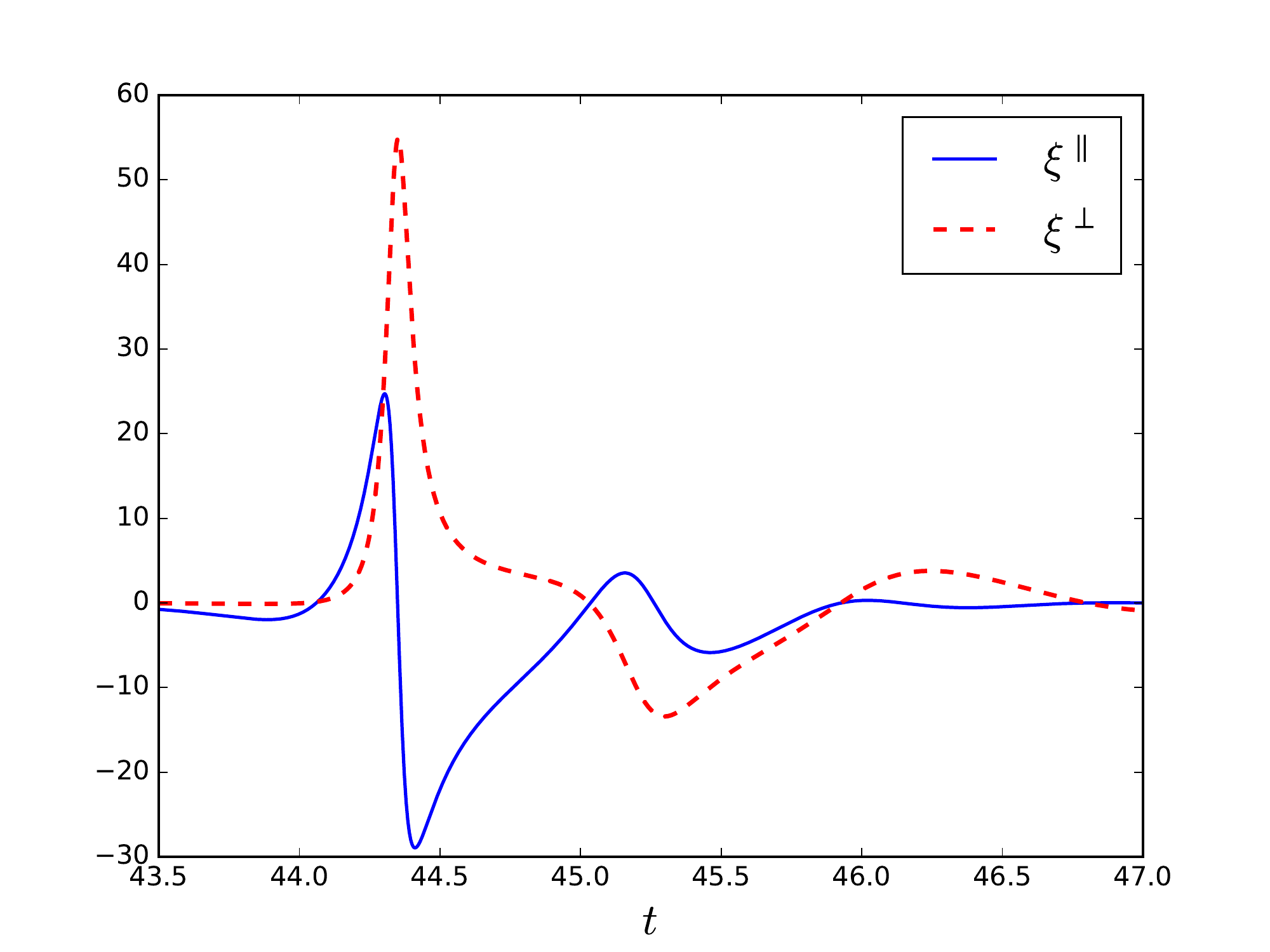}
	\includegraphics[width=0.49\textwidth]{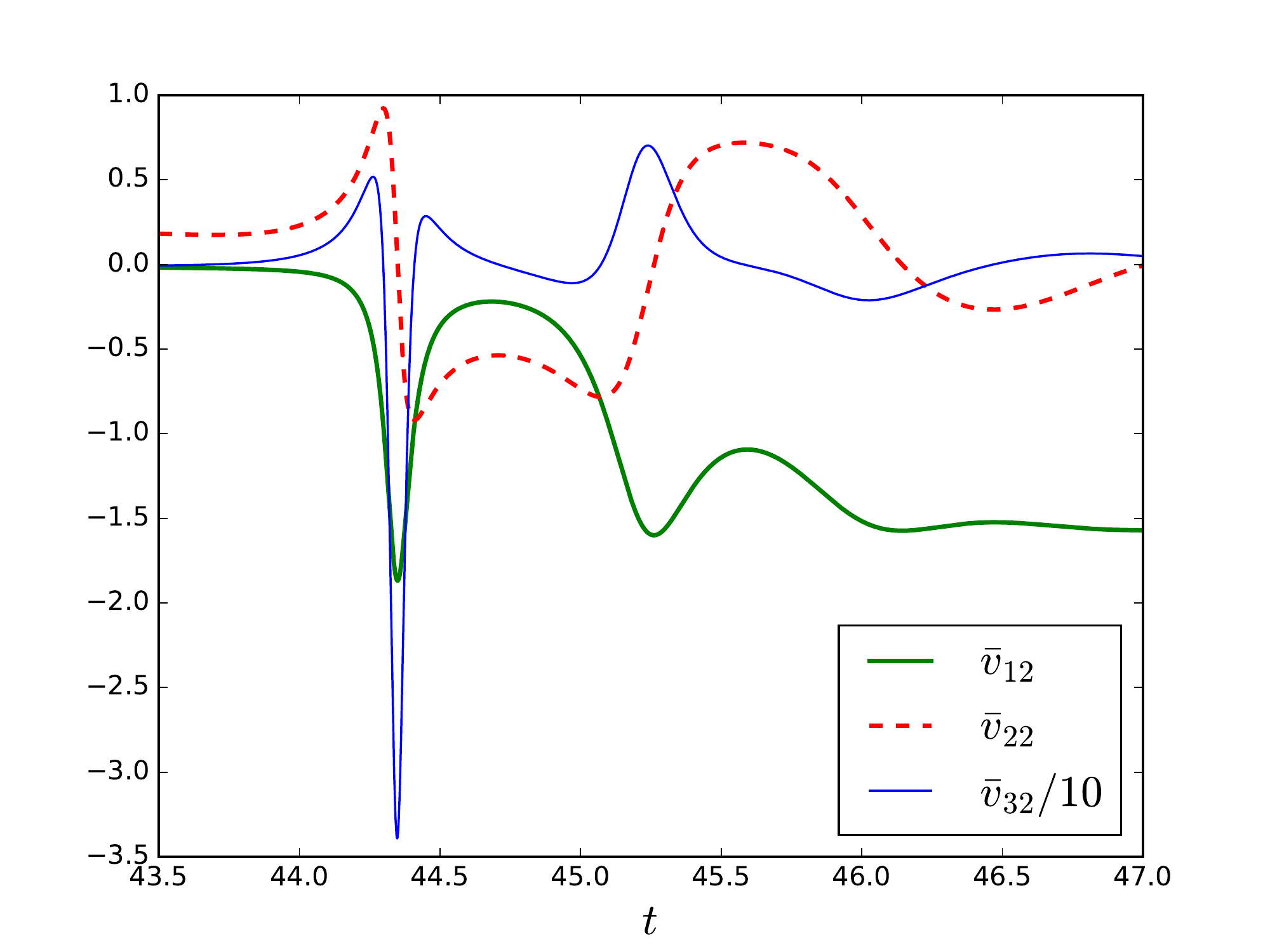}
	\includegraphics[width=0.49\textwidth]{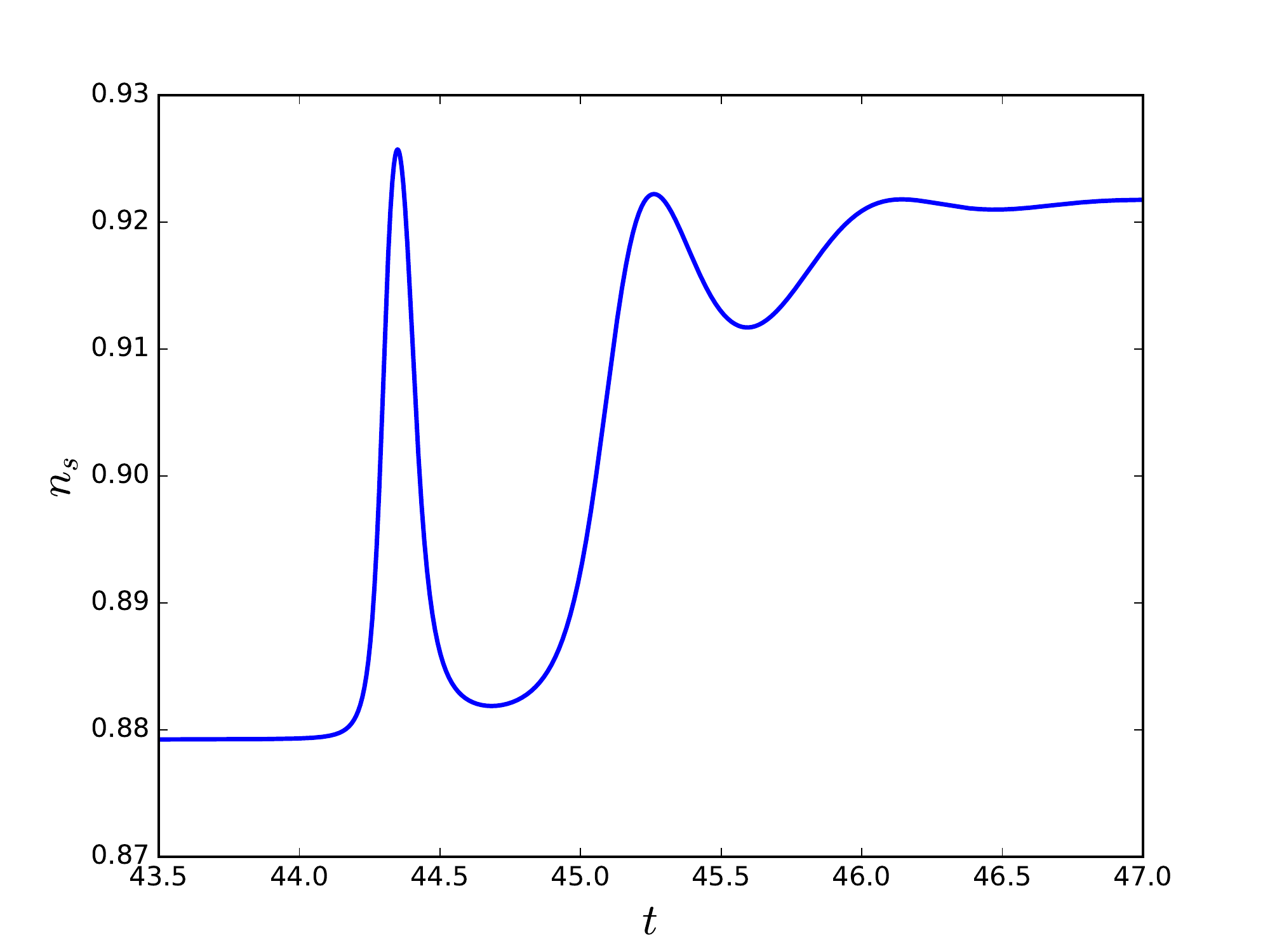}
	\includegraphics[width=0.49\textwidth]{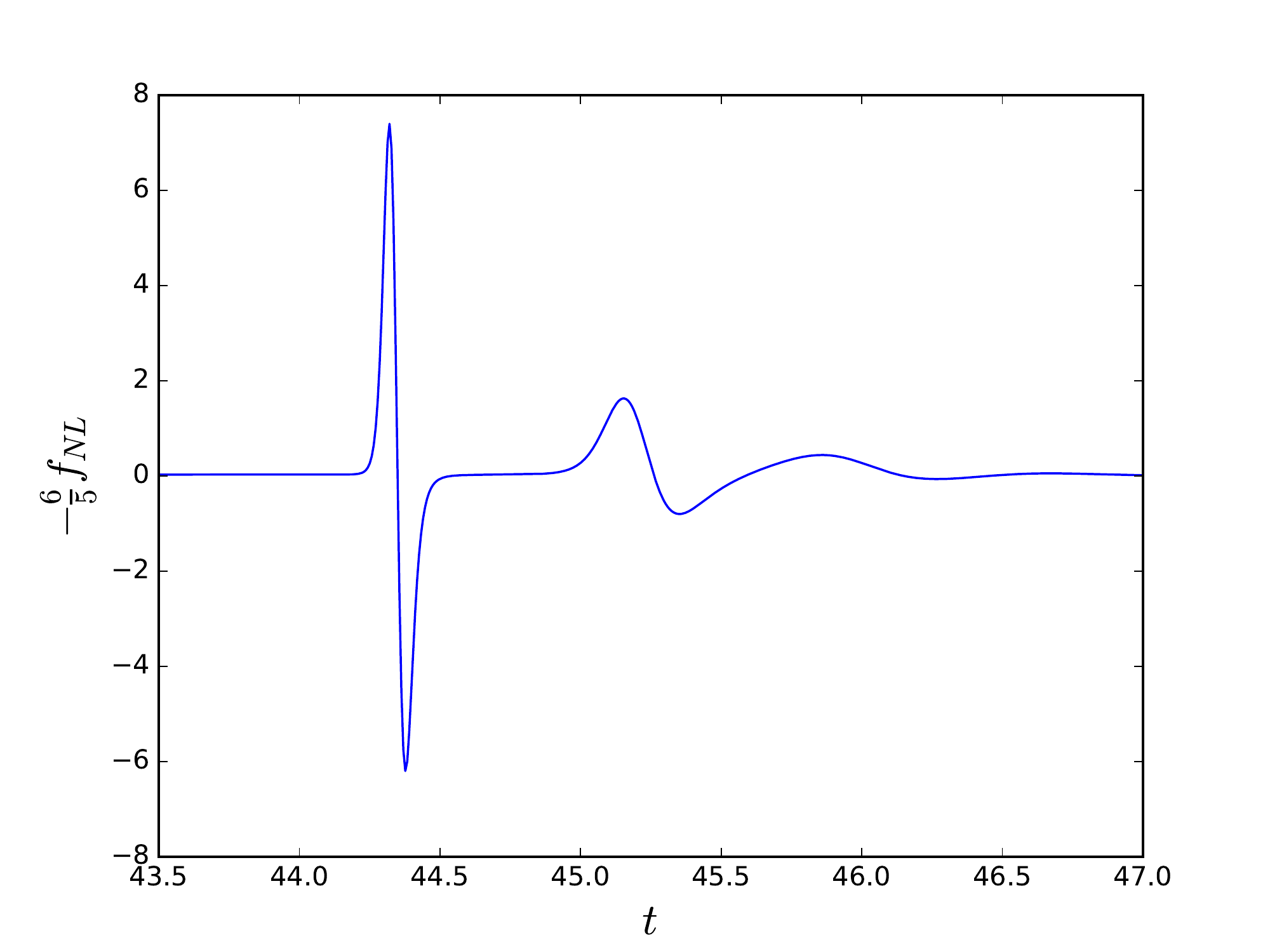}
	\caption{The exact numerical solutions for the different interesting parameters (basis components, slow-roll parameters, Green's functions, the spectral index and $\fnl$) during the turn for the double quadratic potential \eqref{pot quad}.}
	\label{fig:quad}
\end{figure}
In figure \ref{fig:quad} we show how the various relevant quantities evolve during the turn of the field trajectory. First, one can see clearly when the turn occurs: $\getpe$ becomes large and $e_{1\gs}$ becomes of the same order as $e_{1\gf}$. We also see that this example corresponds to the second type of turn where $\gf$ reaches the minimum of its potential and $\ge$ is of order unity before the turn. Another remark is that the second-order parameters $\gxpa$ and $\gxpe$ do not give new information compared to the first-order parameters  $\getpa$ and $\getpe$, at least not by eye.

However, in this model the two most important constraints and goals, concerning the two observables $\ns$ and $\fnl$, are not achieved. The spectral index, which is 0.92, is clearly outside the bounds from the Planck observations. $\fnl$ is slow-roll suppressed and far from the goal of $\fnl$ of order unity. Moreover, $\gv_{12e}$ is only $-1.5$ which is smaller in absolute value than the value 4 needed to use the approximations \eqref{fnllimit} and \eqref{nstwofield} for $\fnl$ and $\ns$.

The main result of the previous section was the validity of the slow-roll expressions in cases beyond slow-roll, like this one, at least to give an estimation of the Green's functions. Hence, we can use this approximation to compute $\gv_{12e}$ to see how the situation can be improved. Using dimensionless fields, \eqref{srparametersflat}, the slow-roll expression for $\gv_{12e}$ given by \eqref{v12e} becomes: 
\be
\gv_{12e}=-\frac{V_*}{W_*e_{1\gf*}e_{1\gs*}}=-\sqrt{2\ge}\frac{V_*}{V_{\gs*}}=-\frac{\gs_*}{\gf_*}.
\ee
This shows that $\gv_{12e}$ can be increased only by changing the initial conditions. Assuming that now we have $\gv_{12e}$ large enough, $\fnl$ takes the form:
\be
-\frac{6}{5}\fnl=-\frac{V_{\gs\gs*}}{V_*}=\frac{2}{\gs_*^2}.
\ee
The value of $\fnl$ becomes smaller if we increase $\gs_*$. Hence, it is impossible to increase both $\gv_{12e}$ and $\fnl$ at the same time. One can also verify there is no optimal value of $\gs_*$ where $\fnl$ would be larger than order slow-roll, meaning that this potential cannot produce large persistent non-Gaussianity. 

Instead of looking directly at $\fnl$, we could also have used the conclusion that for a monomial potential $N_\gs \propto \gs^2$ has to be of order unity to have $\fnl$ large, which requires here to decrease $\gs_*$ and $\gv_{12e}$. The solution is then to add an extra parameter in the potential.

\subsection{How to build a monomial potential model that produces $\fnl$ of order unity}

The form of the potentials we are interested in is:
\be
W(\gf,\gs)= \ga(\gk\gf)^n + C + \gb(\gk\gs)^m \lh + \lambda (\gk \gs)^{m'}\rh,
\ee
which is the one studied in section \ref{Monomial potentials section}. There is an extra term with $m'>m$ inside the parentheses to complete the model (i.e.\ make sure it has a minimum) and we will choose it to be negligible until after the turn. Hence this does not change the different expressions determined for a monomial potential. 

A first step it to choose the value of $m$ and $n$ using figure \ref{fig:etaper1} to be in the region where $\fnl$ of order unity is possible. $\ga$ can be put as an overall factor of the whole potential, hence it does not count in the number of parameters. $\gf_*$ is given by $N_{\gf*}\approx 60$ and this also determines $\ge_*$ because it only depends on $\gf_*$. Once $\ge_*$ is known, it is possible to determine $\gs_*$, $\gb$ and $C$ using the three constraints we have ($\fnl$, $\ns$ and $\gv_{12e}$) as follows.

We can start by choosing the value of $\fnl$ and \eqref{fnllimit} takes the following form for a monomial potential:
\be
\label{constraint1}
-\frac{6}{5}\fnl =  -\frac{V_{\gs\gs*}}{\gk^2 C}.
\ee
Using \eqref{nstwofield} and the lower bound on the spectral index $\ns=0.962$, as this is the easiest way to get a large $\fnl$, we have:
\be
\label{constraint2}
V_{\gs\gs*}= \gk^2 U_*\lh\frac{\ns-1}{2}+\ge_*\rh.
\ee
Finally, we need $\gv_{12e} > 4$. Using the slow-roll expression for $\gv_{12e}$ in \eqref{v12e}, \eqref{srparametersflat} and \eqref{link12}, we get:
\be
\label{constraint3}
\gv_{12e}=-\frac{V_*}{W_*e_{1\gf*}e_{1\gs*}}=-\sqrt{2\ge_*}\frac{\gk C}{V_{\gs*}}=-\sqrt{2\ge_*}\frac{m-1}{\gs_*}\frac{\gk C}{V_{\gs\gs*}}.
\ee
A last step is to determine $\lambda$, this is done using the fact that the minimum of the potential has to be zero. Then it is possible to verify if the last term is really negligible at horizon-crossing, if not it is possible to increase $m'$ to decrease it because $\gs_*$ is small compared to one. We will now apply this to two different potentials with a turn of the first type.

\subsection{First type of turn}
\subsubsection{First example: $n=2$ and $m=4$}
\label{n2m4 section}

This first example corresponds to the case where the turn occurs early enough to have a trajectory with the same direction before and after the turn, see the top right plot of figure \ref{fig:slowrollbroken} for an illustration of the field trajectory. The potential is:
\be
W(\gf,\gs)= \ga \gf^2 + C + \gb \gs^4 + \lambda \gs^{6},
\label{pot n2m4}
\ee
with $\ga=\half\gk^{-2}$, $C= {\textstyle\frac{4}{27}\frac{\gb^3}{\lambda^2}}$, $\gb=-12.5$, and $\lambda=-{\textstyle\frac{4}{3}}\gb\gk^2$. The intial conditions are $\gf_i= 16\gk^{-1}$ and $\gs_i=0.09\gk^{-1}$ and, as usual, $\dot{\gf}_i$ and $\dot{\gs}_i$ are determined by the slow-roll approximation.

\begin{figure}
	\centering
	\includegraphics[width=0.49\textwidth]{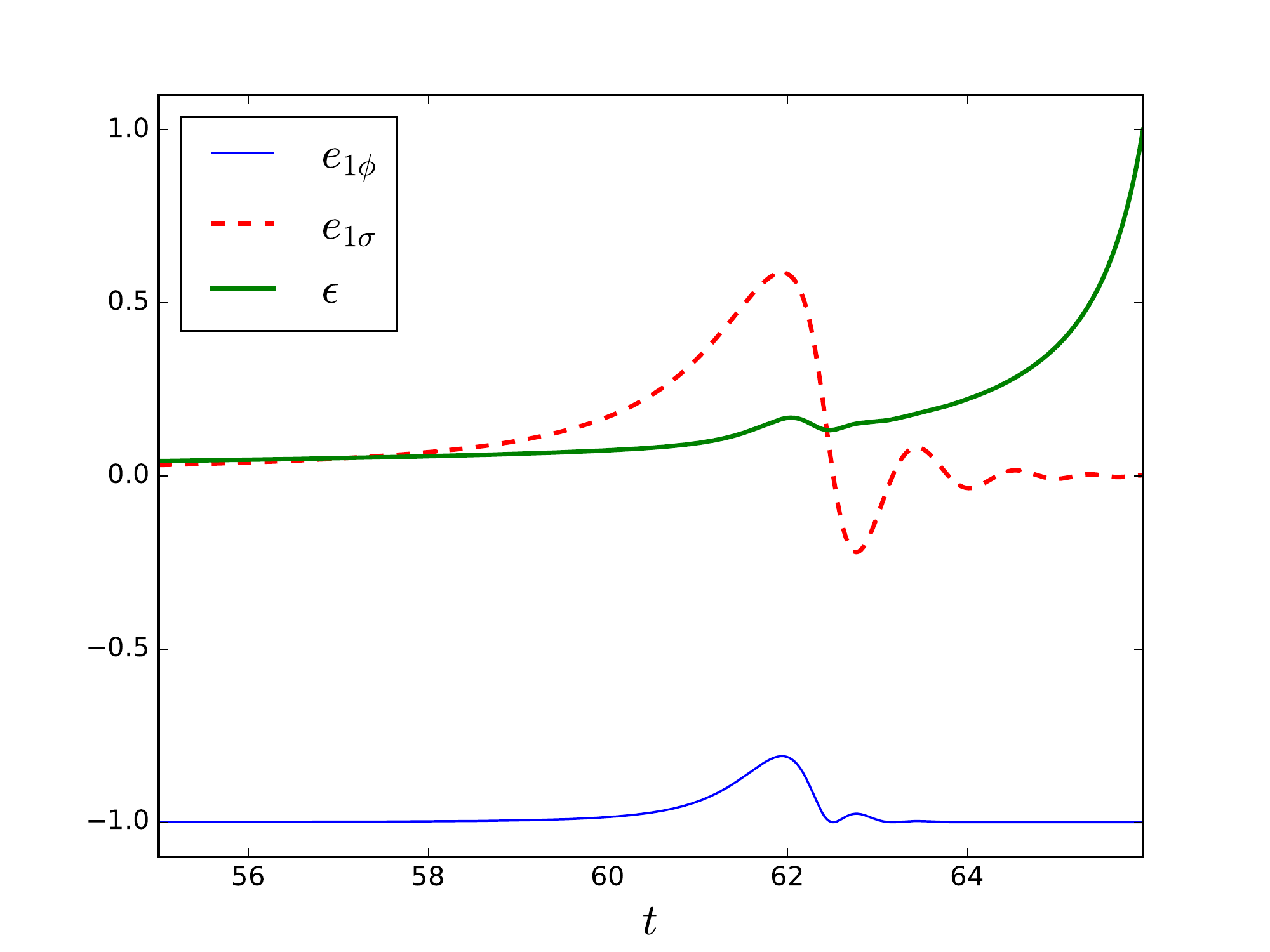}
	\includegraphics[width=0.49\textwidth]{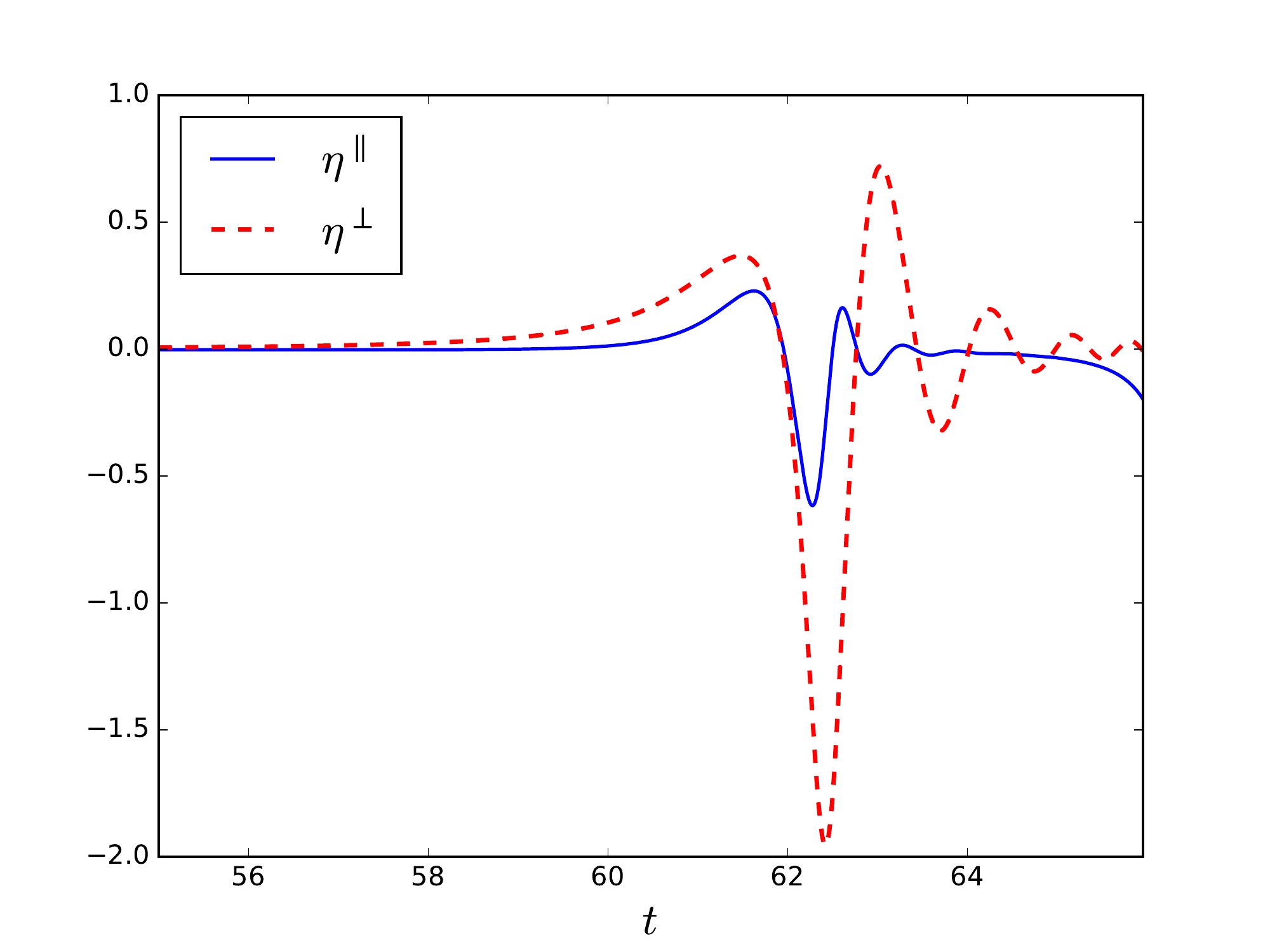}
	\includegraphics[width=0.49\textwidth]{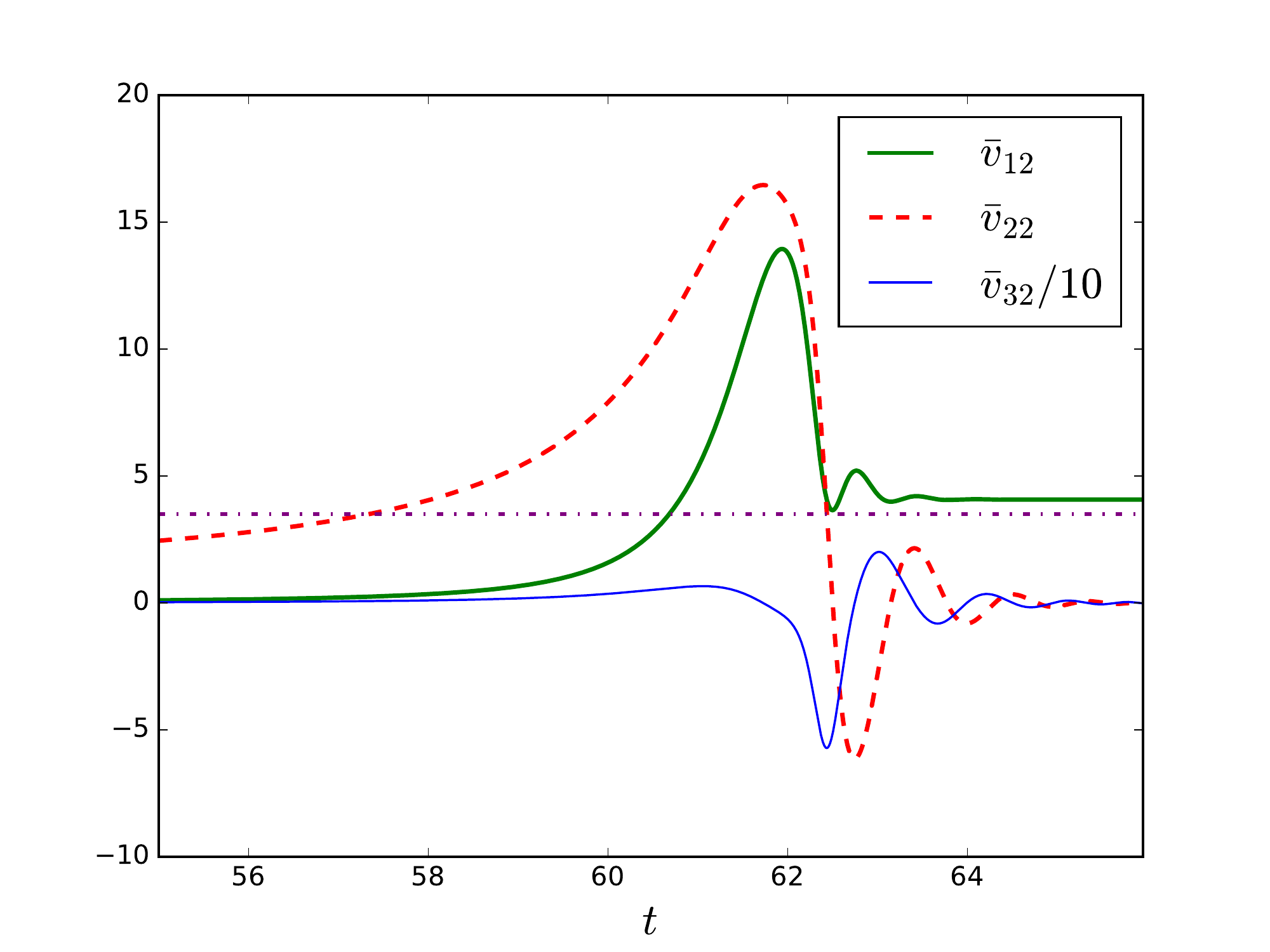}
	\includegraphics[width=0.49\textwidth]{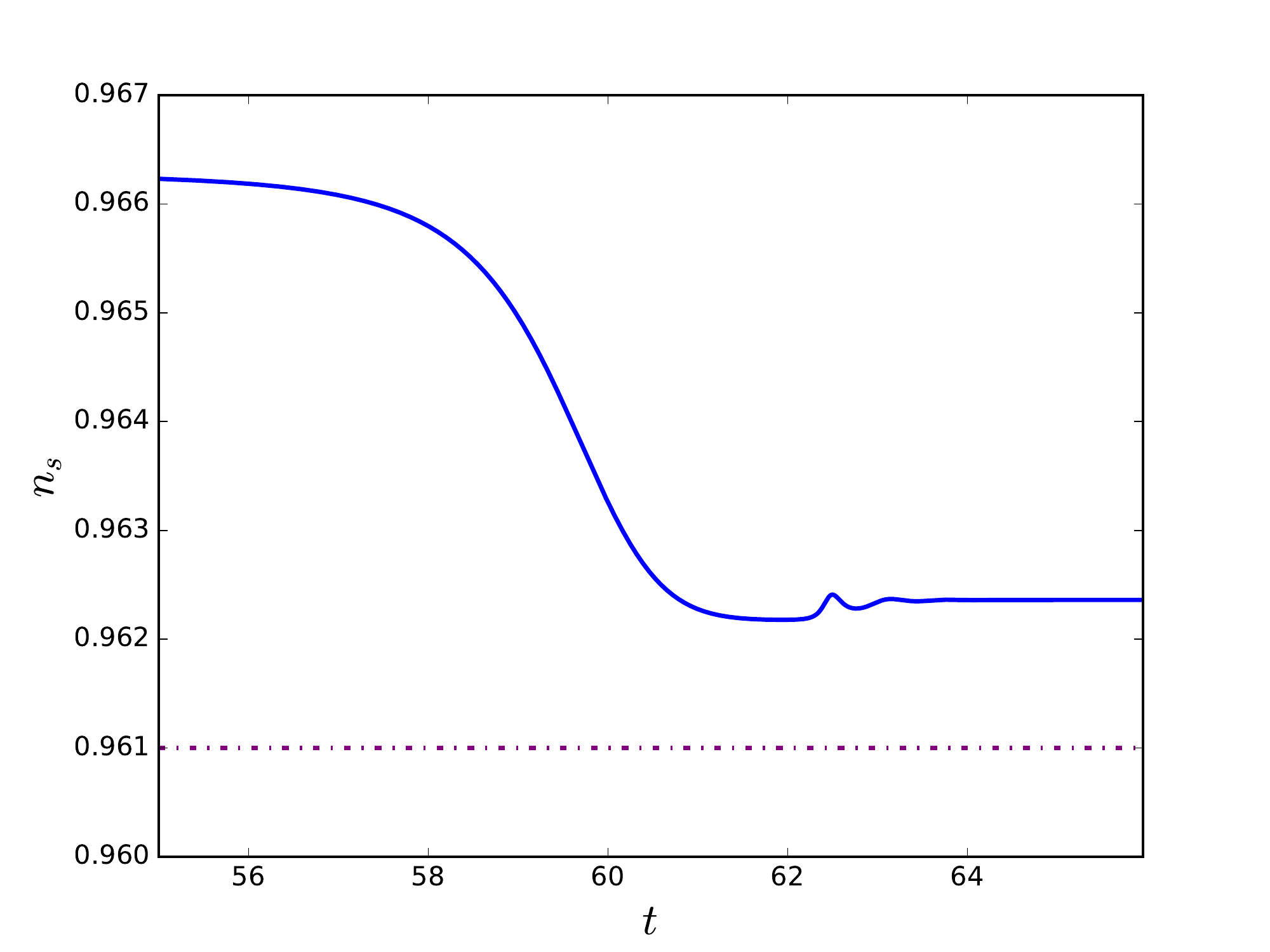}
	\includegraphics[width=0.49\textwidth]{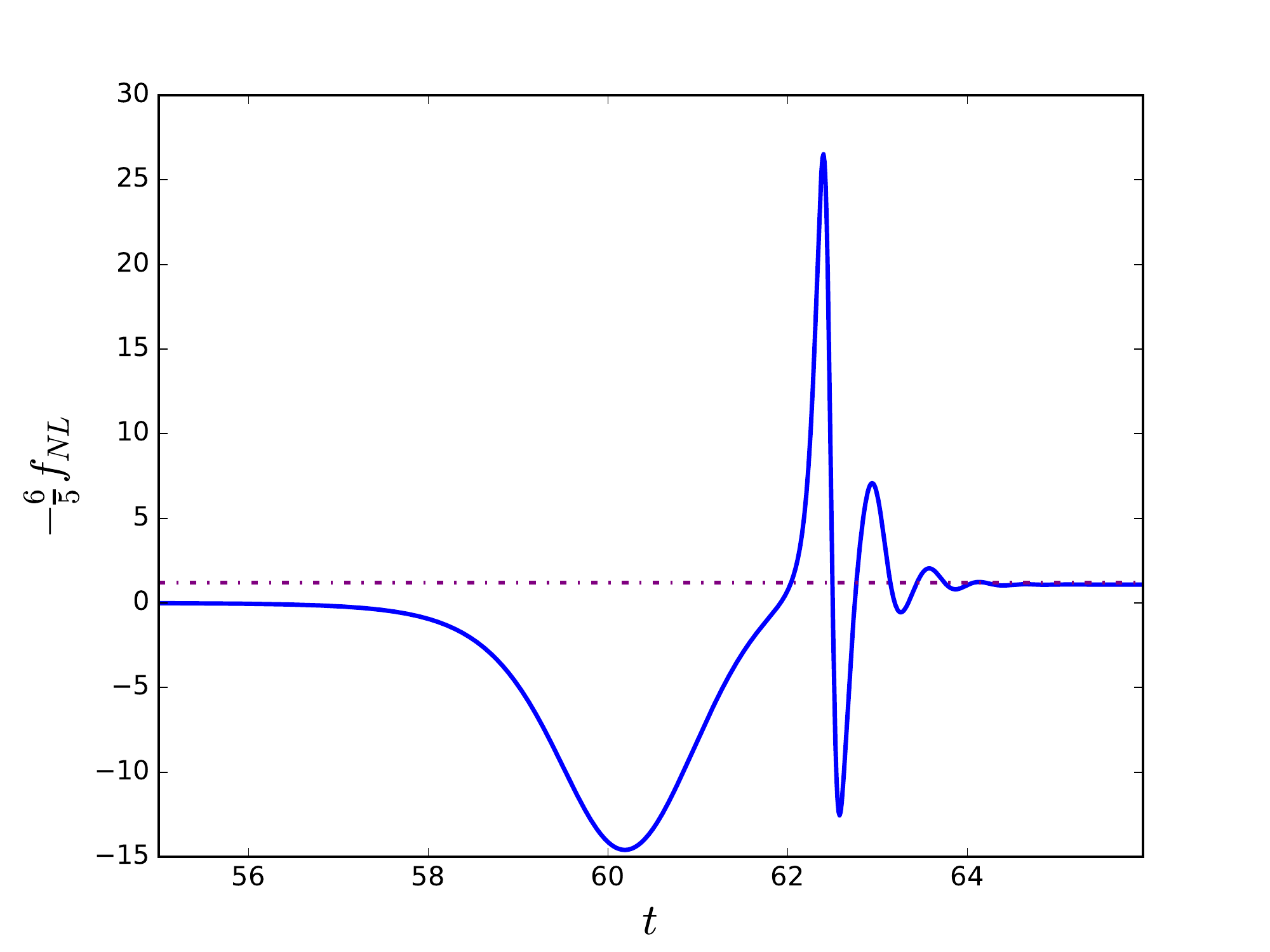}
    \includegraphics[width=0.49\textwidth]{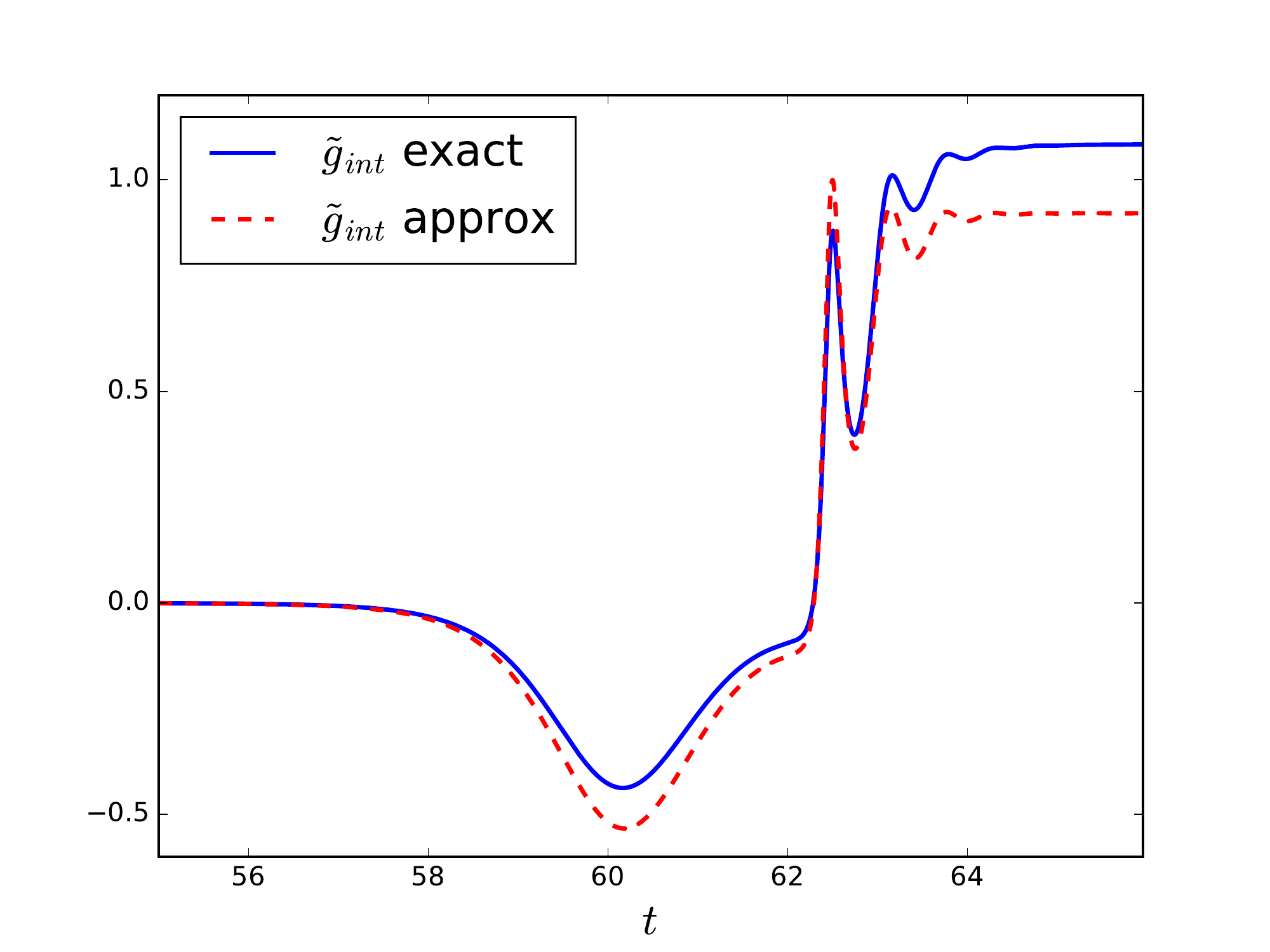}
	\caption{The exact numerical solutions for the different interesting parameters (basis components, slow-roll parameters, Green's functions, the spectral index and $\fnl$) during the turn for the first example of a monomial potential \eqref{pot n2m4} with $n=2$ and $m=4$. The last figure shows both the exact numerical solution for $\tilde{g}_{\mathrm{int}}$ and its analytical approximation. The horizontal purple dash-dot lines are the analytical predictions for $\gv_{12e}$, $\ns$ and $\fnl$.}
	\label{fig:n2m4}
\end{figure}

With this, it is possible to obtain an analytical estimate of the observables. First, we need to compute $\gf_*$ and $\gs_*$, using the solutions of equations \eqref{eqphi(t)} and \eqref{eqsig(t)} determined for monomial potentials. These solutions were computed assuming that $\gf_*$ and $\gs_*$ were the initial conditions, one has just to replace them by $\gf_i$ and $\gs_i$ here. This quick computation gives:
\be
\gf_* = 15.2\gk^{-1} \qquad \text{and} \qquad \gs_* = 0.092\gk^{-1}.
\ee
Using these values and the different expressions \eqref{constraint1}, \eqref{constraint2} and \eqref{constraint3}, we obtain:
\be
\gv_{12e}= -\frac{2}{\gf_*}\frac{3}{\gs_*}\frac{\gk C}{12\gb \gs_*^2}=3.52,\quad \ns =  1-\frac{4}{\gf_*^2}-2\frac{12\gb\gs_*^2}{\gk^2\ga\gf_*^2}=0.961, \quad -\frac{6}{5}\fnl=-\frac{12\gb\gs_*^2}{\gk^2 C}=1.2.
\label{predictions 1}
\ee
In these calculations, there are different approximations. First we use the monomial expressions to compute $\gf_*$ and $\gs_*$ (we refer the reader to section \ref{Monomial potentials section} for the details, but they require the slow-roll approximation and a quasi single-field situation, at least until horizon-crossing). Second, we use the limit of large $\gv_{12e}$ to compute the observables, the validity of this limit is explained in detail in section~\ref{Sum Potential Section}.
Hence, an error of order slow-roll (at horizon-crossing) is expected compared to the exact numerical results, which can be larger here since $\gv_{12e}$ is a little smaller than four.
Figure \ref{fig:n2m4} contains the same plots as shown for the double quadratic potential except that we have removed the plot of $\gxpa$ and $\gxpe$ which does not provide any additional information, and added a plot of $\gint$. The different analytical predictions in \eqref{predictions 1} are reasonable estimations of the different parameters but the difference is larger than expected, especially on the new plot concerning $\gint$. This plot displays both the exact numerical $\tilde{g}_{\mathrm{int}}$ and its analytical prediction from \eqref{gintbsr} (more precisely, the analytical form of the approximated solution, with the different parameters determined numerically), using the definition: 
\be
\tilde{g}_{\mathrm{int}}=-\frac{2(\gv_{12})^2}{\lh 1 + (\gv_{12})^2\rh^2}\gint.
\ee
As one can see, both curves have a similar form, but there is a difference of around 15\%. The reason is that the turn occurs late with $\ge\approx 0.2$ when it starts. This value is already too large to have the slow-roll approximation working perfectly, but not enough for it to totally break down. In fact, this problem is quite general with the monomial potential because the turn has to occur late to get $\fnl$ of order unity, as shown in the section \ref{Monomial potentials section}. 

\begin{figure}
	\centering
	\includegraphics[width=0.49\textwidth]{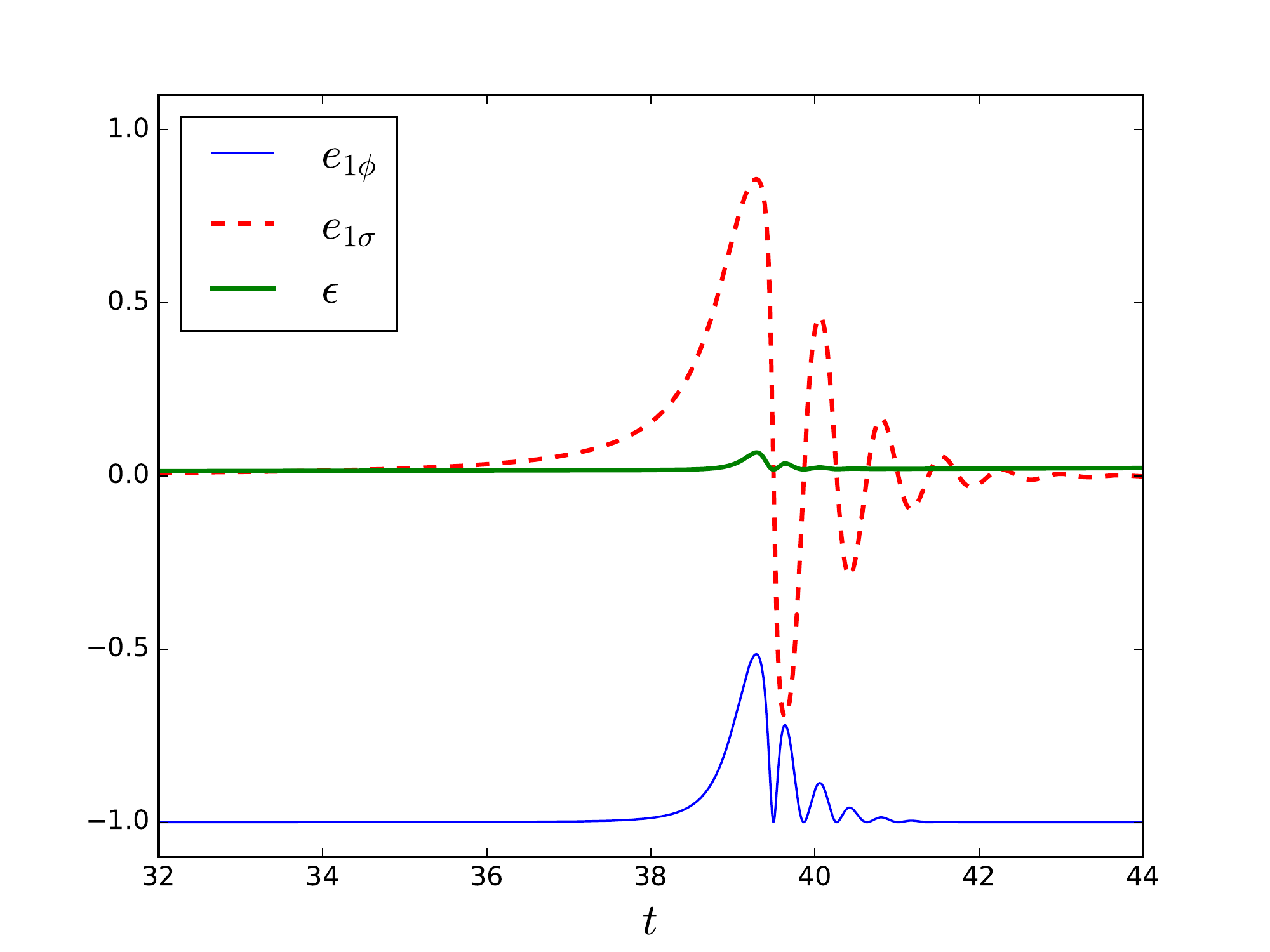}
	\includegraphics[width=0.49\textwidth]{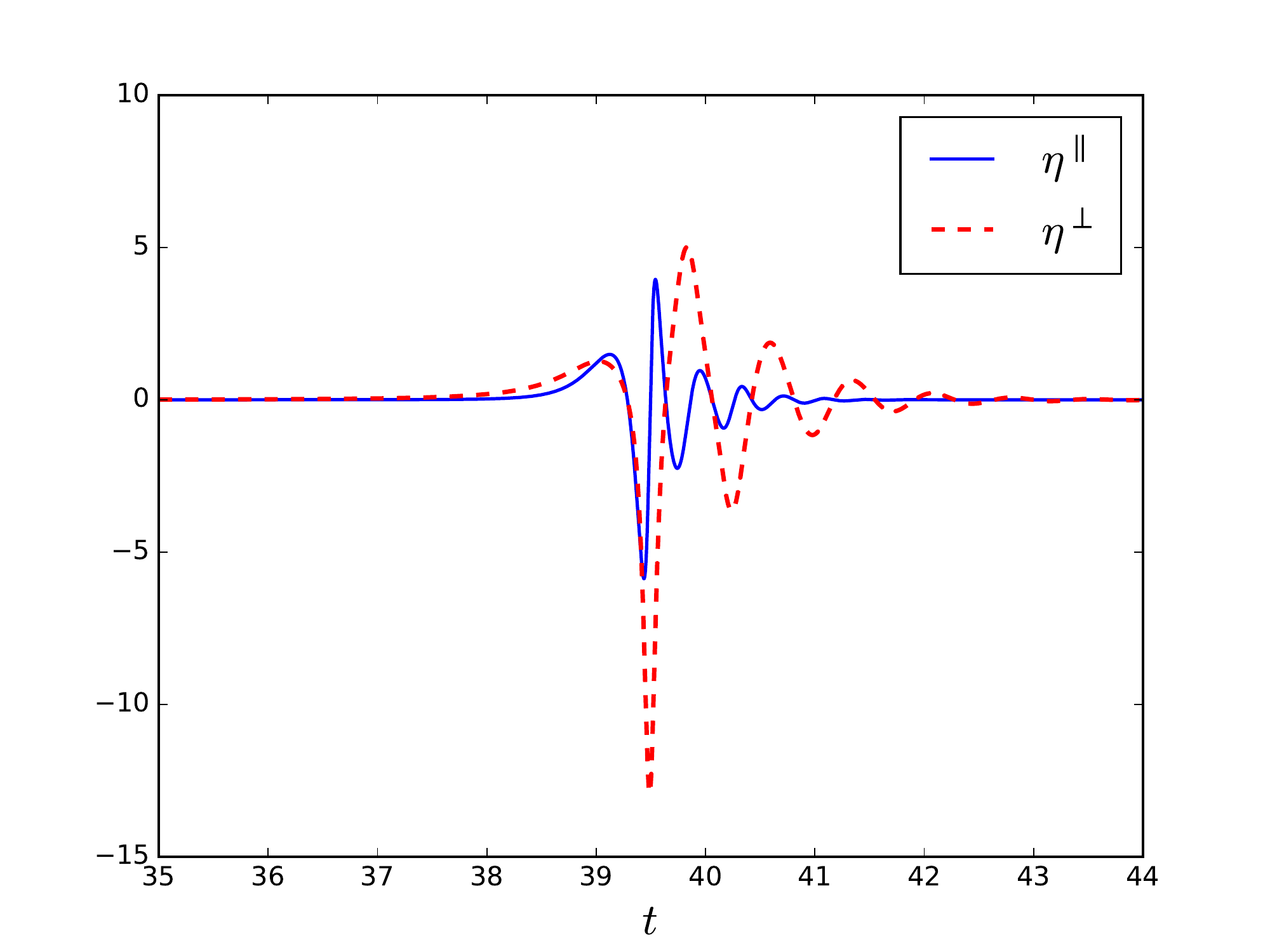}
	\includegraphics[width=0.49\textwidth]{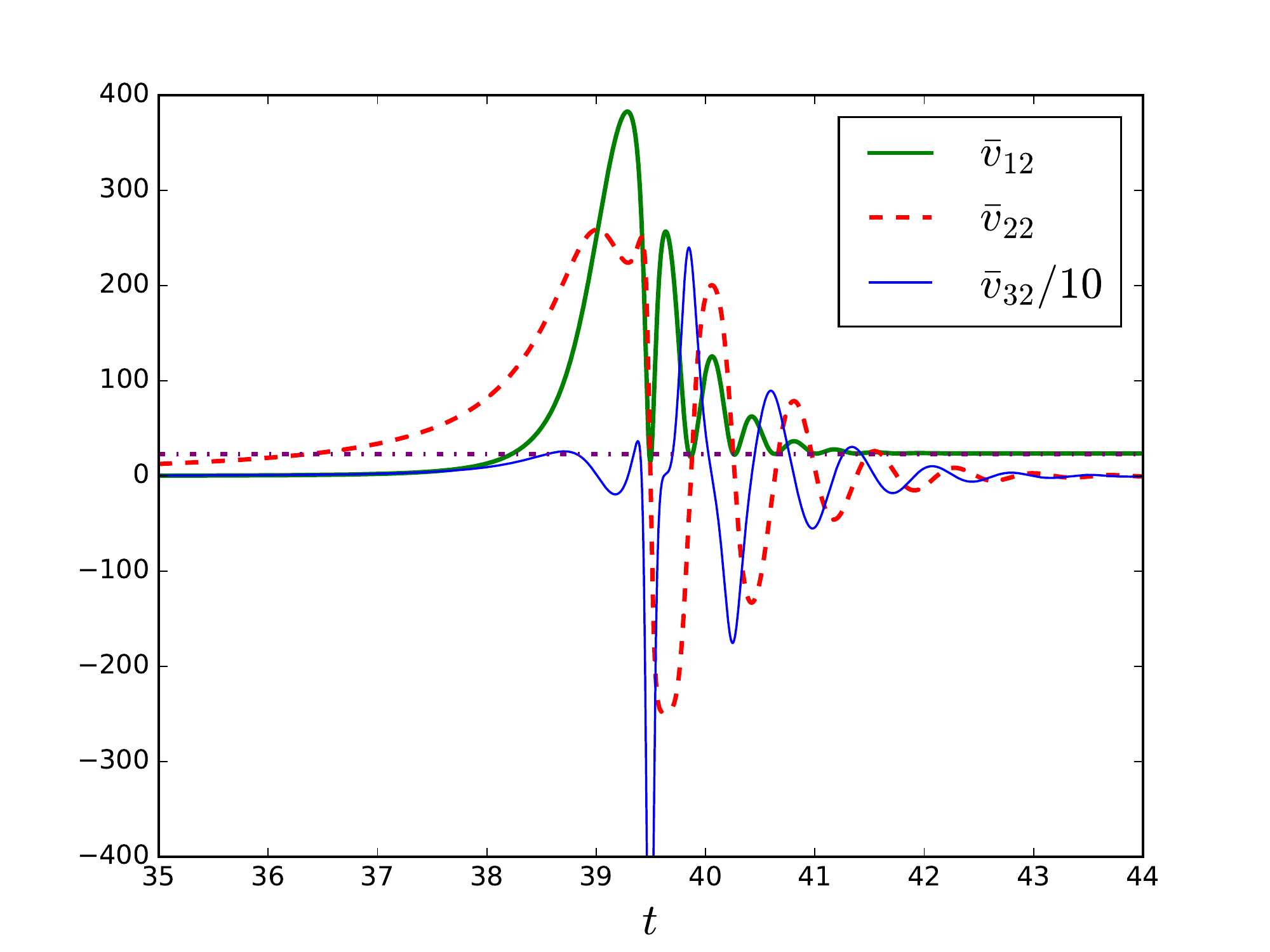}
	\includegraphics[width=0.49\textwidth]{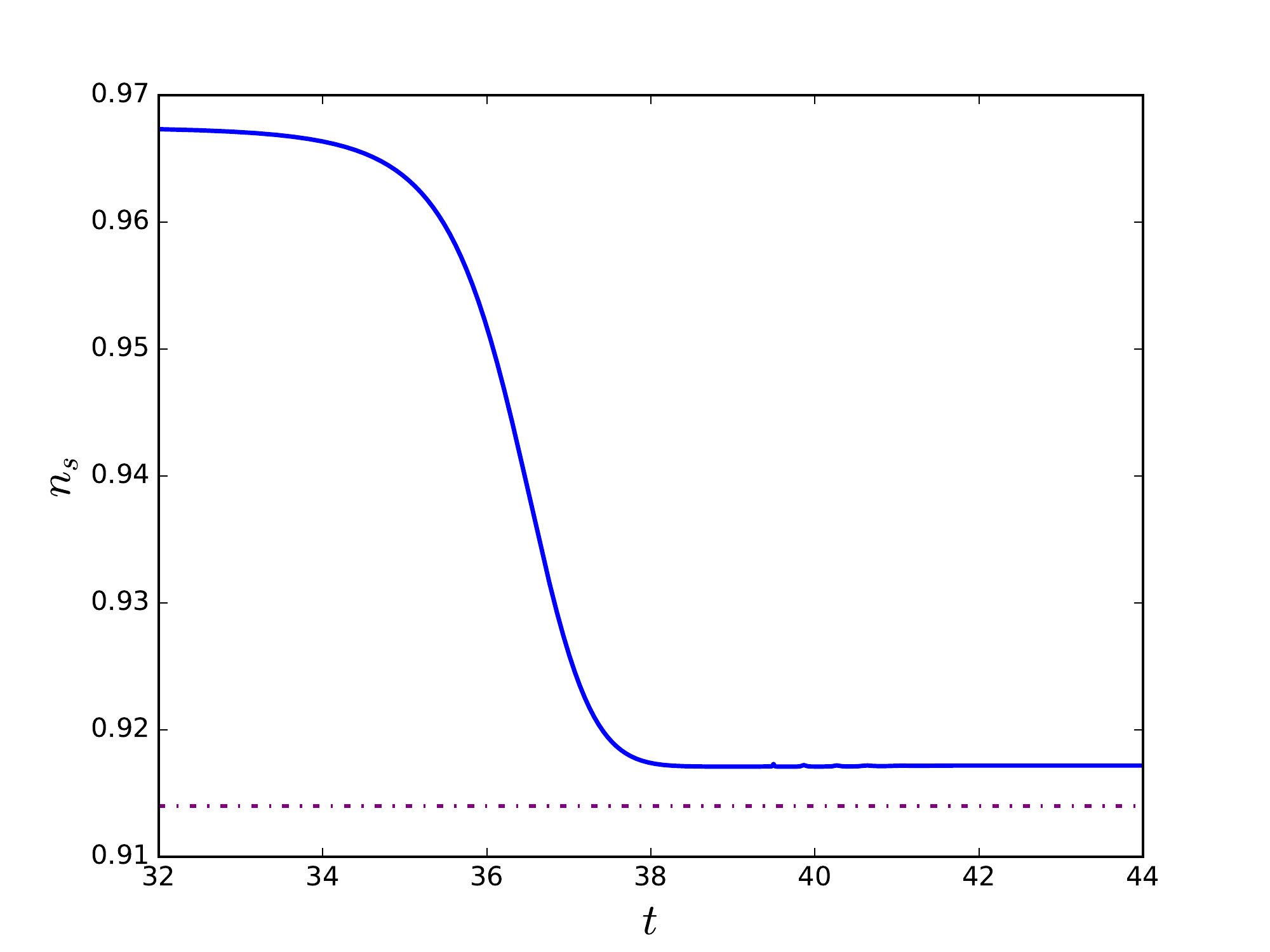}
	\includegraphics[width=0.49\textwidth]{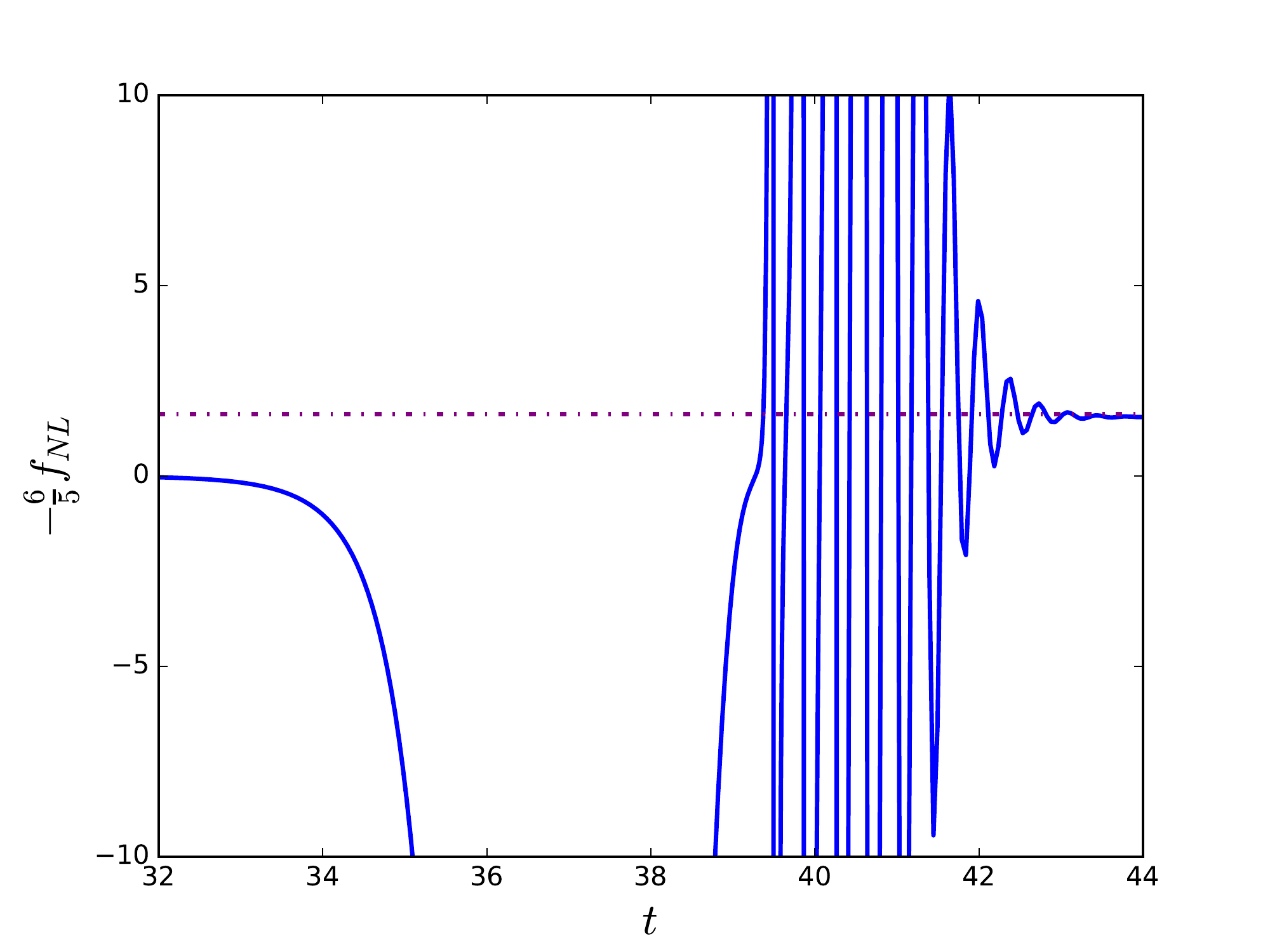}
    \includegraphics[width=0.49\textwidth]{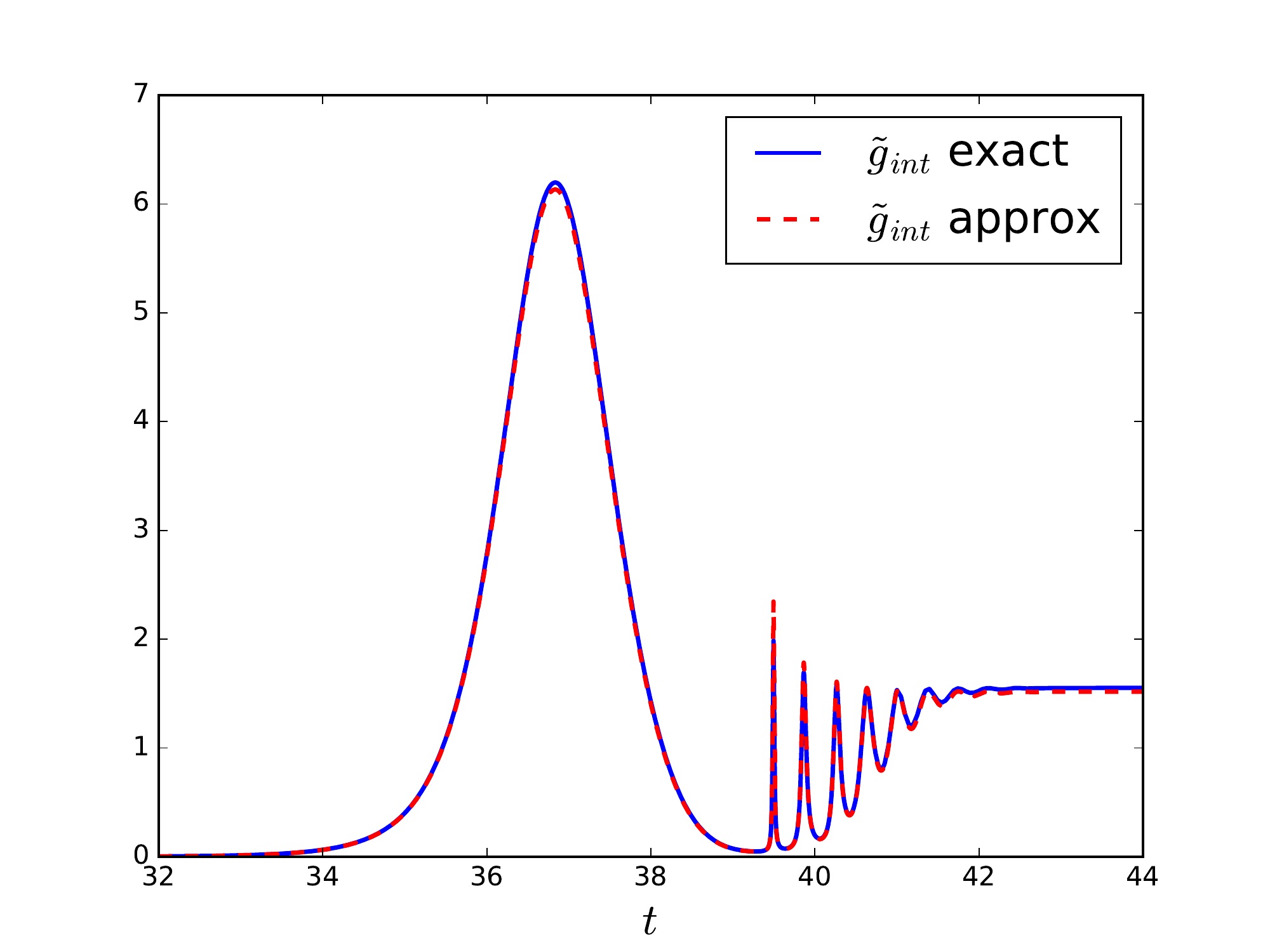}
	\caption{Same as figure \ref{fig:n2m4} but for the second example of monomial potential \eqref{pot n2m4} with $n=2$ and $m=4$ (with the parameter values
given just above (\ref{gfgsstarvalues})).}
	\label{fig:n2m42}
\end{figure}

However, if we forget momentarily about the observational constraint on the spectral index, only for one example to illustrate the validity of the analytical expressions, it is possible to have the turn occuring earlier.
The second set of values is: $\ga=\half\gk^{-2}$, $C= {\textstyle\frac{4}{27}\frac{\gb^3}{\lambda^2}}$, $\gb=-2000$, and $\lambda=-{\textstyle\frac{40}{3}}\gb\gk^2$ with the initial conditions $\gf_i= 16\gk^{-1}$ and $\gs_i=0.01\gk^{-1}$.  This time, the analytical predictions are:
\be
\gf_* = 15.3\gk^{-1} \qquad \text{and} \qquad \gs_* = 0.0106\gk^{-1}.
\label{gfgsstarvalues}
\ee
which leads to:
\be
\gv_{12e}= 23,\quad \ns = 0.914 \quad \text{and} \quad -\frac{6}{5}\fnl=1.62.
\ee
Figure \ref{fig:n2m42}, which contains the same plots as figure \ref{fig:n2m4} but for the new parameters, shows that $\ge$ is of order $10^{-2}$ during the turn, which is in the domain of validity of the main hypothesis $\ge \ll 1$. During the turn, $\getpe$ is of order 10 at most, which shows that the slow-roll regime is broken. As expected, analytical predictions are now a very good estimation. However, the spectral index is 0.917, which is outside the observational bounds. This example is also used in the previous section in figure~\ref{fig:rhs} to illustrate that r.h.s.\ is several orders of magnitude smaller than the left-hand side terms of \eqref{equadiff}.

\subsubsection{Second example: Axion}
\label{Axion section}

The next example is the axion-quartic model originally introduced in \cite{Elliston:2012wm} and discussed more recently in \cite{Dias:2016rjq}. The potential is:
\be
W(\gf,\gs)=\frac{1}{4}g\gf^4 + \Lambda^4 \left[ 1-\cos{\lh \frac{2\pi\gs}{f}\rh}\right],
\label{pot axion}
\ee
with $g=10^{-10}$, $\Lambda^4 = \lh\frac{25}{2\pi}\rh^2 g \gk^{-4}$ and $f=\gk^{-1}$. The initial conditions are $\gf_i = 23.5 \gk^{-1}$ and $\gs_i= \frac{f}{2} - 10^{-3}\gk^{-1}$. Defining $\gs'={\textstyle\frac{f}{2}}-\gs$, we have $\gs'\ll \gk^{-1}$. This will stay true until the turn, hence it is possible to perform an expansion of the potential in terms of this small parameter. At first order, we have $\cos{\lh \frac{2\pi\gs}{f}\rh}= -\cos{\lh \frac{2\pi\gs'}{f}\rh} =-1+ \half\lh {\textstyle\frac{2\pi\gs'}{f}} \rh^2$ which substitued into the potential gives:
\be 
W(\gf,\gs')=\frac{1}{4}g \gf^4 + 2 g \lh \frac{25}{2\pi} \rh^2 \gk^{-4} -\frac{1}{2}g \lh\frac{25}{f}\rh ^2  \gk^{-4} (\gs')^2.
\ee
This is a monomial potential with $n=4$ and $m=2$, hence in the region of parameters where the spectral index constraints cannot be satisfied. This is verified by computing  the analytical predictions like for the previous example. The fields at horizon-crossing are:
\be
\gf_* = 21.8\gk^{-1} \qquad \text{and} \qquad \gs'_* = -1.1 \times 10^{-3}\gk^{-1}.
\ee 
which leads to:
\be 
\gv_{12e}=-\frac{8\gk}{\gf_*\gs'_*}\lh\frac{f}{2\pi}\rh^2=-8.4, \quad \ns = 1 - \frac{16}{\gf_*^2} - 8\frac{25^2}{\gk^2 f^2 \gf_*^4} = 0.944 \quad \text{and} \quad -\frac{6}{5}\fnl = 2\pi^2.
\ee
This model gives $\fnl$ of order ten, however the spectral index is lower than the Planck constraints. 

\begin{figure}
	\centering
	\includegraphics[width=0.49\textwidth]{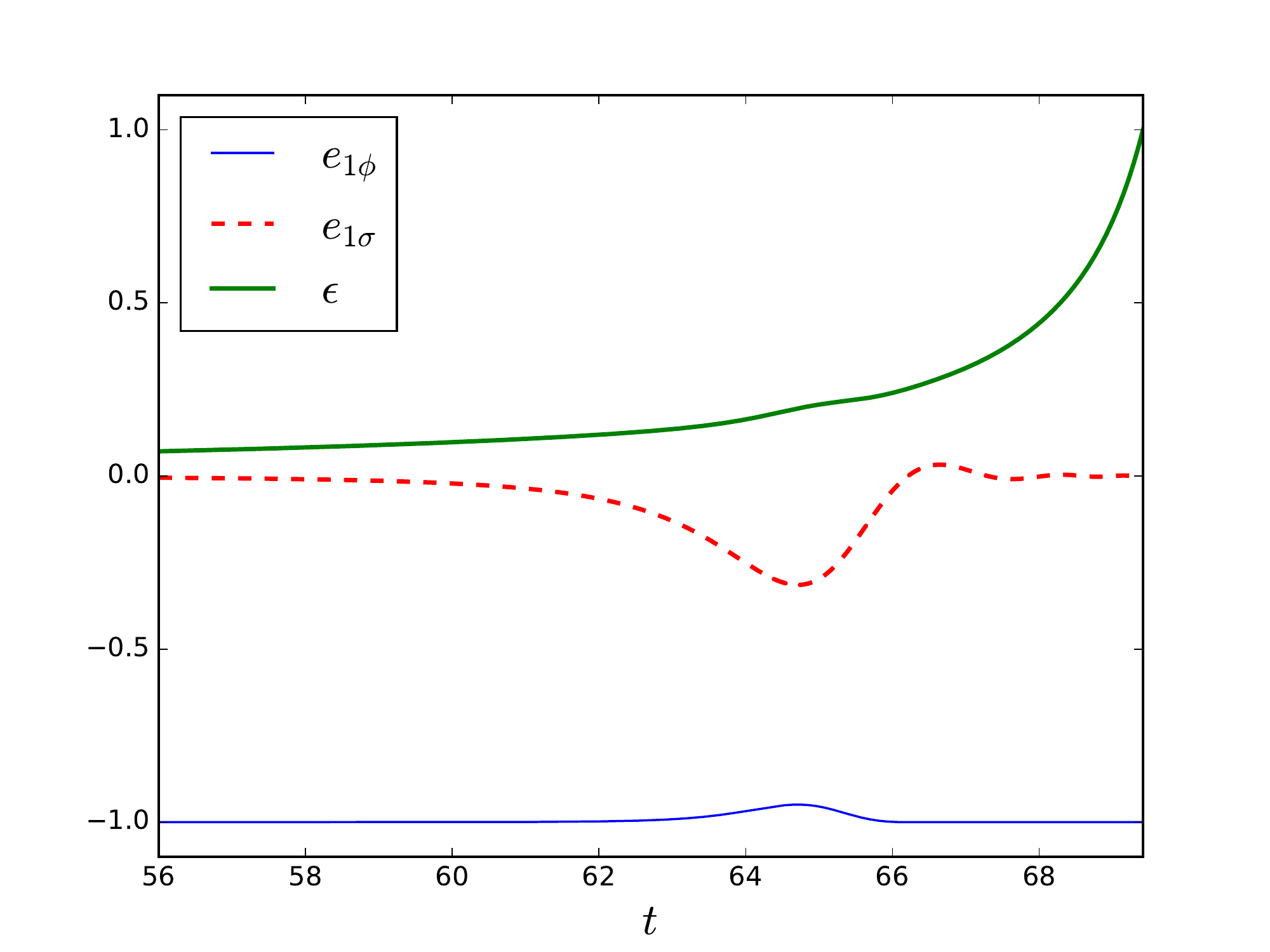}
	\includegraphics[width=0.49\textwidth]{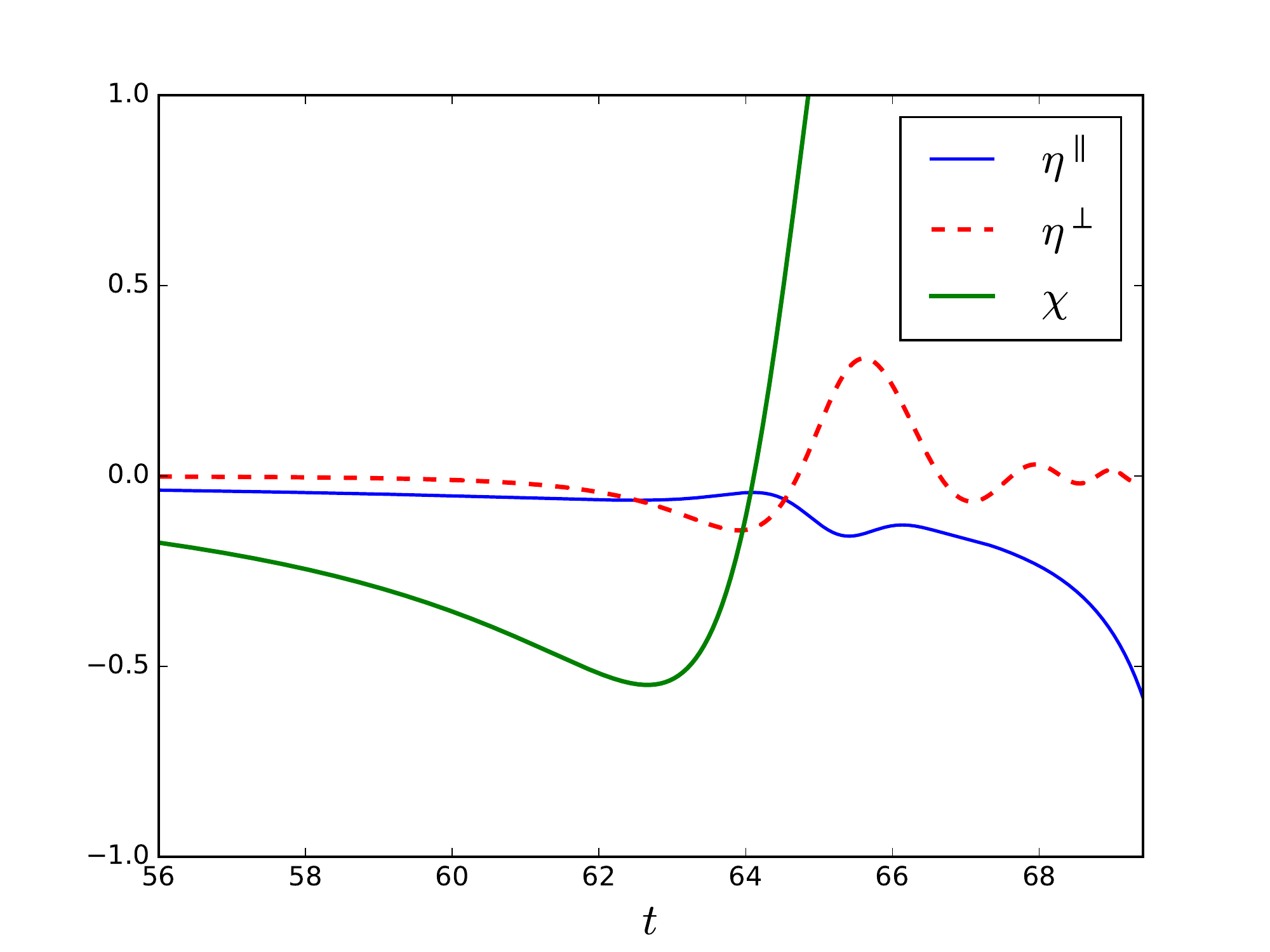}
	\includegraphics[width=0.49\textwidth]{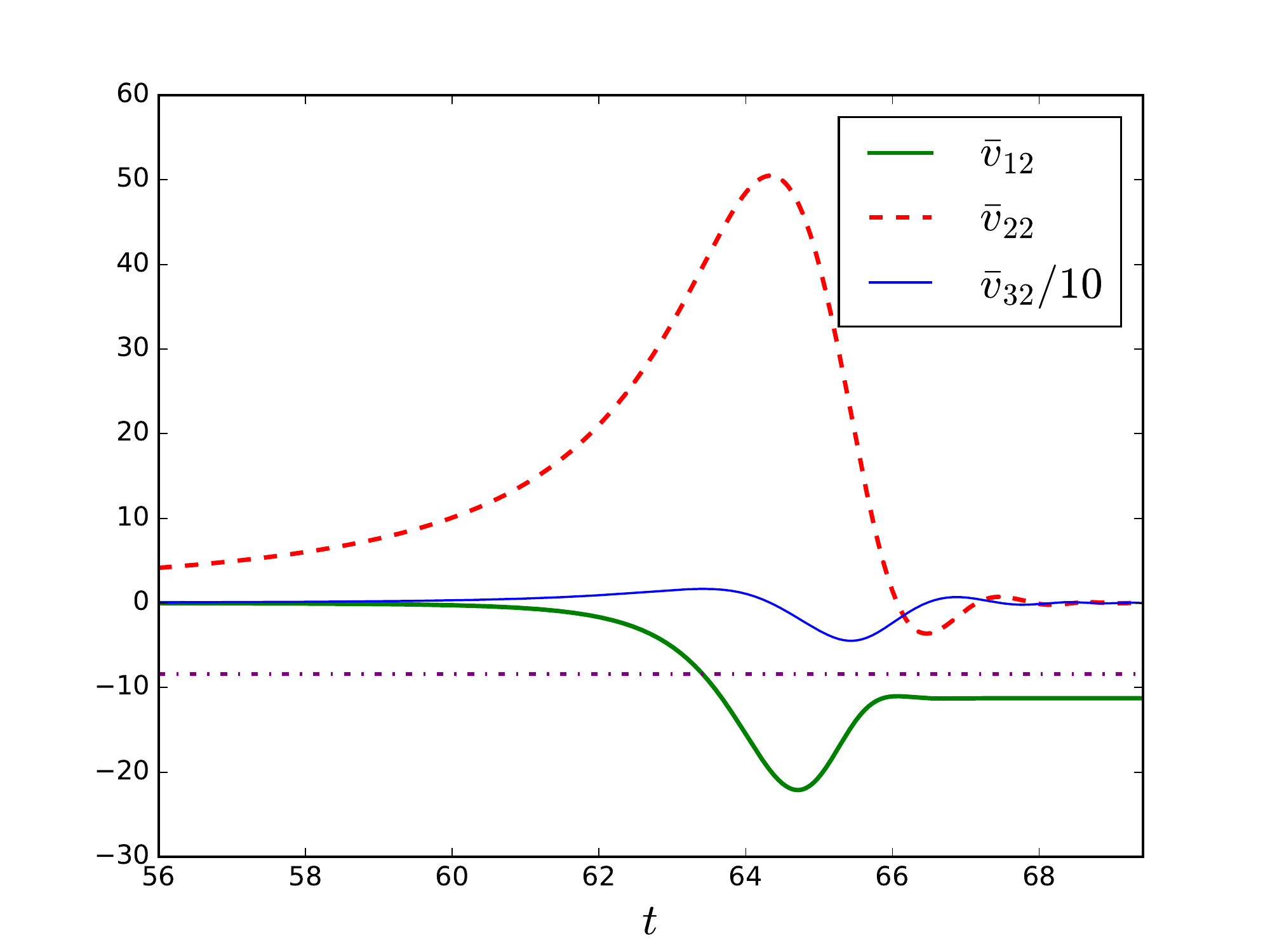}
	\includegraphics[width=0.49\textwidth]{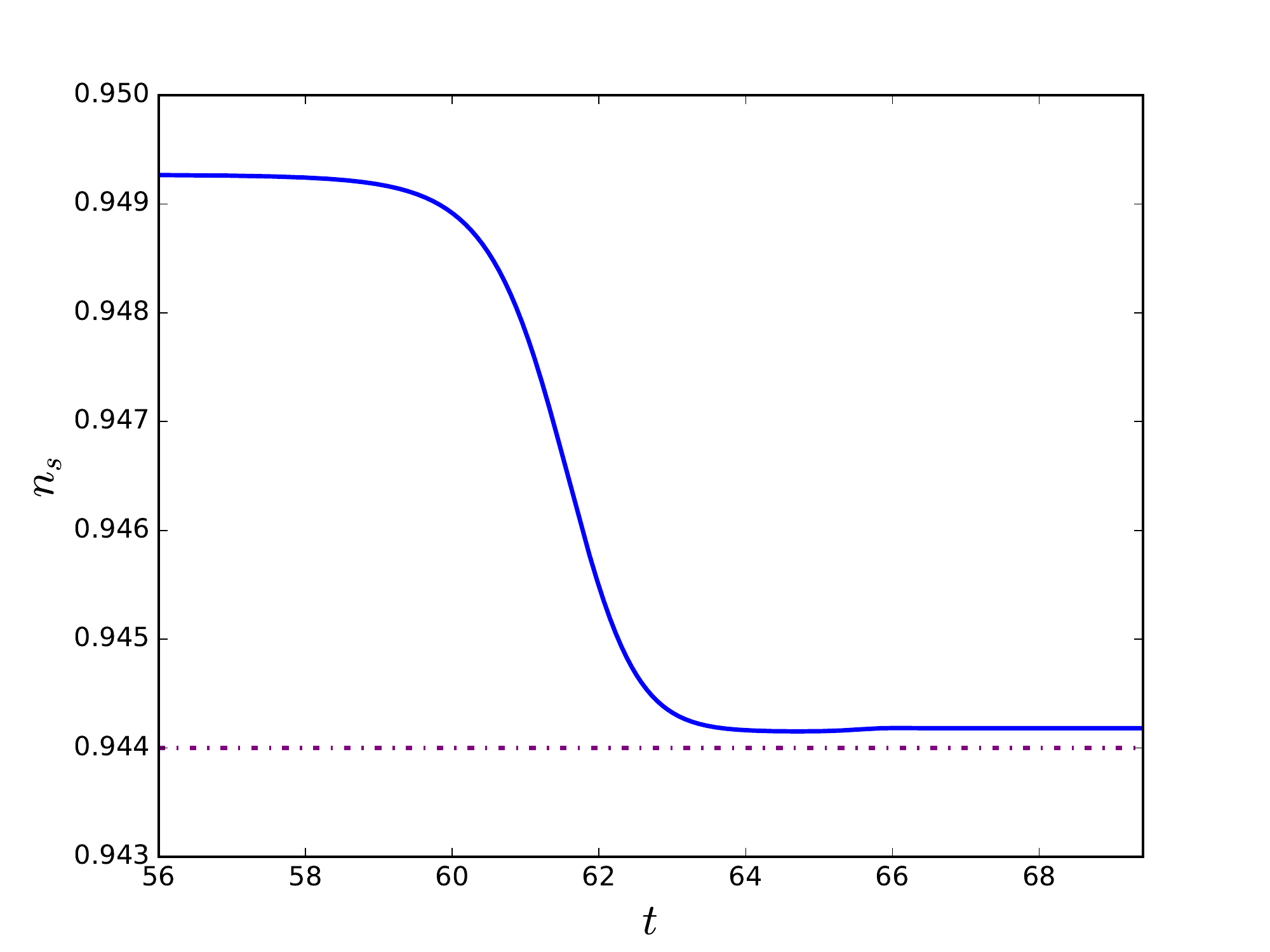}
	\includegraphics[width=0.49\textwidth]{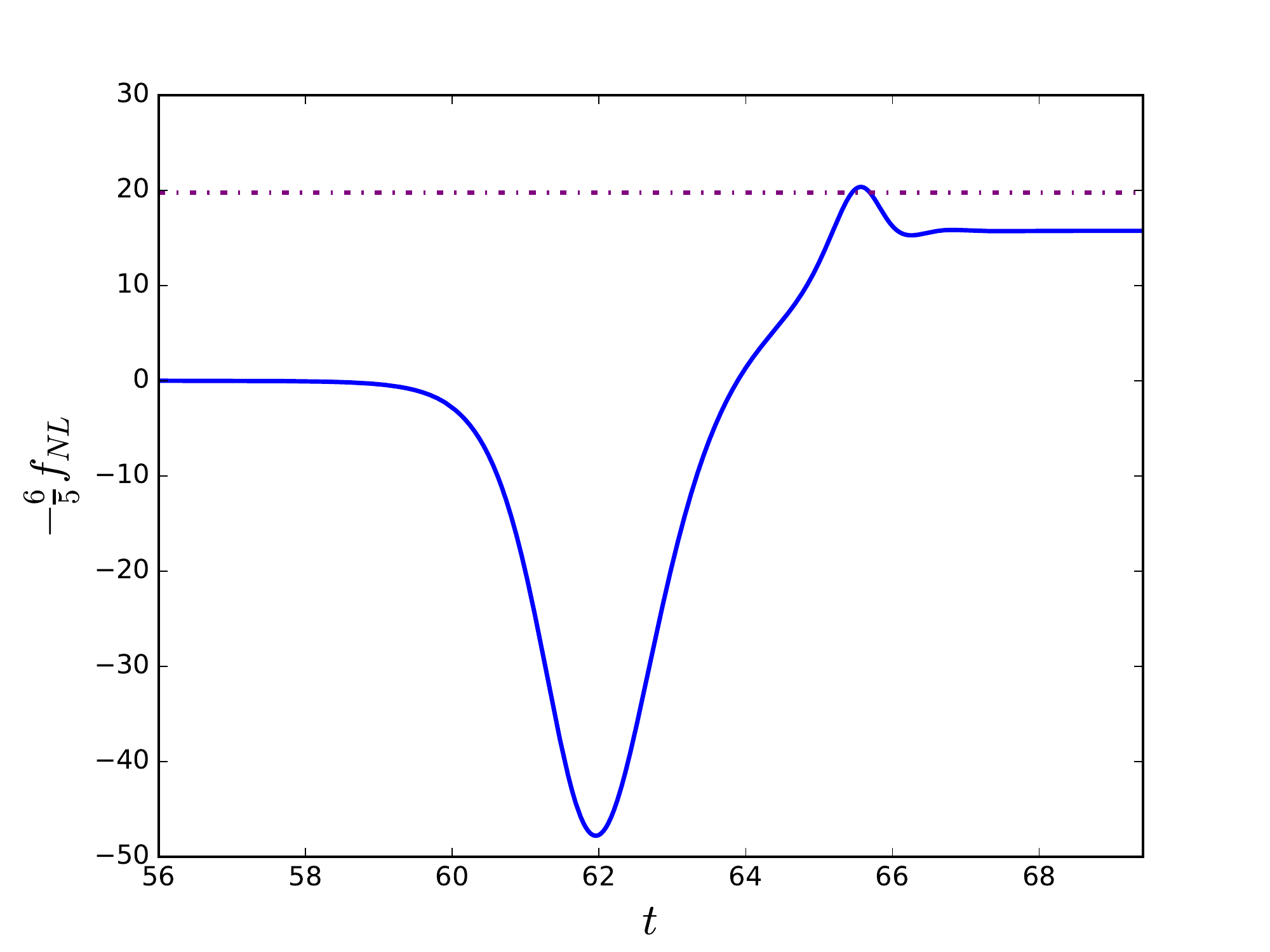}
    \includegraphics[width=0.49\textwidth]{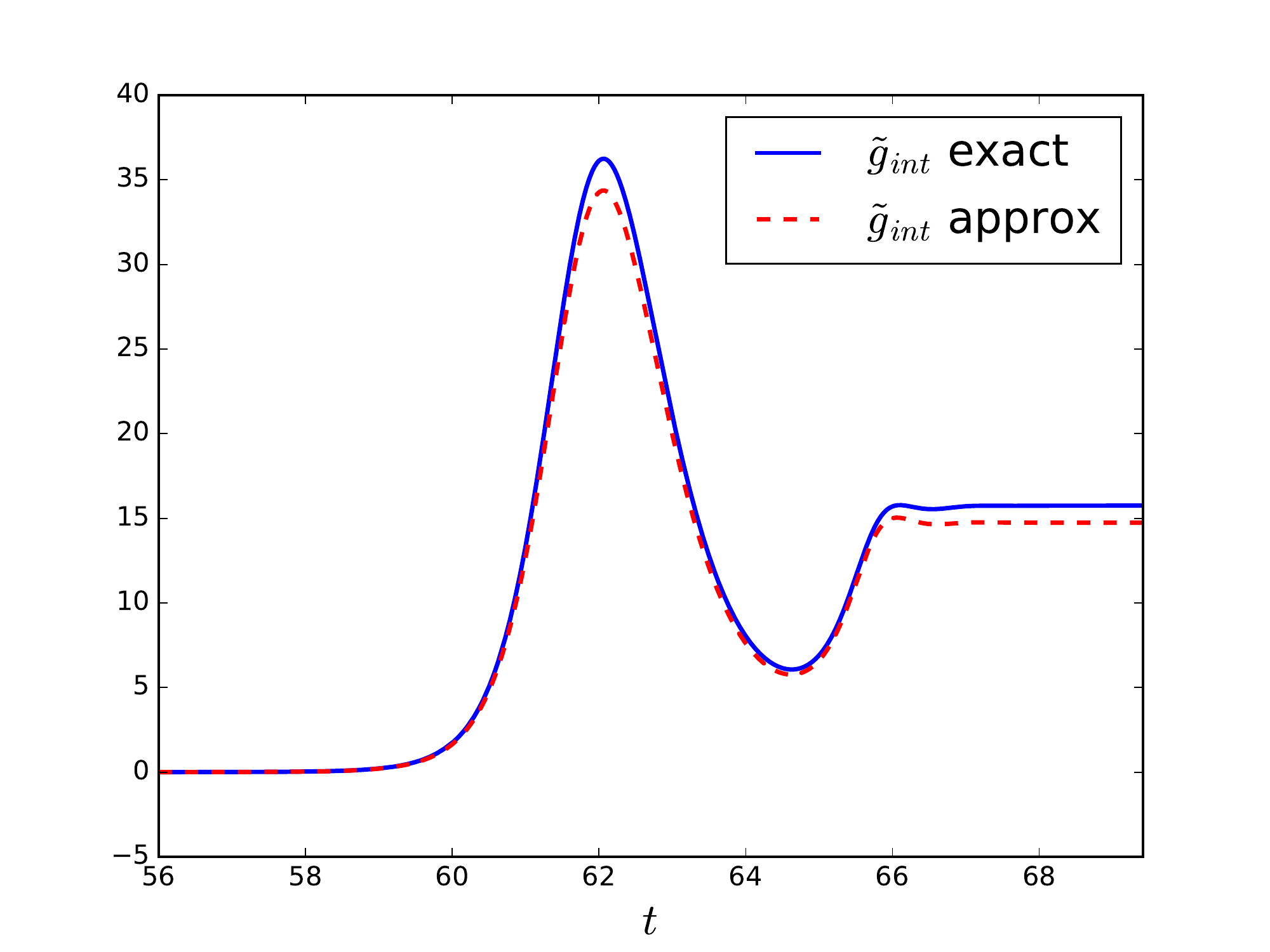}
	\caption{Same as figure \ref{fig:n2m4} but for the quartic-axion potential \eqref{pot axion}.} 
	\label{fig:axion}
\end{figure}

Figure \ref{fig:axion} confirms these results. Again in this model the turn occurs very late and there is a shift between the prediction and the exact result even if $\ge$ is still small enough during the turn. Moreover, $\getpa$ and $\getpe$ stay smaller than one during the turn, but $\gc$, which is displayed on the same plot, becomes large. This is another regime than the ones studied in section \ref{Green's functions section}. This has a direct impact on the Green's functions because $\gc$ appears in \eqref{G22eq} which explains the difference between the slow-roll prediction for $\gv_{12e}$ and the exact value. However, one interesting point is that the analytical form of $\gint$ stays valid. This case of large $\gc$ when other slow-roll parameters are small is not common and is due here partially to the fact that $\gV_{\gs\gs*}$ is too large to respect the Planck constraint (because as discussed in section \ref{Sum Potential Section}, $\gc_*=\ge_* + \getpa_* + \gV_{\gs\gs*}$).

\subsection{Second type of turn}
\subsubsection{$m=2$ and $n=2$}
\label{n2m2 section}

Figure \ref{fig:etaper1} shows that a turn of the first type respecting observational constraints is not possible for a monomial potential with $n=2$ and $m=2$. However, if we do not keep the constraint that the turn must start before the end of the slow-roll regime, this model can have a turn of the second type. This example was published originally in \cite{Tzavara:2010ge} and is here adapted to be in agreement with the latest Planck constraints. See the second line of plots in figure \ref{fig:slowrollbroken} for an illustration of the field trajectory. The potential has the form:
\be
W(\gf,\gs) = \ga \gf^2 + C + \gb \gs^2 + \lambda \gs^4,
\label{pot n2m2}
\ee
with $\ga=20 \gk^{-2}$, $C= {\textstyle\frac{\gb^2}{4\lambda}}$, $\gb= -9\gk^{-2}$ and $\lambda=2$. The initial conditions are $\gf_i=18 \gk^{-1}$ and $\gs_i = 0.01 \gk^{-1}$ with $\dot{\gf}_i$ and $\dot{\gs}_i$ determined by the slow-roll approximation. At horizon-crossing, we have:
\be
\gf_* = 14.9\gk^{-1} \qquad \text{and} \qquad \gs_* = 0.011 \times 10^{-3}\gk^{-1}.
\ee
Substituted into \eqref{constraint1}, \eqref{constraint2} and \eqref{constraint3} this gives:
\be 
\gv_{12e} = -\frac{2\gk}{\gf_*\gs_*}\frac{C}{2\gb} = 6.9, \quad \ns = 1 - \frac{4}{\gf_*^2} +\frac{4\gb}{\gk^2 \ga\gf_*^2} = 0.974 \quad \text{and} \quad -\frac{6}{5}\fnl = -\frac{2\gb}{\gk^2 C} = 1.8 .
\ee

\begin{figure}
	\centering
	\includegraphics[width=0.49\textwidth]{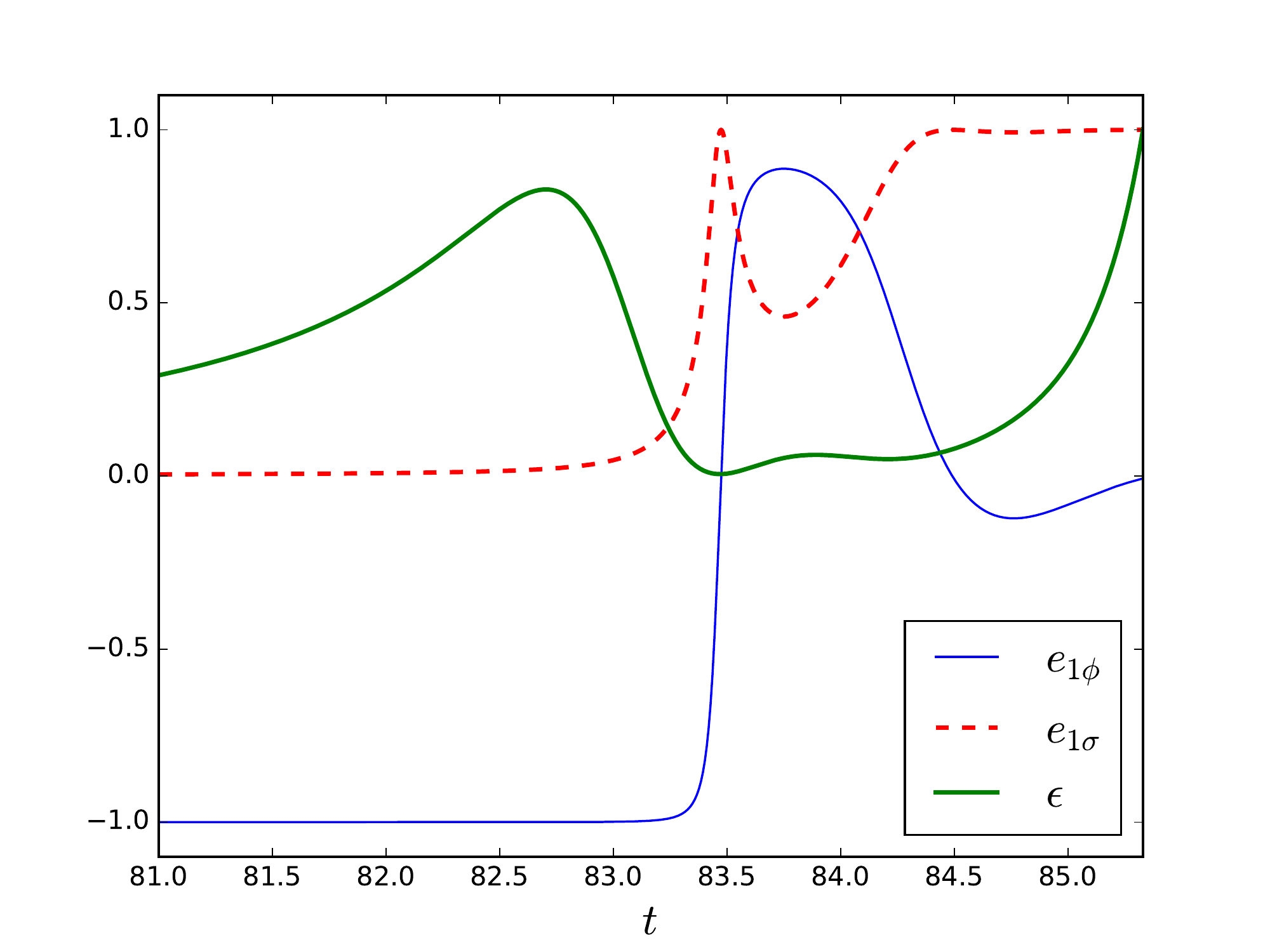}
	\includegraphics[width=0.49\textwidth]{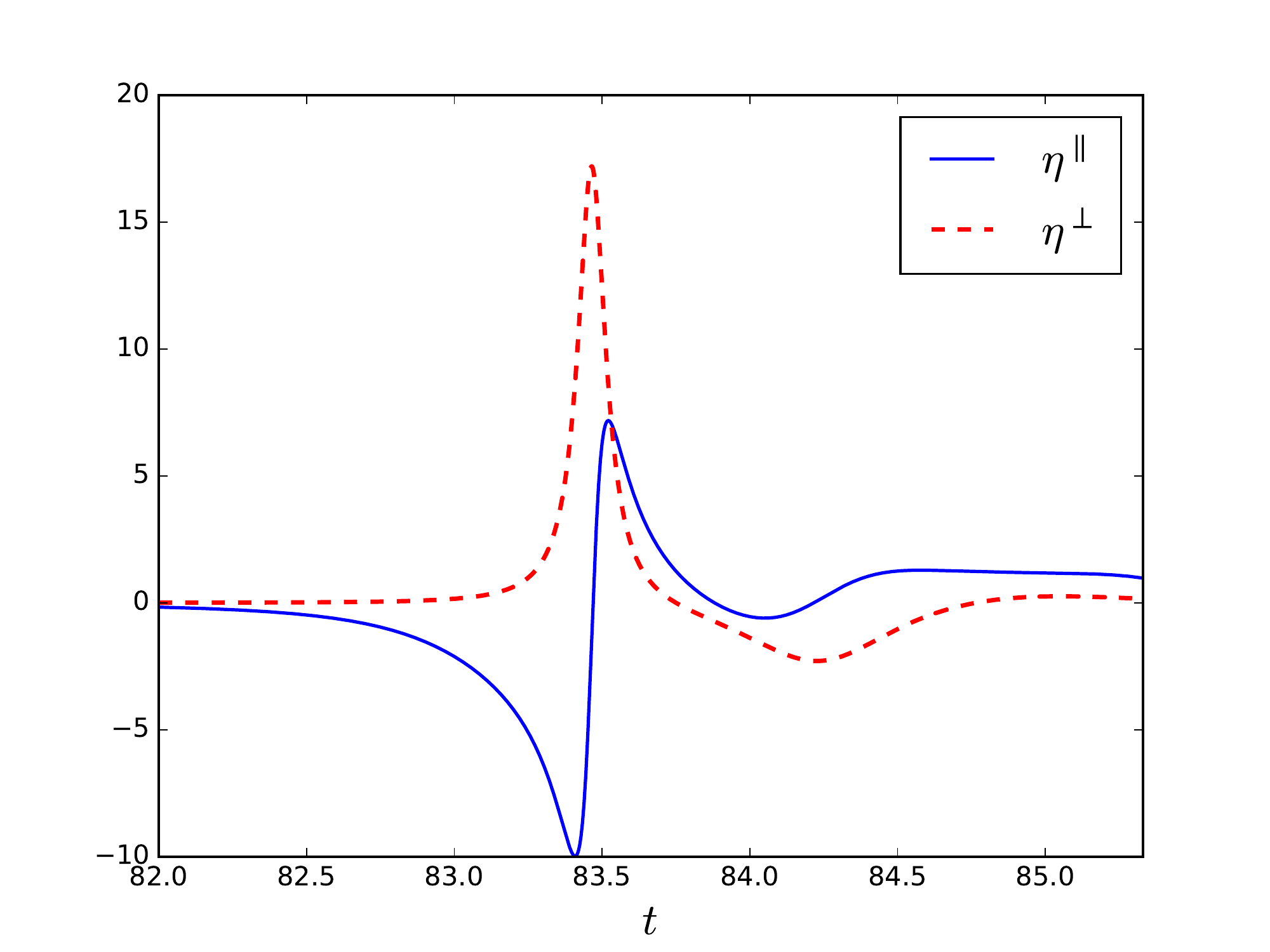}
	\includegraphics[width=0.49\textwidth]{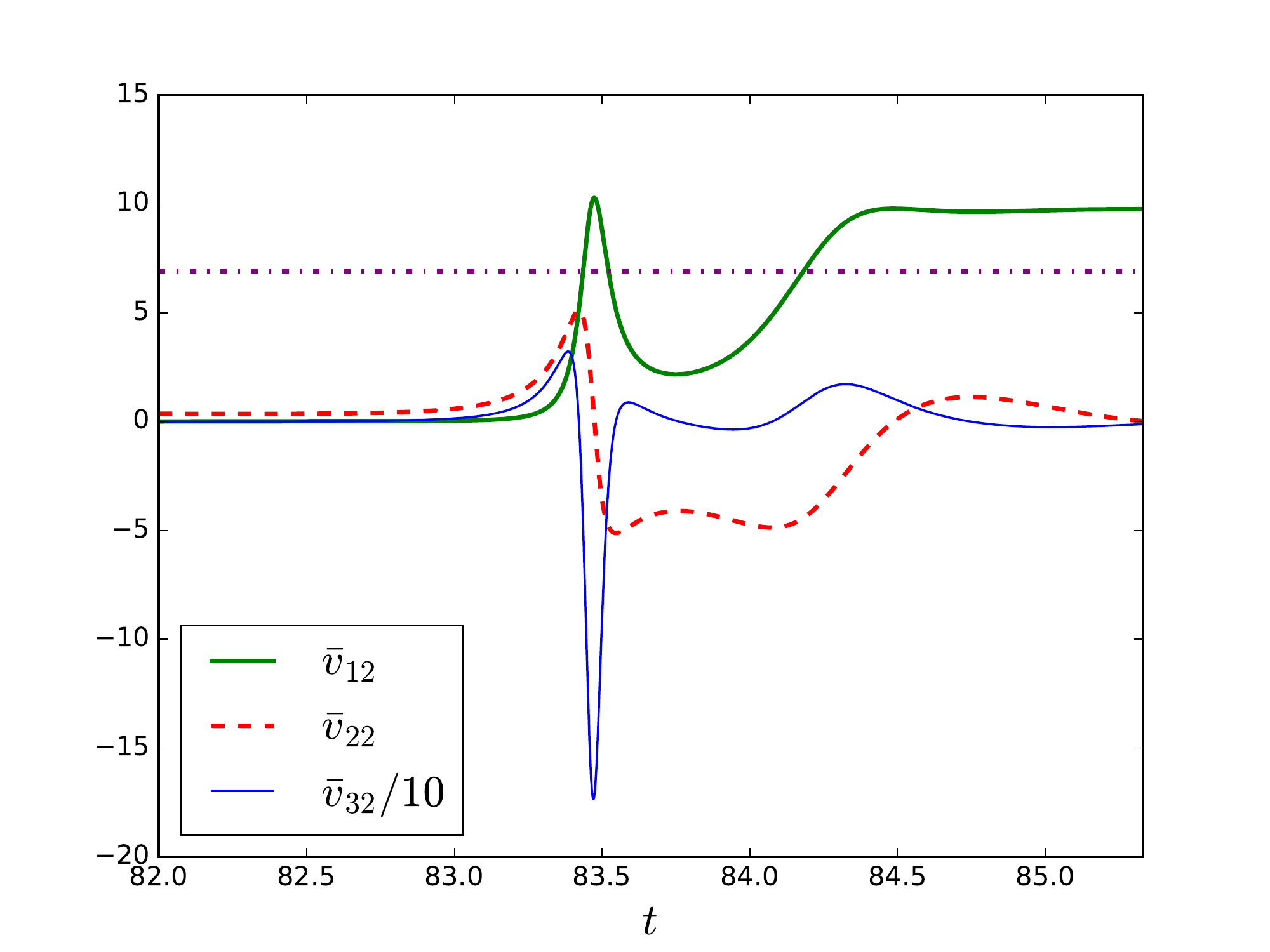}
	\includegraphics[width=0.49\textwidth]{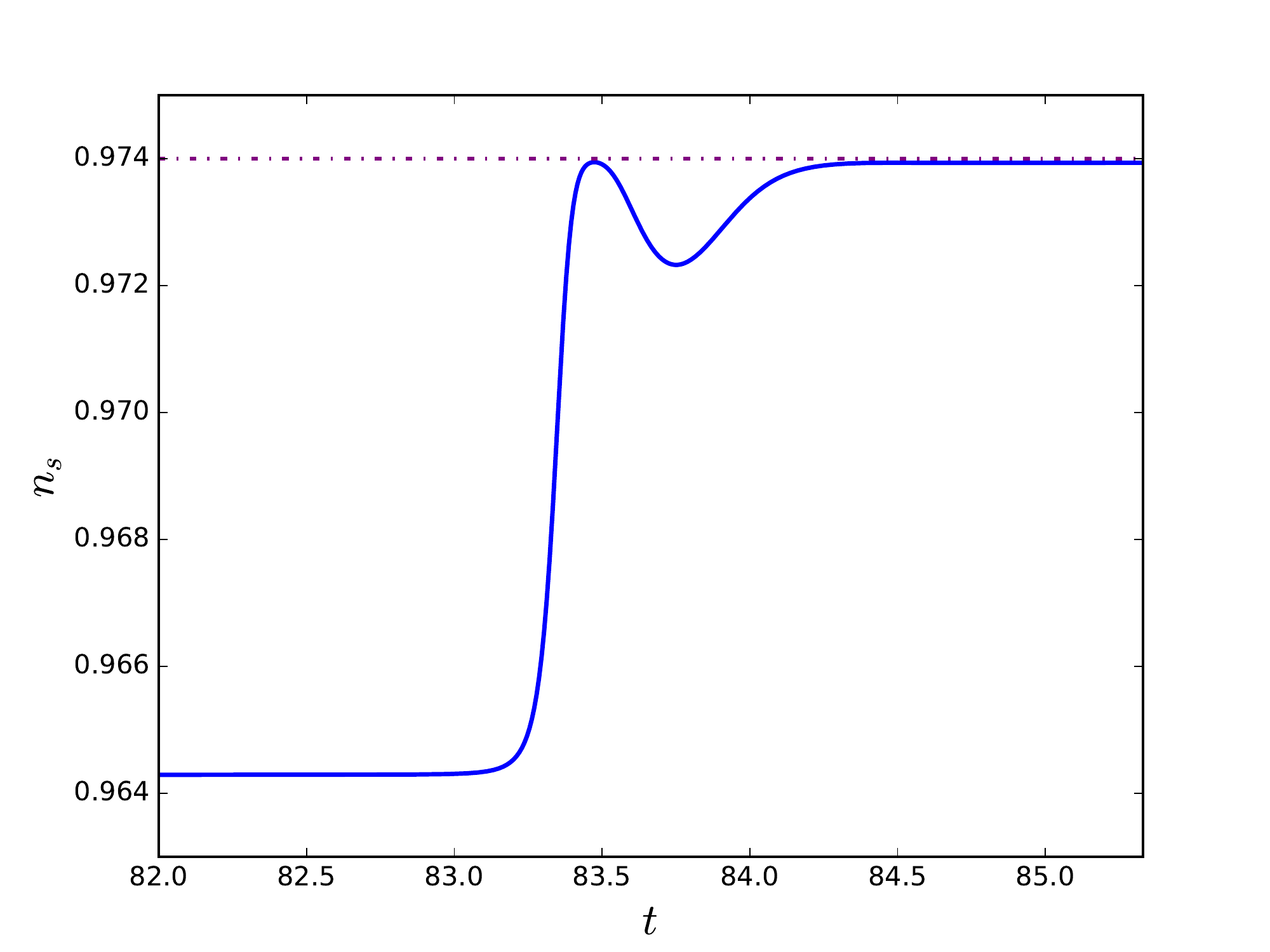}
	\includegraphics[width=0.49\textwidth]{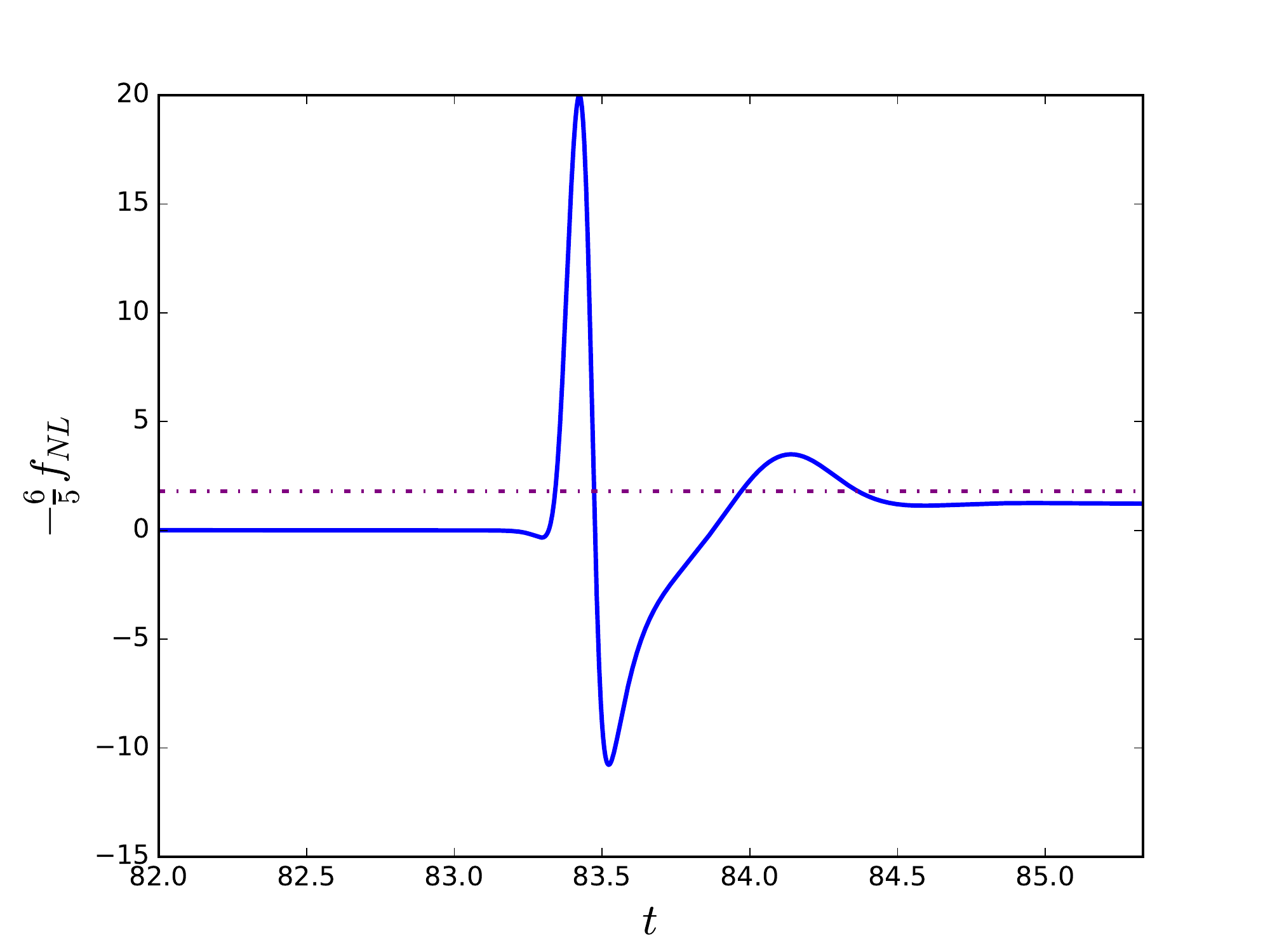}
    \includegraphics[width=0.49\textwidth]{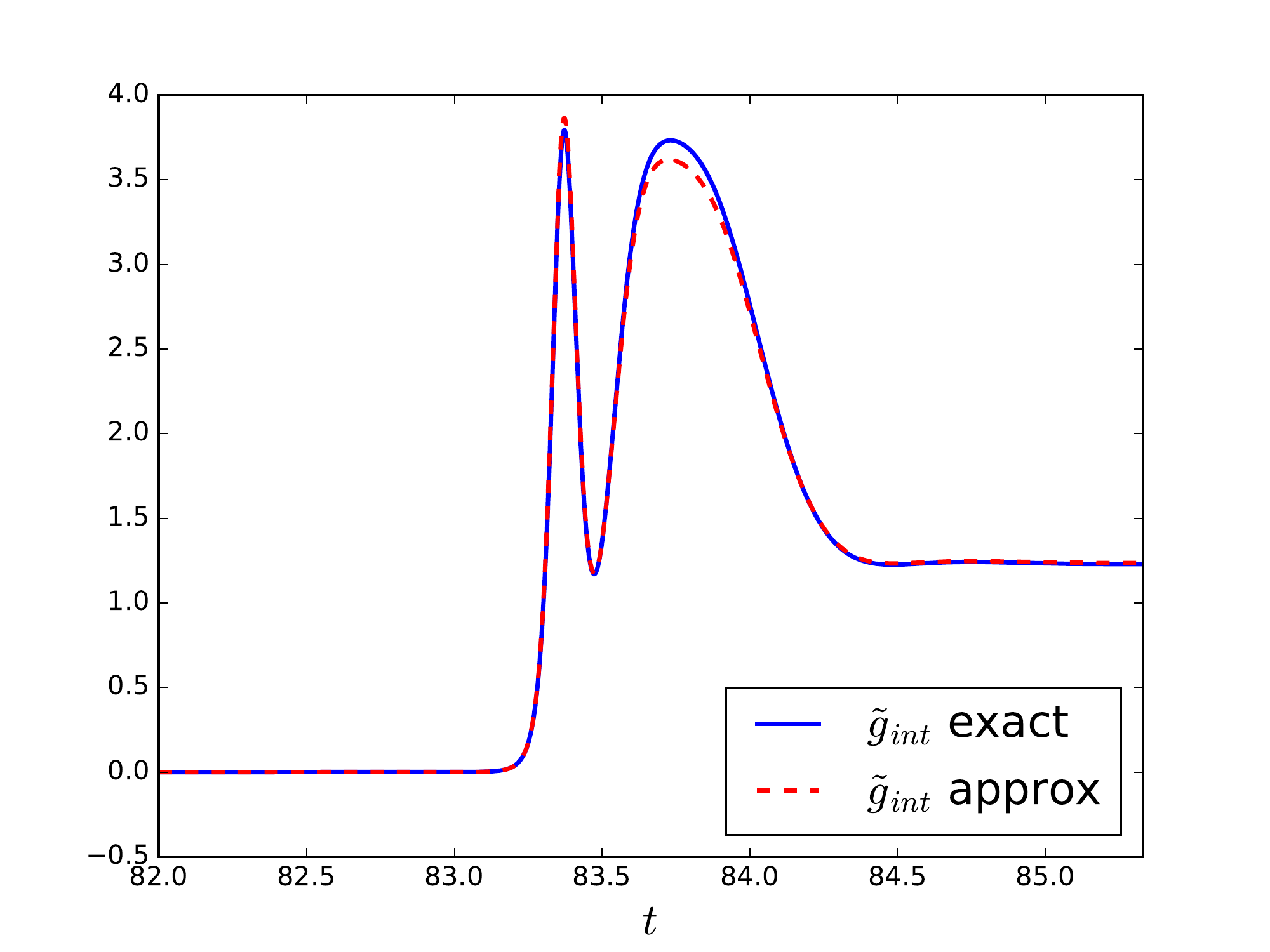}
	\caption{Same as figure \ref{fig:n2m4} but for the monomial potential with $n=2$ and $m=2$ \eqref{pot n2m2}.} 
	\label{fig:n2m2}
\end{figure}

Figure \ref{fig:n2m2} confirms that in this example the turn occurs after the field $\gf$ reaches the minimum of its potential. The Green's function $\gv_{12e}$ is larger than the slow-roll value, hence $\fnl$ is a little smaller than expected. This is in agreement with the discussion of the second type of turn in section \ref{end of inflation section}. However, this does not have any impact on the spectral index because the dependence on $\gv_{12e}$ disappears when it is larger than 4. Hence, this model is allowed by the Planck constraints.

\subsubsection{A non-monomial example}
\label{last section}

This last example is in the vein of the previous one in terms of the form of the field trajectory. However, there are several supplementary terms to show the validity of some analytical results beyond simple monomial potentials. The model has the following potential:
\be
W(\gf,\gs)= \frac{1}{4}\lambda \lh \gf^4 + \gs^4 + m^4 - 2 m^2  \gf^2 - 2 m^2  \gs^2 \rh + \nu  (m-\gf)^3 + W_0,
\label{last}
\ee
with $\lambda = 1200$, $\nu = 100\gk^{-1}$, $m=2\gk^{-1}$ and $W_0={\textstyle\frac{1}{4}}\lambda m^4$. The initial conditions are $\gf_i=25\gk^{-1}$ and $\gs_i=0.05 \gk ^{-1}$. We cannot use the monomial potential equations to determine $\gf_*$ and $\gs_*$, however the slow-roll estimation of $\fnl$ does not require them:
\be
-\frac{6}{5}\fnl=\frac{4}{\gk^2 m^2} = 1.
\ee

\begin{figure}
	\centering
	\includegraphics[width=0.49\textwidth]{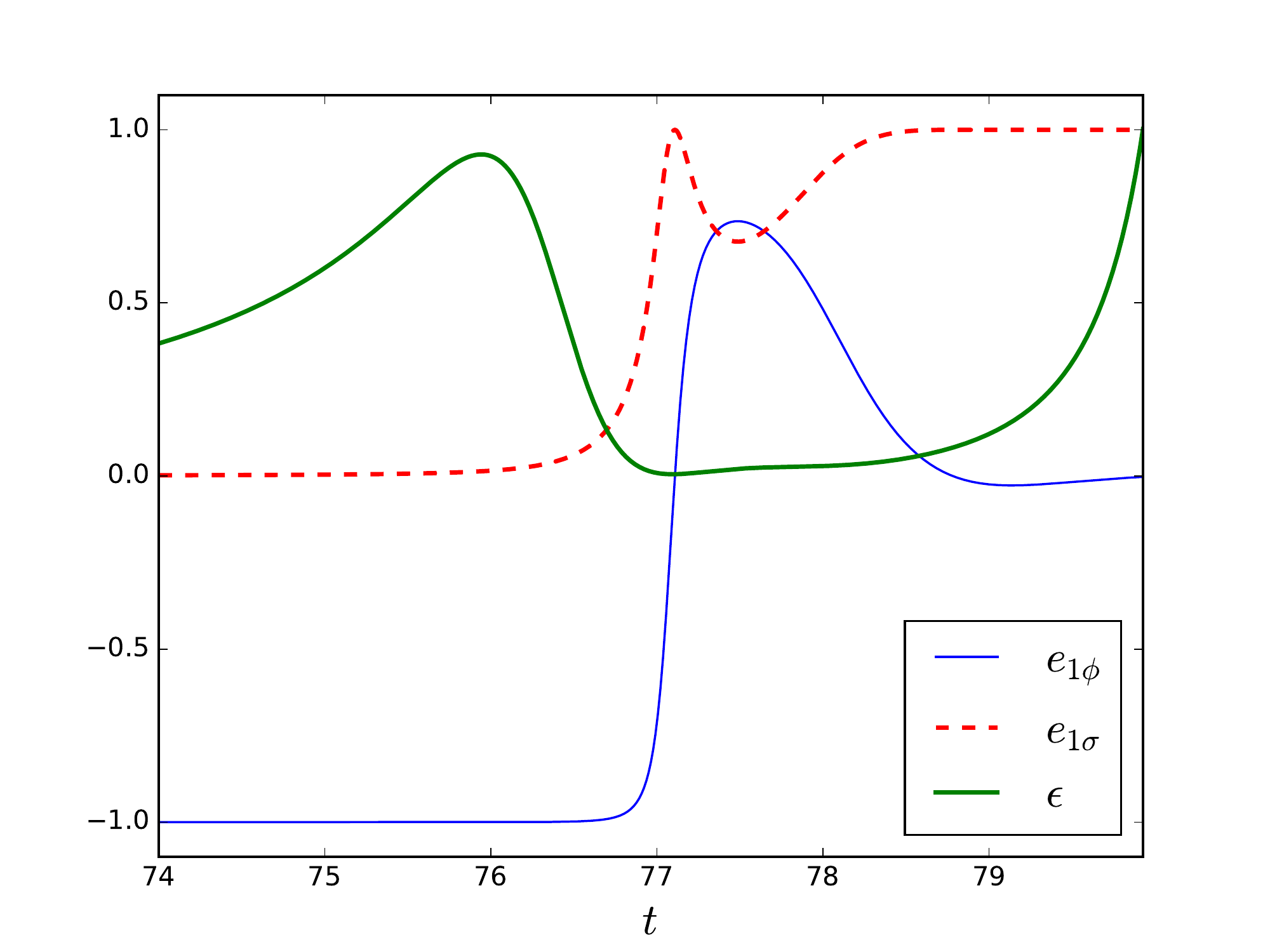}
	\includegraphics[width=0.49\textwidth]{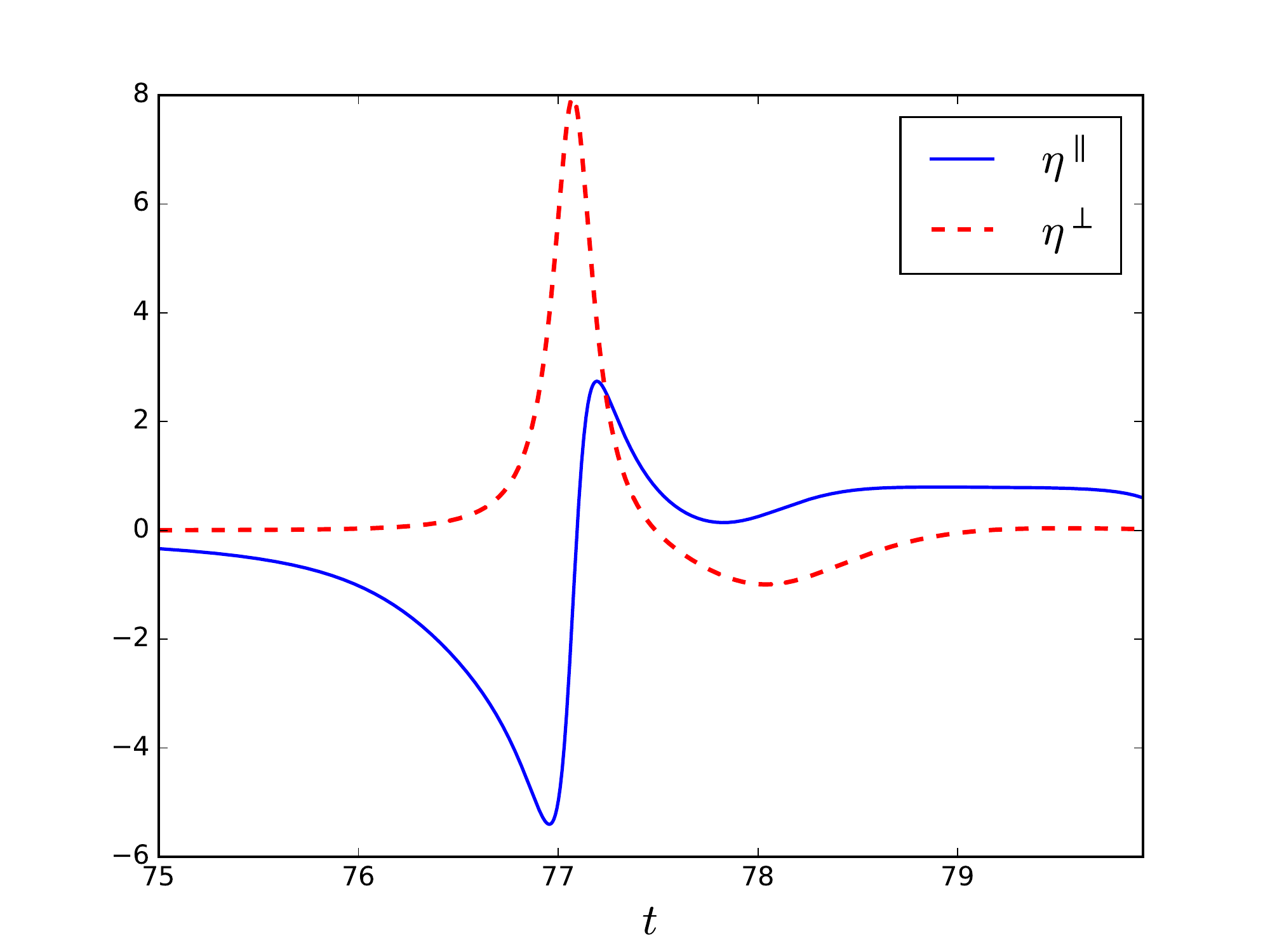}
	\includegraphics[width=0.49\textwidth]{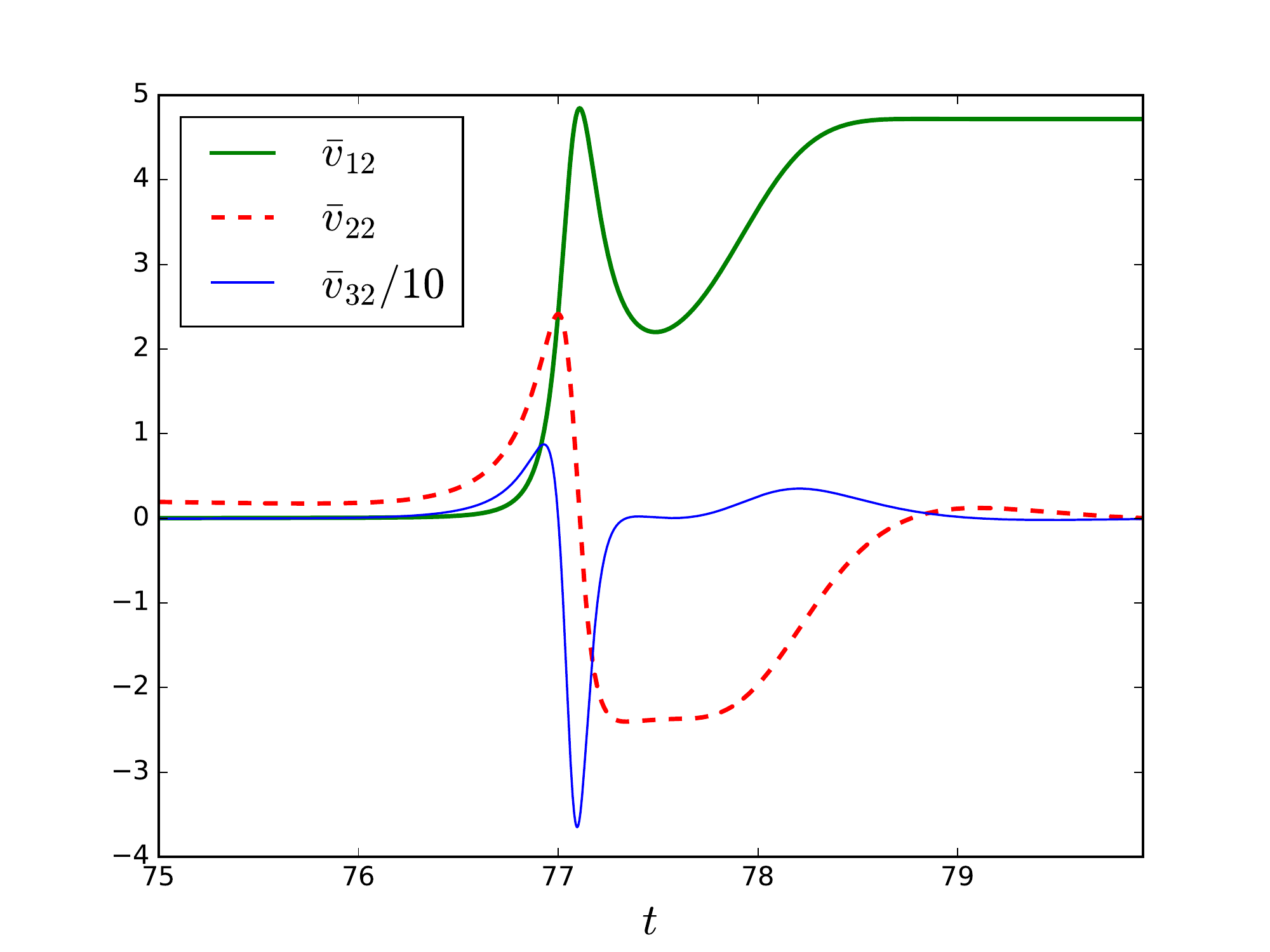}
	\includegraphics[width=0.49\textwidth]{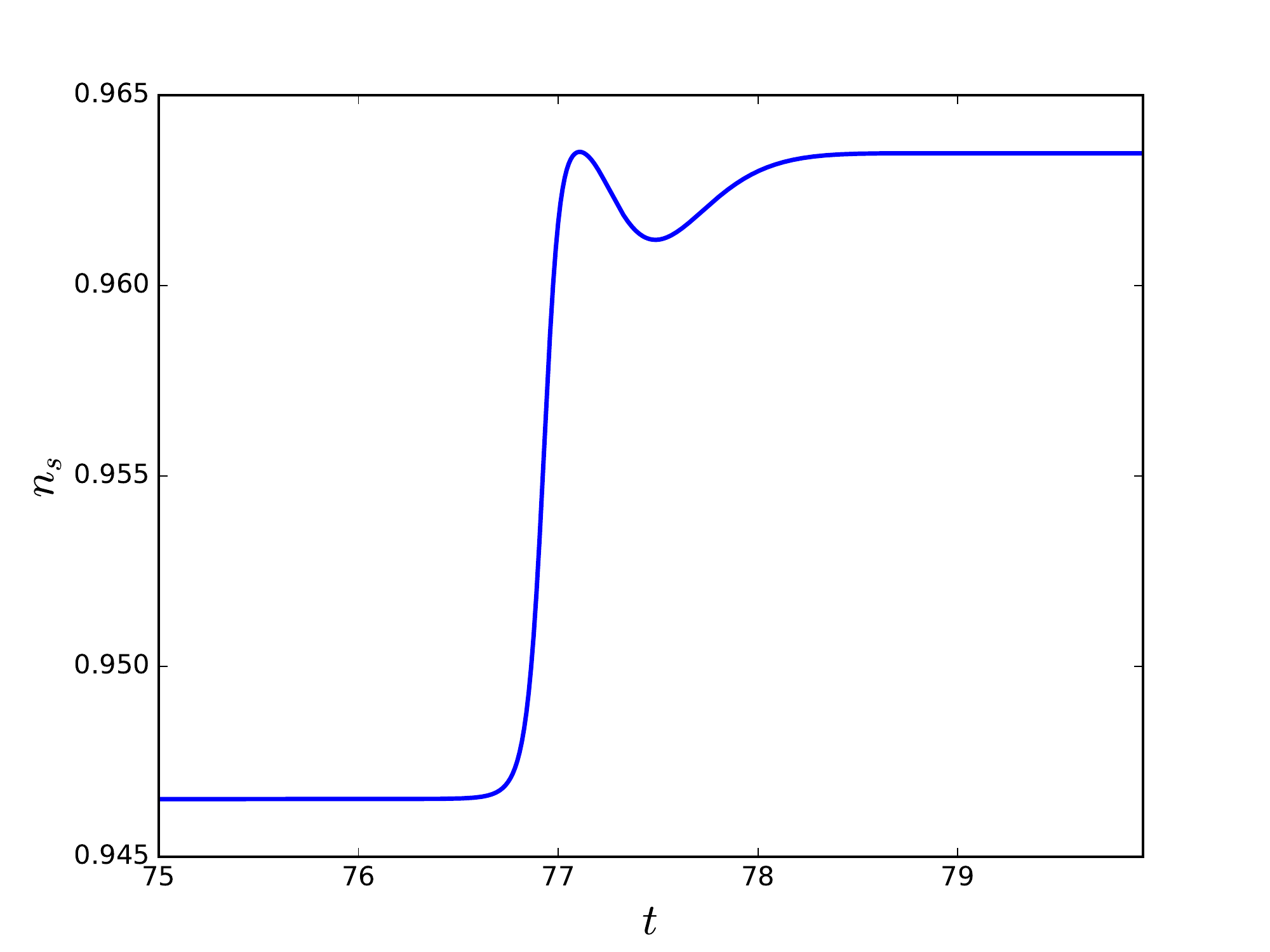}
	\includegraphics[width=0.49\textwidth]{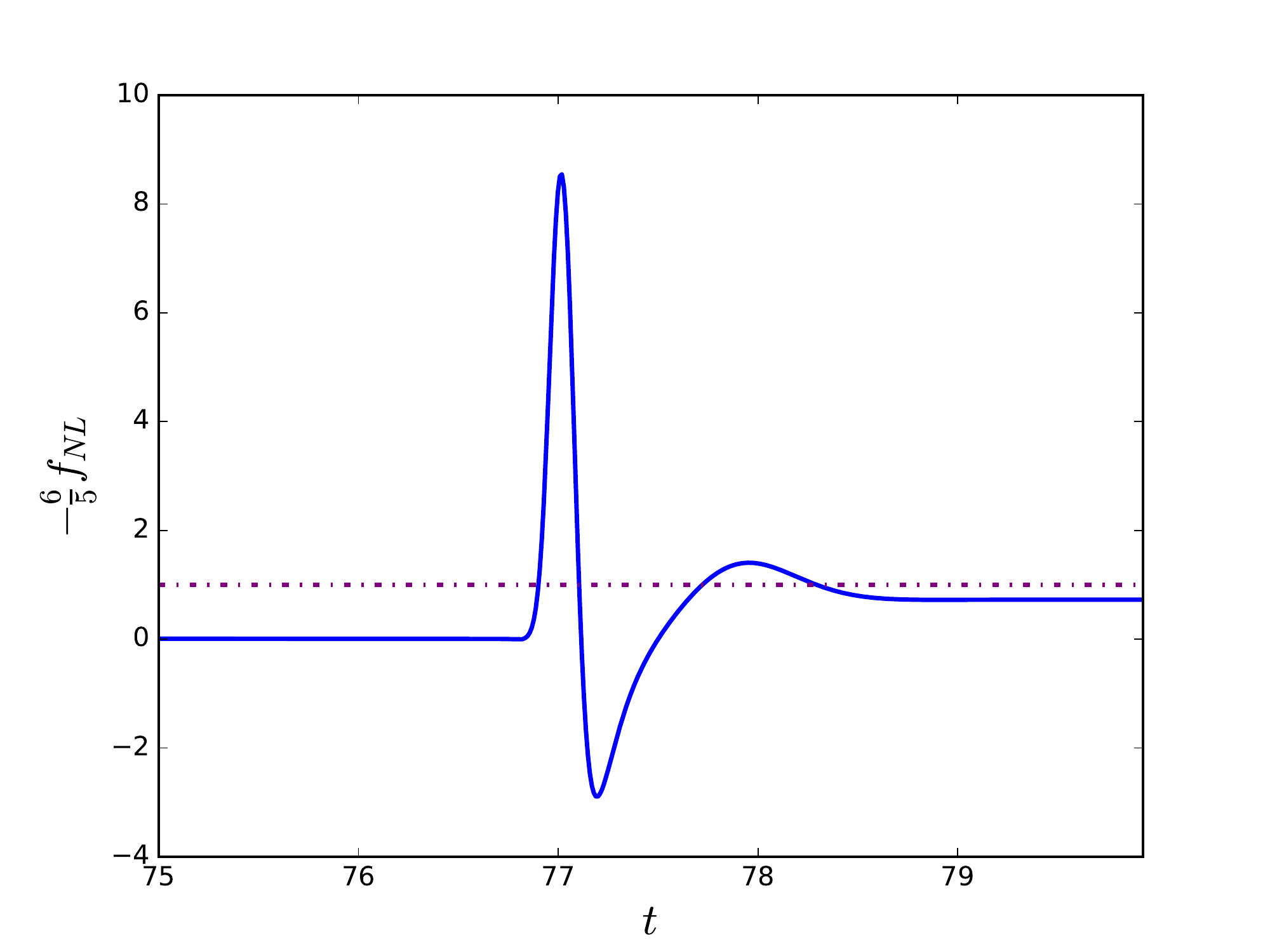}
    \includegraphics[width=0.49\textwidth]{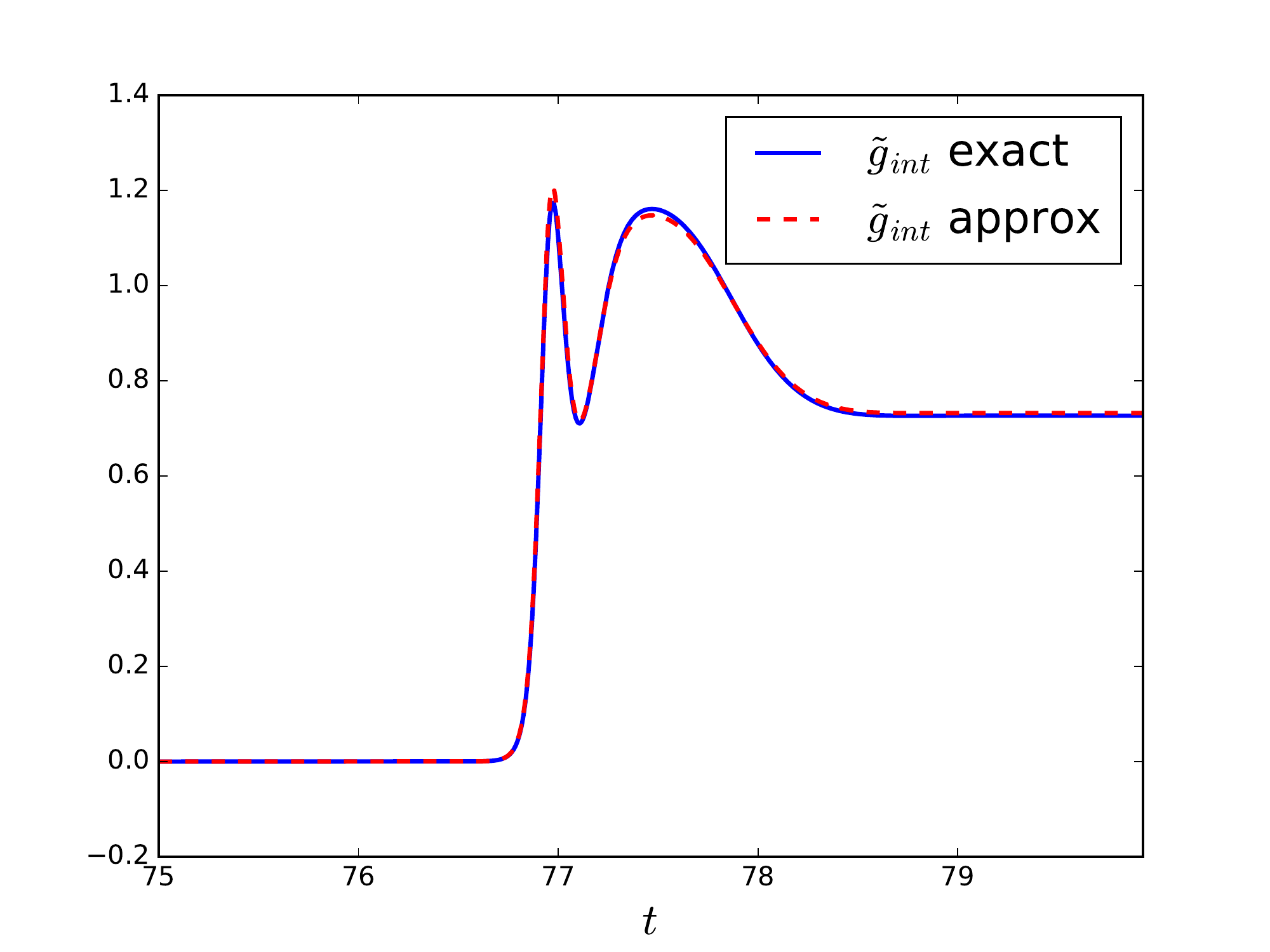}
	\caption{Same as figure \ref{fig:n2m2} but for the non-monomial potential \eqref{last} (without analytical predictions for $\ns$ and $\gv_{12e}$).} 
	\label{fig:last}
\end{figure}

Figure \ref{fig:last} shows a similar behaviour as for the previous example. Again $\fnl$ is smaller than its slow-roll prediction. The reason is still the same, the period of large $\ge$ makes $\gv_{12e}$ larger by a factor of order unity than in the slow-roll approximation and the direct consequence is that $\fnl$ is reduced by the same factor.

\section{Conclusion}
\label{conclusion}

In this article, we discussed the levels of non-Gaussianity produced
in two-field inflation with a sum potential\footnote{For comparison we also
looked at the case of a product potential in appendix~\ref{appendix:product}.
As was shown before, in that case one cannot get large non-Gaussianity at
all in the slow-roll approximation and with a vanishing isocurvature mode
at the end of inflation.}
$W(\phi,\sigma)=U(\phi)+V(\sigma)$ and standard kinetic terms. We
looked both at the case where the (strong) slow-roll approximation is valid
throughout inflation (meaning that all slow-roll parameters, even the
perpendicular ones, are small), and at the case where slow roll is broken 
during the turn of the field trajectory. An important assumption in our
models is that we impose that the isocurvature mode that is present
during inflation (and whose interaction with the adiabatic mode on
super-Hubble scales generates the non-Gaussianity) has disappeared by
the end of inflation. In that case the super-Hubble adiabatic mode is
constant after inflation and we can extrapolate the results at the end
of inflation directly to the time of recombination and observations of
the CMB without knowing any details about the evolution of the universe
in between. Without this assumption it would be much easier to
create large non-Gaussianity, simply by ending inflation in the middle
of the turn, but the result at the end of inflation would be
meaningless from the point of view of CMB observations without a
proper treatment of the transition at the end of inflation and the
consecutive period of (p)reheating.

We use the long-wavelength formalism for our computations of
non-Gaussianity.  In this formalism \cite{Tzavara:2010ge}, under the
assumption mentioned above, any large (meaning order unity or more, so
non slow roll suppressed) contribution to $\fnl$ can only come from an
integral expression called $g_\mathrm{int}$. The original formulation
of this expression contains an integral over Green's functions, which
depend on two different times, making it hard to interpret the
expression and see which types of potentials will lead to large
non-Gaussianity. In this paper we have found another expression for
$g_\mathrm{int}$, as the solution (\ref{solgint}) of a differential
equation (\ref{equadiff}), which can be written as the sum of a
homogeneous and a particular solution. This expression is very useful,
since for the homogeneous solution we have an exact analytic
expression that does not require any slow-roll approximation, while
for the particular solution we have seen (within the context of the
class of models studied) that we can either compute it explicitly, or
show that it is negligible.  We also derived some relations of
proportionality between the different Green's functions which are
useful in the computations.

We have highlighted the tension between a large $\fnl$ (of order unity
or more) and the current observational bounds on the spectral index
$\ns$, both being linked to the second derivative of the potential
$V_{\gs\gs}$, where $\sigma$ is the sub-dominant field at horizon
crossing and until the turn of the field trajectory.  We evaluated
these tensions (within the slow-roll approximation) for monomial
potentials, where it would otherwise be easy, with some fine-tuning,
to reach the requirements for a large $\fnl$.  We have shown that a
large part of the parameter space for $\fnl$ of order unity
is simply forbidden because of the constraints on $\ns$. However, we
found that these constraints are very sensitive to the value of $\ns$:
if the lower bound were only smaller by 0.02 ($\ns$ of order 0.94),
the situation would be completely different and most of the parameter
space would be allowed.\footnote{One might argue that we were rather strict
in using the $1\sigma$ error bars on $\ns$ and not the $2\sigma$ ones.
However, while that would change the exact numerical values, the
general conclusion would remain the same.} 
This analysis of the monomial potential also
revealed that the duration of inflation after horizon-crossing is
important: a value around fifty e-folds is much more constraining than
the usual sixty e-folds. This also indicates that in the rare working
models, the turn of the field trajectory occurs near the end of
inflation. This raises several issues, the main one being that at that
time, slow-roll parameters generally stop to be small compared to one
and the slow-roll approximation does not work anymore. Moreover, if
the turn occurs too close to the end of inflation, the isocurvature mode
may not have time to vanish. By studying turns where the slow-roll
parameter $\ge$ is still small compared to one we avoid this problem:
the time $\ge$ needs to increase to one and end inflation can give
enough time for the isocuvature mode to vanish.

The natural continuation of this study was to consider what would
happen if we abandoned the slow-roll approximation during the turn and
allowed the slow-roll parameters $\getpa$ and $\getpe$ to become large
there.  On the other hand, we still assume that $\ge$ remains small
during the turn, for several reasons: because of the issue regarding
the vanishing of the isocurvature mode mentioned
above, because we saw numerically in the models we looked at that this
was a good approximation, and because this approximation allowed us to
derive some very interesting analytical results (a potential period of
large $\ge$ right before the turn was taken into account though).  We
identified two different types of models where such a turn can happen, shown
in figure~\ref{fig:slowrollbroken}. Substituting the slow-roll
expression for $\dot{g}_\mathrm{int}$ into \eqref{equadiff}, we were able to
show (using simple comparisons of the different terms of the
differential equation) that it is also a very good approximation even
if the slow-roll parameters $\getpa$ and $\getpe$ become large during
the turn. The main idea is the one mentioned above: as long as the
slow-roll approximation is valid, we can compute the particular
solution explicitly, while when it is broken, we can show that the
particular solution becomes negligible, even though we cannot compute
an analytic expression for it in that case (the fact that $\ge$
remains small is a crucial ingredient in this proof). For the
homogeneous solution we have an analytic expression that is valid
everywhere. We were also able to show that adding the slow-roll
particular solution to the homogeneous solution in the regions where
the exact particular solution is negligible does not introduce a
significant error, which means that we do not have to perform an explicit
matching of the solutions at each transition between a slow-roll and a
non-slow-roll region.

This led us to the conclusion that, within the context of the models studied
and the assumptions mentioned above, the slow-roll expression for $\fnl$
is a very good approximation for the exact value, even in models where
$\getpa$ and $\getpe$ become large during the turn of the field trajectory
and break slow roll. Hence the implications of this expression for having
large non-Gaussianity, discussed in the context of the slow-roll approximation,
mostly apply to this case as well. In particular, the constraints due
to the spectral index $\ns$ remain very important.
A two-field sum potential with large $\fnl$ requires a lot of fine-tuning
(and we showed explicitly in the section with numerical examples how to
construct such a model). Reducing the error bars on the measurements of 
the spectral index could even shrink the parameter region of these models 
where $\fnl$ is of order unity more than reducing the error bars on $\fnl$.

\appendix
\section{Derivation of the $g_\mathrm{int}$ equation (\ref{equadiff})}
\label{appendix}

In this appendix we present the derivation of the differential equation
(\ref{equadiff}) for $\dot{g}_\mathrm{int}$.
A direct computation of the first, second, and third derivatives of
the definition of $g_\mathrm{int}$ in (\ref{defgint}) with respect to $t$ using \eqref{srderivatives} and \eqref{greent}
gives:
\be
\dot{g}_{\mathrm{int}}=  - 2 (\getpe)^2 (\gv_{22})^2 - (\ge + \getpa) \gv_{22}\gv_{32} - (\gv_{32})^2 + 2\getpe \int_{t_{*}}^t \d t' \,  \gv_{22} G_{23} \lh \Xi \gv_{22}+9 \getpe \gv_{32}\rh,
\label{derivgint}
\ee
\begin{align}
\ddot{g}_{\mathrm{int}} = & 2\lh\gxpe+\getpe (\ge -2 \getpa)\rh \int_{t_{*}}^t \d t' \,  \gv_{22} G_{23} \lh \Xi \gv_{22}+9 \getpe \gv_{32}\rh \nonumber\\ 
& + 2\getpe\int_{t_{*}}^t \d t' \,  \gv_{22} G_{33} \lh \Xi \gv_{22}+9 \getpe \gv_{32}\rh \nonumber\\
& + (\gv_{22})^2 \lh 3 (\ge +\getpa) \gc + 2 \ge^3 + 6\ge^2 \getpa + 4 \ge (\getpa)^2 +12 \getpa(\getpe)^2 + (\ge +\getpa) \gxpa - 4\getpe\gxpe \rh\nonumber\\
& + \gv_{22} \gv_{32} \lh3\ge + 3\getpa + 6\gc + 3\ge^2 +8\ge\getpa + 3(\getpa)^2 + 3 (\getpe)^2 + \gxpa\rh +(\gv_{32})^2 (6 +\ge+ 3 \getpa),
\end{align}
\begin{align}
\dddot{g}_{\mathrm{int}} = & -(3\getpe-\ge\getpe+6\getpa\getpe-2\gxpe) \int_{t_*}^t \d t' \,  \gv_{22} G_{33} \lh \Xi \gv_{22}+9 \getpe \gv_{32}\rh\nonumber\\
& + \lh 9\ge\getpe + 6\getpa\getpe - 6\getpe\gc -3 \gxpe - 3\gw_{211} + \ge^{2}\getpe - 8 \ge \getpa \getpe + 6 (\getpa)^2\getpe - 6 (\getpe)^3 \right.\nonumber\\
&\left. ~~~ - 4 \getpe \gxpa +(3\ge-2\getpa)\gxpe \rh \int_{t_{*}}^t \d t' \,  \gv_{22} G_{23} \lh \Xi \gv_{22}+9 \getpe \gv_{32}\rh\nonumber\\
& + (\gv_{22})^2 \lh32 \getpa \getpe \gxpe-60 (\getpa)^2 (\getpe)^2-36 \getpa (\getpe)^2-4 (\getpa)^2 \gxpa-3 \getpa \gxpa-12 (\getpa)^2 \gc -9 \getpa \gc \right.\nonumber \\
& + 6 (\getpe)^2 \gxpa  +12 \getpe \gxpe+6 (\getpe)^2 \gc +12 (\getpe)^4-6 \gxpa \gc -4 (\gxpe)^{2}-18 \gc ^2-3 \getpa \gw_{111}-3 \ge \gw_{111} \nonumber   \\
& +9 \getpe \gw_{211}+3 \getpa \gw_{221} +3 \ge \gw_{221}  - 3 \getpe \gw_{222}+\getpa \gxpa \ge -33 \getpa \gc  \ge +14 \getpa \ge ^3-4 (\getpa)^2 \ge ^2-6 \getpa \ge ^2 \nonumber  \\
&\left. -12 (\getpa)^3 \ge -8 \getpe \gxpe \ge -12 (\getpe)^2 \ge ^2-36 (\getpe)^2 \ge +5 \gxpa \ge ^2-3 \gxpa \ge -9 \gc  \ge ^2-9 \gc  \ge +6 \ge ^4-6 \ge ^3\rh \nonumber\\
& + \gv_{22}\gv_{32}\lh-12 \getpa (\getpe)^2 - 24 \getpa \gc -12 (\getpa)^3 -21 (\getpa)^2-9 \getpa+4 \getpe \gxpe-15 (\getpe)^2-9 \gxpa - 54 \gc \right. \nonumber\\
& \left. -3 \gw_{111} +6 \gw_{221}+14 \getpa \ge ^2-21 (\getpa)^2 \ge -57 \getpa \ge +3 (\getpe)^2 \ge +9 \gxpa \ge +6 \gc  \ge + 9 \ge ^3 - 30 \ge ^2 - 9 \ge \rh \nonumber\\
& + (\gv_{32})^2 \lh-12 (\getpa)^2-39 \getpa+6 (\getpe)^2+4 \gxpa+6 \gc +3 \getpa \ge +3 \ge ^2-15 \ge -36\rh.
\end{align}
Taking the specific combination of the three expressions above that eliminates
all the terms with integrals then gives the differential equation 
(\ref{equadiff}), with $K_{22}, K_{23}, K_{33}$ given by
\be
\begin{split}
K_{22} = & -18 (\getpe)^2 \gc ^2 + 2 (\getpa)^2 (\getpe)^2 \gxpa - 6 \getpa \getpe \gxpe \gc + 6 (\getpa)^2 (\getpe)^2 \gc- 6 (\getpe)^2 \gc \gxpa -2 (\getpe)^4 \gxpa  \\
& - 6 (\getpe)^4 \gc - 18\ge\getpa(\getpe)^2 \gc  + 12\ge^2\getpa (\getpe)^2+ 12 \ge(\getpa)^2 (\getpe)^2  -6 \ge\getpe \gxpe \gc - 12\ge^2 (\getpe)^2 \gc \\
& - 12 \ge(\getpe)^4 - 3 \getpa (\getpe)^2 \gw_{111}-3 \ge(\getpe)^2 \gw_{111}  +3 (\getpe)^3 \gw_{211}+3 \getpa (\getpe)^2 \gw_{221} \\
& +3\ge (\getpe)^2 \gw_{221} -3 (\getpe)^3 \gw_{222} -2 \getpa \getpe \gxpa \gxpe + 6\ge\getpa (\getpe)^2 \gxpa  - 12\ge ^2\getpa \getpe \gxpe \\
& + 20\ge^3 \getpa (\getpe)^2 + 28 \ge^2(\getpa)^2 (\getpe)^2   - 12\ge \getpa (\getpe)^4  +12\ge(\getpa)^3 (\getpe)^2   -2 \ge\getpe \gxpa \gxpe  \\
& - 8\ge(\getpa)^2 \getpe \gxpe + 4\ge^2 (\getpe)^2 \gxpa  - 4 \ge^3\getpe \gxpe -4 \ge(\getpe)^3 \gxpe    + 4\ge^4(\getpe)^2  - 12 \ge^2(\getpe)^4,
\end{split}
\nonumber
\ee
\be
\begin{split}
K_{23} = &-36(\getpe)^2\gc - 6\ge\getpa(\getpe)^2 - 12\ge^2(\getpe)^2 - 6(\getpe)^4 -6 \ge (\getpe)^2 \gc + 6(\getpa)^2(\getpe)^2  - 6 \getpa \getpe \gxpe    \\ 
& + 6 \getpa (\getpe)^2 \gc - 6 (\getpe)^2 \gxpa  -6\ge \getpe \gxpe - 12 \getpe \gc \gxpe - 3(\getpe)^2 \gw_{111}-3 \getpa \getpe \gw_{211} \\
& -3 \getpe \ge \gw_{211}  +6 (\getpe)^2 \gw_{221} -2\getpe\gxpa \gxpe  -2 \getpa (\gxpe)^2 + 2 (\getpa)^2 \getpe \gxpe + 2 \getpa (\getpe)^2 \gxpa   \\
& -2 (\getpe)^3 \gxpe -8 \ge\getpa\getpe\gxpe  + 18\ge(\getpa)^2(\getpe)^2 + 24\ge^2\getpa (\getpe)^2   + 4\ge(\getpe)^2\gxpa - 6 \ge^2\getpe\gxpe \\
&  - 6 (\getpe)^4 \ge+ 6\ge^3(\getpe)^2 -2 \ge(\gxpe)^{2}  ,
\end{split}
\nonumber
\ee
\be
\begin{split}
K_{33} = &-18 (\getpe)^2 -6 \ge(\getpe)^2 + 6 \getpa (\getpe)^2 -12 \getpe \gxpe - 3 \getpe \gw_{211}  +6 \ge \getpa (\getpe)^2  +2 \ge ^2 (\getpe)^2  \\
& + 2 \getpa \getpe \gxpe -2 \ge\getpe \gxpe  -2 (\gxpe)^{2} .
\end{split}
\label{defKxy}
\ee

To obtain \eqref{equadiffsr}, the slow-roll approximation of \eqref{equadiff}, several steps have to be followed. First, on the right-hand side of the equation, one can use \eqref{greensr2} to eliminate $\gv_{32}$. Then one sees that the lowest-order terms (the first of each $K$ in \eqref{defKxy}) cancel each other. The remaining terms are one or two orders higher than the ones which cancel, so that in the leading-order slow-roll aproximation we only have to keep those one order higher. On the left-hand side of the equation, we also use the fact that a time derivative adds an order in slow roll, so that $\dddot{g}_{\mathrm{int}}$ is one order higher in slow-rol than $\ddot{g}_{\mathrm{int}}$. Hence, we see that the $\dddot{g}_\mathrm{int}$ term disappears completely from the equation. Finally, it is possible to substitute the second line of \eqref{sreta} into the two sides of \eqref{equadiff} to eliminate $\gw_{111}$ and $\gw_{211}$, and after simplifying the common factor $3\getpe$ the result is given in \eqref{equadiffsr}.

\section{Product potential}
\label{appendix:product}

In this section, we study the case of product potentials, which take the form $W(\gf,\gs)=U(\gf)V(\gs)$. This case was solved analytically in \cite{Choi:2007su}. Here, we show that the slow-roll version of the $\gint$ equation \eqref{equadiffsr} takes a simple and nice form which is easy to deal with.

As for the sum-separable case, we start by using the specific form of the potential to find some new relations concerning its derivatives without assuming any approximation. A simple one is $W_{\gf\gs}=\frac{W_{\gf}W_{\gs}}{W}$ which links the second-order mixed derivative of the potential to the first-order ones. Then using the field equation \eqref{fieldeq} and the definitions of $\ge$, $\getpa$ and $\getpe$ given in \eqref{defeps} and \eqref{defeta}, this relation can be rewritten in terms of slow-roll parameters:
\begin{equation}
(3-\ge)\gw_{\gf\gs}=\frac{2}{3}\ge \left[ e_{1\gf}e_{1\gs}\lh(\getpa+3)^2-(\getpe)^2\rh+\getpe(\getpa+3)(e_{1\gf}^{2}-e_{1\gs}^{2})\right].
\end{equation}
We also need the generalized version of \eqref{W11eq}, valid for any two-field potential, which is:
\be
e_{1\gf}e_{1\gs}(\gw_{11}-\gw_{22})=(e_{1\gf}^{2}-e_{1\gs}^{2})\gw_{21} + \gw_{\gf\gs}.
\ee 
Combining the two previous equations and using \eqref{srpareq}, we obtain:
\begin{equation}
\begin{split}
&e_{1\gf}e_{1\gs}\left[-3\gc -\gxpa +\ge\gc-2\ge^2-4\ge\getpa+ \frac{\ge}{3}\lh \gxpa - 2(\getpa)^2 + 2(\getpe)^2\rh\right]\\
&=(e_{1\gf}^{2}-e_{1\gs}^{2})\left[3\getpe+\gxpe+\ge\getpe-\frac{\ge}{3}\lh \gxpe +2\getpa\getpe\rh\right].
\end{split}
\label{product1}
\end{equation}
Similar computations can be done for the third-order derivatives $W_{\gf\gf\gs}=\frac{W_{\gf\gf}W_{\gs}}{W}$ and $W_{\gf\gs\gs}=\frac{W_{\gs\gs}W_{\gf}}{W}$ to show that:
\begin{equation}
\begin{split}
&(3-\ge)\gw_{\gf\gf\gs}=-2\ge\left[(\getpa+3)e_{1\gs}-\getpe e_{1\gf}\right]\gw_{\gf\gf},\\
&(3-\ge)\gw_{\gf\gs\gs}=-2\ge\left[(\getpa+3)e_{1\gf}+\getpe e_{1\gs}\right]\gw_{\gs\gs}.
\end{split}
\label{product2}
\end{equation}
Finally, using the definitions of $\gw_{221}$ and $\gw_{222}$ in terms of third-order derivatives and basis components, substituting them into \eqref{product1} and \eqref{product2} and performing a first-order expansion in terms of slow-roll parameters gives:
\begin{equation}
\begin{split}
\gw_{221} = &-\ge\getpa -\ge\gc +(\getpa)^2 -2\getpa\gc +\gc^2 +(\getpe)^2-\gxpa +\frac{\gc}{\getpe}\gxpe,\\
\gw_{222} = &-\gxpe -\getpe(\ge-2\getpa+2\gc) \\
&-\frac{\gc}{\eta^\perp}\lh-2\ge^2-3\ge\getpa+(\getpa)^2-\gxpa - \ge \gc -2\getpa \gc +\gc^2\rh - \lh\frac{\gc}{\getpe}\rh^2\gxpe.
\end{split}
\end{equation}

These equations can then be used to simplify the right-hand side of \eqref{equadiffsr}, and one easily finds that in fact the right-hand side completely vanishes. Hence, the slow-roll solution consists only of the homogeneous solution and using the initial condition $\dot{g}_{\mathrm{int}*}=-2(\getpe_*)^2 + (\ge_*+\getpa_*-\gc_*)\gc_*$ (from the slow-roll approximation of \eqref{derivgint}) we find:
\begin{equation}
\gint=-\left[\getpe_{*} - \frac{1}{2\getpe_{*}}(\ge_{*}+\getpa_{*}-\gc_{*})\gc_{*}\right]\gv_{12} 
=- \frac{e_{1\gf*}^2\gV_{\gs\gs*} - e_{1\gs*}^2\gU_{\gf\gf*}}{2e_{1\gf*}e_{1\gs*}} \gv_{12}.
\end{equation}
The most important thing to note here is that the second expression has exactly the same form as the homogeneous part of the sum potential case in \eqref{gint}, without the particular solution. As discussed in section \ref{Sum Potential Section}, it is that term which can give a large contribution to $\fnl$. The natural question is then if the situation is the same for the product potential. The similarity of the expressions makes it possible to use exactly the same method to answer this question as for the treatment of the sum potential. 

First, we define:
\be
\tilde{g}_{\mathrm{int}}=\frac{-2(\gv_{12})^2}{(1+(\gv_{12})^2)^2} \, \gint = \frac{e_{1\gf*}^2\gV_{\gs\gs*} - e_{1\gs*}^2\gU_{\gf\gf*}}{e_{1\gf*}e_{1\gs*}} \frac{(\gv_{12})^3}{(1+(\gv_{12})^2)^2},
\label{gint product}
\ee 
which is the entire term depending on $\gint$ in $\fnl$ \eqref{fNLexpression}. As for the sum potential (see section \ref{Sum Potential Section}), the only possibility of having this expression larger than order slow-roll is to have one field dominating at horizon crossing: $e_{1\gf*}^2 \approx 1 \gg e_{1\gs*}^2$. But at the same time, it is required that $\gv_{12}$ is at least of order unity (and at least four to obtain the largest $\fnl$). The main difference with the sum potential case comes in fact from the expression for $\gv_{12}$. In the slow-roll approximation, it is possible to solve the Green's function equations. The computation is similar to the sum potential case and is detailed in \cite{Tzavara:2010ge} where it is shown that:
\be
\gv_{12}=\frac{S-S_*}{2e_{1\gs*}e_{1\gf*}},\qquad \gv_{22}=\frac{e_{1\gf}e_{1\gs}}{e_{1\gf*}e_{1\gs*}},
\ee
with $S\equiv e_{1\gf}^2 - e_{1\gs}^2$. These expressions are quite different from \eqref{Green srsolution} for the sum potential.

At horizon crossing, $e_{1\gf*}^2 \approx 1$ meaning $S_* \approx 1$. For the value of $S$ at the end of inflation there are two different situations. As discussed several times in this paper, we want that $\gv_{22}$ goes to zero at the end of inflation to get rid of the isocurvature mode, meaning that the situation is far closer to single-field inflation at the end of inflation than at horizon crossing. Hence, if at the end $\gf$ also dominates (same direction of the field trajectory), $|e_{1\gs}|\ll |e_{1\gs*}|$. This means that $S-S_*\approx e_{1\gs*}^2$, which leads to the fact that $\gv_{12}$ is small compared to 1. In that case $\gint$ cannot give a large $\fnl$. However, if $\gs$ dominates at the end of inflation (different direction of the field trajectory), we have:
\be
\gv_{12}=\frac{-1}{e_{1\gf*}e_{1\gs*}},
\ee
which is large compared to 1. We can then use that $\frac{(\gv_{12})^3}{(1+(\gv_{12})^2)^2}\approx \frac{1}{\gv_{12}}$ if $|\gv_{12}|\gg 1$ and \eqref{gint product} to write:
\be
\tilde{g}_{\mathrm{int}} \approx  \frac{e_{1\gf*}^2\gV_{\gs\gs*} - e_{1\gs*}^2\gU_{\gf\gf*}}{e_{1\gf*}e_{1\gs*}} \times \frac{e_{1\gf*}e_{1\gs*}}{-1} \approx -\gV_{\gs\gs*},
\ee
which is of order slow roll. Hence also in this case $\fnl$ is small. This is in agreement with the known conclusion that a product potential cannot give a large $\fnl$ in the slow-roll approximation with vanishing isocurvature mode at the end of inflation \cite{Byrnes:2008wi, Tzavara:2010ge}.

\section{Influence of the choice of the value of $\fnl$ on the start of a turn of the second type}
\label{appendix case 2}

In the second type of turn, we supposed that at horizon-crossing the situation was so close to single-field inflation that it is still quasi single-field at the time when $\ge$ becomes large and the slow-roll approximation breaks down for the first time. That time would have been the end of inflation in a purely single-field situation, the only difference being that here the potential goes to $W=V_*$ instead of $W=0$. As we have seen, $\ge$ becomes small again soon after, but the remaining question is how much time is needed to break the quasi single-field situation and have the turn start? In other words, how much time is there typically between the moment when $\ge$ becomes small again and the start of the turn? We will see that is in particular related to the value of $\fnl$.

First, for this argument we do not need to know the potential $U$ as it is already supposed to be almost zero because $\gf$ is near the minimum. For $\gs$, we will keep the simple monomial potential $V$ \eqref{potential V} because, as already discussed, it is quite general when seen as an expansion in terms of $\gs$ around its local extremum. As discussed, $\gf$ is near the minimum of its potential while $\gs$ has not evolved much since horizon-crossing, hence $W$ is simply $C$. Moreover, $\ge$ decreases again just before the turn, meaning that the slow-roll regime is back (see figure \ref{fig:slowrollbroken}), with only one exception: $\getpa$ can be large (order unity or more). However, this concerns only the field $\gf$ and the only possible effect on $\gs$ is through $\ge$ which is small. Then, the field equation for $\gs$ is:
\be
\frac{\d \gs}{\gs^{m-1}}=-\frac{m \gb}{C} \d t.
\ee

For $m=2$, we have:
\be
\gs = \gs_1 \exp{\left[ -\frac{2\gb}{C} (t-t_1)\right]},\qquad \ge\approx \frac{1}{2} \dot{\gs}^2 = 2 \lh \frac{\gs_1 \gb}{C}\rh^2 \exp{\left[-\frac{4\gb}{C} (t-t_1)\right]},
\ee
where $t_1$ is the time when $\ge$ has become small compared to one again. The factor ${-\textstyle\frac{2\gb}{C}}$ in the exponential is in fact $-{\textstyle\frac{6}{5}}\fnl$ in slow-roll \eqref{fnllimit}. As we supposed that it is of the same order as the real value of $\fnl$, it is of order unity and positive because of the form of the potential. The parameter $\ge$ increases exponentially and unless the initial value $\gs_*$ is ridiculously small compared to one, a very small number of e-folds after $t_1$ will be needed to reach $\ge=1$, which is the end of inflation. However, in general the slow-roll regime will be broken again before that time. But here we are interested in the time when the turn starts, that is to say when $\dot{\gf}$ and $\dot{\gs}$ are of the same order. As $\half\dot{\gf}^2\ll 1$ (because $\ge\ll 1$ and $\gf$ dominates at $t_1$), this will occur before the end of inflation when $\ge=1$. Hence, this period of quasi single-field inflation between $t_1$ and the start of the turn will only last a very few e-folds at most, because $\fnl$ is of order unity.

For $m \neq 2$, the solution is
\be
\gs = \gs_1 \left[ 1 -  \frac{(2-m)m\gb}{C \gs_1^{2-m}} (t-t_1)\right]^\frac{1}{2-m}.
\ee
Then, we obtain:
\be
\ge= \frac{1}{2} \lh\frac{m \gb}{C}\rh^2 \gs_1^{2m-2}\left[1-\frac{(2-m)m\gb}{C \gs_1^{2-m}} (t-t_1)\right]^{\frac{2m-2}{2-m}}.
\ee 
When $m>2$ and $m-2$ not small compared to 1, the end of inflation will be reached when ${\textstyle\frac{(2-m)m\gb}{C \gs_1^{2-m}}}(t-t_1) \approx 1$ which implies that 
\be
\label{epsilon}
t_e-t_1=\frac{C \gs_1^{2-m}}{(2-m)m \gb}= \lh\frac{V_{\gs\gs*}}{V_*}\rh^{-1}\lh\frac{\gs_1}{\gs_*}\rh^{2-m}\frac{(m-1)}{(m-2)}.
\ee
The first factor is the inverse of $-{\textstyle\frac{6}{5}}\fnl$ in slow-roll, hence this is of order unity. The two ratios $\gs_1/\gs_*$ and $(m-1)/(m-2)$ are also of order unity. Hence, $t_e-t_1$ is small and is of order one.
When $m$ is close to 2, we do an expansion at first order using the small parameter $m-2$ to obtain:
\be
\begin{split}
\ge&=\frac{1}{2} \lh\frac{m \gb}{C}\rh^2 \gs_1^{2m-2}\exp\left[-2\frac{m(m-1)\gb\gs^{m-2}}{C}\lh\frac{\gs_1}{\gs_*}\rh^{m-2}(t-t_1)\right]\\
&=\frac{1}{2} \lh\frac{m \gb}{C}\rh^2 \gs_1^{2m-2}\exp\left[-2\lh\frac{V_{\gs\gs*}}{V_*}\rh \lh \frac{\gs_1}{\gs_*} \rh^{m-2}(t-t_1)\right].
\end{split}
\ee
This gives back the formula for the $m=2$ case. Again, we can see the factor $-{\textstyle\frac{V_{\gs\gs*}}{V_*}}$ which is the slow-roll expression for $-{\textstyle\frac{6}{5}}\fnl$, while the whole expression in the exponential is positive because of the sign of $\gb$. It is of order one, for the same reason as in the other cases. Hence, in a matter of a few e-folds $\ge$ is large enough to say that $\dot{\gs}$ is at least of the same order as $\dot{\gf}$ and that the turn has started.

A last important remark is that if $\fnl$ is too large (more than order unity), inflation will end even faster. One has to verify that the turn has enough time to finish so that the isocurvature mode can vanish before the end of inflation.

\bibliographystyle{JHEP}
\bibliography{biblio}

\end{document}